\documentclass{elsart}
\usepackage{ifpdf}
\usepackage{epsfig,graphicx,amssymb,lineno}

\def\etal{{et~al.}}
\def\eg{{e.g.}}
\def\ie{{i.e.}}
\def\pdif#1#2{\mathchoice{\partial{#1}\over\partial{#2}}
{\partial{#1}/\partial{#2}}{\partial{#1}/\partial{#2}}
{\partial{#1}/\partial{#2}}}
\newcommand{\bv}[1]{\mbox{$\bf #1$}}
\mathcode`\|="8000      
{\catcode`\|=\active \gdef|#1{_{\rm #1}}}

\begin{document}

\begin{frontmatter}
\title{A new Godunov-type numerical code with a non-linear Riemann solver for equations of relativistic hydrodynamics}

\author{Pavlo V. Tytarenko},
\address{Astronomical Observatory of Kyiv Shevchenko University, 3
Observatorna St., Kiev, 04053, Ukraine}

\ead{pvtyt@gala.net, pvtyt@observ.univ.kiev.ua}

\author{Iurii A. Karpenko},
\ead{karpenko@bitp.kiev.ua}
\author{Yury M. Sinyukov}

\address{Bogolyubov Institute for Theoretical Physics,
14-b Metrolohichna St., Kiev, 03680, Ukraine}

\ead{sinyukov@bitp.kiev.ua}

\label{firstpage}
\maketitle

\begin{abstract}
We present a second-order upwind numerical scheme for equations of relativistic hydrodynamics with a source term. A new non-linear Riemann solver is constructed. Solution of a Riemann problem on a cells boundary is based on exact relations in case of zero tangential velocities. In this sense our solver is "exact". In case of non-zero tangential velocities a reasonable approximation is made that allows to avoid solution of very complicated exact equations. The scheme is tested on several one- and three-dimensional solutions demonstrating a good performance for both strong shocks and strong rarefactions.
\end{abstract}

\end{frontmatter}

\section{Introduction}

Hydrodynamic flows with relativistic velocities are essential feature of modern astrophysics and physics of
ultra-relativistic mucleus-nucleus collisions. Description of such objects as active galactic nuclei, quasars and microquasars, supernova collapse and explosion, gamma ray bursts, pulsar winds, extragalactic jets is impossible without using of relativistic hydrodynamics (see \eg{} Refs. \cite{krolik,woosley,lewinetal95}). Also relativistic hydrodynamics is widely used for description of multiparticle production in high energy nucleus-nucleon collisions in the way firstly proposed by Landau \cite{landau}. Since then 3D hydrodynamics is applied to study the hadronic spectra and momentum correlations, radial and elliptic flows in relativistic processes of nuclear collisions (see, e.g. review by Kolb \& Heinz \cite{kolbheinz}). Diversity of complex micro- and astro- physical relativistic processes requires development of reliable and universal  (\ie{} applicable to all range of initial thermodynamic conditions and velocity distributions) numerical methods for equations of relativistic hydrodynamics

Relativistic hydrodynamics numerical calculations have already a rich history. The first methods of numerical treatment of equations of relativistic hydrodynamics were based on introduction into equations an artificial viscosity \cite{wilson}. This allows to make initially hyperbolic equations of hydrodynamics parabolic and as a result discontinuous features like shock waves become continuous. But in such methods viscosity should be chosen separately for every particular task introducing thus an element of art in science and making corresponding code difficult for application. Another difficulties are loosing energy due to viscosity and problems with ultrarelativistic regime (see Ref. \eg{} \cite{centrellawilson}). And finally there is a danger of losing causality when one passes to the relativistic equations of parabolic type \cite{muller,israelstewart}.

Eventually artificial viscosity methods were replaced by more advanced ones. One of well-known and successful methods is SPH (Smooth Particle Hydrodynamics), see Refs. \cite{lucy,gingoldmonaghan}. The obvious disadvantages of SPH are the use of Lagrangian coordinates, problems with resolution at low densities and and still need for an artificial viscosity at high Mach numbers. Another class is HRSC (High Resolution Shock Capturing) schemes based on Godunov method. Our method is an HRSC one. The main idea of Godunov method is introduction of Riemann problems - ones of the simplest possible initial value problems for hyperbolic systems - at cell boundaries in order to compute fluxes. But complete solution of the Riemann problem is a non trivial task. Thus many methods have been developed in order to avoid solution of the Riemann problem. A comprehensive review can be found in Ref. \cite{martimuller03}.

The most obvious way of simplifying the task is a proper linearization of initially non-linear system. Roe-type solvers (following the original work by Roe \cite{roe}) reduce the nonlinear system of hydrodynamics equations to a locally linear system where the linear Jacobian matrix is chosen in such a way that a non-linear shock is a single discontinuity for the linear system too. Generalization of the method to equations of relativistic hydrodynamics was made in Ref. \cite{eulderinkmellema}. Another method of local linearization based on using primitive variables instead of conservative ones and applying either HLLE-type or Roe-type solver was proposed in Refs. \cite{falle,fallekomissarov}. Though linearized methods are numerically efficient and robust, they can generate unphysical states with negative energy density. It means that additional efforts should be made to handle strong enough rarefactions (see Ref. \cite{fallekomissarov}).

HLLE method \cite{hartenetal83,einfeldt} is based on approximate handling of the Riemann problem by introduction of an intermediate state and two waves propagating with some signal velocities. Extension of the method to relativistic hydrodynamics was made in Refs. \cite{schneideretal93,duncanhughes}. Quality of the scheme depends on choice of signal velocities thus making it a bit of art. The scheme also does not take into account presence of a contact discontinuity in real solution of the Riemann problem. It leads to a bad resolution of contact discontinuities in particular and to high numerical diffusion in general. This problem is partially lifted in recently developed HLLC numerical scheme \cite{toroetal94,mignonebodo} where contact discontinuity is present in solution of the Riemann problem. HLLE method was applied for relativistic heavy ion collisions by Rischke \etal{} \cite{rischke}. A higher order extension of HLLE metod, PPM (Piecewise Parabolic Method) was introduced in Ref. \cite{colella} and used for description of ultra-relativistic A+A collisionsby \cite{hirano,hiranoetal}.

Another approach to approximate solution of the Riemann problem is two-shock approximation \cite{balsara,daiwoodward,mignoneetal05}. This method ignores presence of rarefactions and take solution of the Riemann problem consisting only of shock waves. Such approach works quite well for shock configurations but starts having problems for configurations with strong enough rarefactions due to violation of entropy condition. Additional work like making entropy fixes should be done to repair this.

Finally, an approach that does not consider the Riemann problem at all, is also possible. This is flux splitting method (see \eg{} Ref. \cite{donatetal98}). The method splits the flux into right going and left going parts in a way that satisfies conservation laws and conforms with the wave structure of the flow. One of the best realizations of this method is Marquina flux \cite{marquina}. Marquina-type flux is the base of the code GENESIS \cite{aloyetal99}.

Another problem in constructing a numerical code is the choice of accuracy. Accuracy is achieved by proper reconstruction of quantities inside a numerical cell or a proper reconstruction of fluxes if a flux splitting method is used. There are high order accuracy ENO (essentially non-oscillatory) \cite{hartenetal87,dolezalwong,delzannabucc} and WENO (weighted ENO) schemes \cite{liuetal94,jiangshu}. The recent achievements here are the RAM code \cite{zhangmacfadyen} and the WHAM code \cite{tchekhovskoyetal} both of 5-th order accuracy. It is thought that high accuracy helps in simulating flows with different energy scales.

We use a second order accuracy scheme by Falle \& Komissarov \cite{falle,fallekomissarov} with a non-linear Riemann solver. This solver is constructed in such a way that it is an approximate solver in case of non-zero tangential velocities and an exact solver in case of zero ones (including one-dimensional flows). This gives us an advantage of equally good treating of difficult situations such as ultrarelativistic flows, strong shocks and rarefactions. In our general scheme we incorporate the source term in right hand side of hydrodynamics equations aiming utilization of the code for description of the physical processes where such a source is required. It concerns, in particular, the matter evolution in ultra-relativistic nucleus-nucleus collisions.Our paper is organized as follows.

In Section 2 the basic equations of relativistic hydrodynamics and their numerical treatment are presented.
In Section 3 an exact Riemann solver for one dimensional flows is constructed. Section 4 contains description of treatment the special case of equation of state $p\left( \varepsilon, n \right)=0$. In Section 5 we present how our scheme interacts with vacuum. In Section 6 we give the approximate method of including tangential velocities. Section 7 describes how to convert conservative variables into primitive ones. In Section 8 the method of including a source term is given. Section 9 is devoted to testing our code on several examples of one-dimensional flows. In Section 10 we test our code on three-dimensional flows. Section 11 summarizes the results of the paper.

\section{Basic equations}

\label{s:basic}

Our task is to numerically simulate the system of equations of
relativistic hydrodynamics. It can be written as
\begin{equation}
\pdif{\mbox{T}^{\mu\nu}}{\mbox{x}^{\nu}}=0, \label{cov_num}
\end{equation}
where $\mbox{T}^{\mu\nu}=\left( \varepsilon+\mbox{p} \right)
\mbox{u}^{\mu} \mbox{u}^{\nu}-\mbox{p}\mbox{g}^{\mu\nu}$ is
energy-momentum tensor of ideal fluid, $\varepsilon$ is energy
density taken in the rest frame of a fluid element,
$\mbox{p}=\mbox{p}\left( \varepsilon,\mbox{n} \right)$ is
pressure, $\mbox{n}$ is baryonic number density taken in the rest
frame of a fluid element, $\displaystyle \mbox{u}^{\mu}=\left(
\frac{1}{1-\mbox{v}^2}, \frac{\bv{v}}{1-\mbox{v}^2} \right)$ is 
4-vector of fluid velocity and $\mbox{g}^{\mu\nu}$ is the metric
tensor that in case of flat space-time has the form
$\mbox{g}^{\mu\nu}=\mbox{diag} \left( 1,-1,-1,-1 \right)$. We use
such notations that Greek indices take values from 0 to 3 and
Latin ones from 1 to 3 (are the space coordinates).

Equations (\ref{cov_num}) must be supplemented by conservation equation for
baryonic number
\begin{equation}
\pdif{\mbox{nu}^{\mu}}{\mbox{x}^{\mu}}=0. \label{bar}
\end{equation}
The system (\ref{cov_num})-(\ref{bar}) is full and represents
five equations for the set of five independent variables
$\bv{P}=\left( \varepsilon, \mbox{n}, \mbox{v}_{x}, \mbox{v}_{y},
\mbox{v}_{z} \right)$. This set is called primitive variables.

The system (\ref{cov_num})-(\ref{bar}) is a system of conservation
laws and can be rewritten as
\begin{equation}
\pdif{\bv{Q}}{\mbox{t}}+\pdif{\bv{F(Q)}}{\mbox{x}}+
\pdif{\bv{G(Q)}}{\mbox{y}}+\pdif{\bv{H(Q)}}{\mbox{z}}=0,
\label{cons}
\end{equation}
where
\begin{eqnarray}
\bv{Q}=\left( \frac{\varepsilon+\mbox{p}\mbox{v}^2}{1-\mbox{v}^2},
\frac{\left( \varepsilon+\mbox{p}
\right)\mbox{v}_x}{1-\mbox{v}^2}, \frac{\left(
\varepsilon+\mbox{p} \right)\mbox{v}_y}{1-\mbox{v}^2},
\frac{\left( \varepsilon+\mbox{p}
\right)\mbox{v}_z}{1-\mbox{v}^2},
\frac{\mbox{n}}{\sqrt{1-\mbox{v}^2}} \right)= \nonumber & \\
\left( \mbox{Q}_1,
\mbox{Q}_x, \mbox{Q}_y, \mbox{Q}_z, \mbox{Q}_n \right) \label{Q}&,
\end{eqnarray}
 is the vector of different five independent variables that are
called conservative variables and
\begin{equation}
\bv{F}=\left( \frac{\left( \varepsilon+\mbox{p} \right)
\mbox{v}_x}{1-\mbox{v}^2}, \frac{\left( \varepsilon+\mbox{p}
\right) \mbox{v}^2_x}{1-\mbox{v}^2}+\mbox{p}, \frac{\left(
\varepsilon+p \right) \mbox{v}_x \mbox{v}_y}{1-\mbox{v}^2},
\frac{\left( \varepsilon+\mbox{p} \right) \mbox{v}_x
\mbox{v}_z}{1-\mbox{v}^2},
\frac{\mbox{n}\mbox{v}_x}{\sqrt{1-\mbox{v}^2}} \right), \label{F}
\end{equation}
\begin{equation}
\bv{G}=\left( \frac{\left( \varepsilon+\mbox{p} \right)
\mbox{v}_y}{1-\mbox{v}^2}, \frac{\left( \varepsilon+\mbox{p}
\right) \mbox{v}_y \mbox{v}_x}{1-\mbox{v}^2}, \frac{\left(
\varepsilon+\mbox{p} \right) \mbox{v}_y^2}{1-\mbox{v}^2}+\mbox{p},
\frac{\left( \varepsilon+\mbox{p} \right) \mbox{v}_y
\mbox{v}_z}{1-\mbox{v}^2},
\frac{\mbox{n}\mbox{v}_y}{\sqrt{1-\mbox{v}^2}} \right), \label{G}
\end{equation}
\begin{equation}
\bv{H}=\left( \frac{\left( \varepsilon+\mbox{p} \right)
\mbox{v}_z}{1-\mbox{v}^2}, \frac{\left( \varepsilon+\mbox{p}
\right) \mbox{v}_z \mbox{v}_x}{1-\mbox{v}^2}, \frac{\left(
\varepsilon+\mbox{p} \right) \mbox{v}_z \mbox{v}_y}{1-\mbox{v}^2},
\frac{\left( \varepsilon+\mbox{p} \right)
\mbox{v}_z^2}{1-\mbox{v}^2}+\mbox{p},
\frac{\mbox{n}\mbox{v}_z}{\sqrt{1-\mbox{v}^2}} \right), \label{H}
\end{equation}
are the vectors of fluxes in $\mbox{x}$, $\mbox{y}$ and $\mbox{z}$
directions respectively.

There are numerous numerical methods for simulation of conservation systems. As a basis we implement a Godunov-type method, described by Falle \& Komissarov \cite{falle,fallekomissarov} in one- and
two-dimensional cases.

In order to construct a numerical scheme for system (\ref{cons}) we
divide numerical domain into cells with size $\mbox{h}_x$ along
the x-axis, $\mbox{h}_y$ along the y-axis and $\mbox{h}_z$ along
the z-axis. Knowing solution at time $\mbox{t}_n$ we want to
find solution at time $\mbox{t}_{n+1}=\mbox{t}_n+\Delta
\mbox{t}$.

Integrating (\ref{cons}) in time from $\mbox{t}$ to
$\mbox{t}_n+\Delta \mbox{t}$ and in space over domain
$(i-1)\mbox{h}_x \leq x \leq i\mbox{h}_x$, $(j-1)\mbox{h}_y \leq y
\leq j\mbox{h}_y$, $(k-1)\mbox{h}_z \leq z \leq k\mbox{h}_z$ we
obtain the following exact relation
\begin{eqnarray}
\bv{Q}^{n+1}_{ijk}=\bv{Q}^{n}_{ijk}-\frac{\Delta
\mbox{t}}{\mbox{h}_x} \left( \bv{F}^{n}_{ijk}-\bv{F}^{n}_{(i-1)jk}
\right) - \frac{\Delta \mbox{t}}{\mbox{h}_y} \left(
\bv{G}^{n}_{ijk}-\bv{G}^{n}_{i(j-1)k} \right) - & \nonumber \\
\frac{\Delta
\mbox{t}}{\mbox{h}_z} \left( \bv{H}^{n}_{ijk}-\bv{H}^{n}_{ij(k-1)}
\right) \label{exact} &
\end{eqnarray}
Here
\begin{equation}
\bv{Q}^{n}_{ijk}=\frac{1}{\mbox{h}_x \mbox{h}_y \mbox{h}_z}
\int^{i\mbox{h}_x}_{(i-1)\mbox{h}_x}
\int^{j\mbox{h}_y}_{(j-1)\mbox{h}_y}
\int^{k\mbox{h}_z}_{(k-1)\mbox{h}_z} \bv{Q} \left( \mbox{x},
\mbox{y}, \mbox{z}, \mbox{t}_n \right) \mbox{dx dy dz},
\label{Q_num}
\end{equation}
\begin{equation}
\bv{F}^{n}_{ijk}=\frac{1}{\Delta \mbox{t} \mbox{h}_y \mbox{h}_z}
\int_{\mbox{t}_n}^{\mbox{t}_{n+1}}
\int^{j\mbox{h}_y}_{(j-1)\mbox{h}_j}
\int^{k\mbox{h}_z}_{(k-1)\mbox{h}_z} \bv{F} \left( \bv{Q} \left(
i\mbox{h}_x, \mbox{y}, \mbox{z}, \mbox{t} \right) \right) \mbox{
dt dy dz}, \label{F_num}
\end{equation}
\begin{equation}
\bv{G}^{n}_{ijk}=\frac{1}{\Delta \mbox{t} \mbox{h}_x \mbox{h}_z}
\int_{\mbox{t}_n}^{\mbox{t}_{n+1}}
\int^{i\mbox{h}_x}_{(i-1)\mbox{h}_x}
\int^{k\mbox{h}_z}_{(k-1)\mbox{h}_z} \bv{G} \left( \bv{Q} \left(
\mbox{x}, j\mbox{h}_y, \mbox{z}, \mbox{t} \right) \right) \mbox{
dt dx dz}, \label{G_num}
\end{equation}
\begin{equation}
\bv{H}^{n}_{ijk}=\frac{1}{\Delta \mbox{t} \mbox{h}_x \mbox{h}_y}
\int_{\mbox{t}_n}^{\mbox{t}_{n+1}}
\int^{i\mbox{h}_x}_{(i-1)\mbox{h}_x}
\int^{j\mbox{h}_y}_{(j-1)\mbox{h}_y} \bv{H} \left( \bv{Q} \left(
\mbox{x}, \mbox{y}, k\mbox{h}_z, \mbox{t} \right) \right) \mbox{
dt dx dy}, \label{H_num}
\end{equation}
Note that in the right side of (\ref{F_num})-(\ref{H_num}) there
is integrating in time and conservative numerical methods differ
from each other just in ways of estimating these integrals.

As a first order scheme we use the following method. We put that
inside of a cell there is a uniform distribution of conservative
quantities equal to $\bv{Q}^{n}_{ijk}$. In this case on every cell
boundary we have a discontinuity that sets up a Riemann problem.
We have to solve it in order to estimate fluxes
(\ref{F_num})-(\ref{H_num}). For this we
solve a one-dimensional Riemann problem in every space direction. 
Indeed, to obtain, for
instance, $\bv{F}^{n}_{ijk}$ we suppose that in a neighborhood
of discontinuity the flow depends only on $\mbox{x}$ and not on
$\mbox{y}$, $\mbox{z}$, \ie
\begin{equation}
\bv{F}^{n}_{ijk}=\frac{1}{\Delta \mbox{t}}
\int^{\mbox{t}_{n+1}}_{\mbox{t}_n} \bv{F} \left( \bv{Q} \left(
i\mbox{h}_x, j\mbox{h}_y, k\mbox{h}_z, \mbox{t} \right) \right)
\mbox{dt}, \label{F_numone}
\end{equation}
which is true as the distribution of quantities inside a cell is
uniform. After we have solved the Riemann problem, on the place of a
discontinuity we obtain a state $\bv{Q}^{\star} \left(
\bv{Q}^{n}_{ijk}, \bv{Q}^{n}_{(i-1)jk} \right)$ and the
corresponding flux is
\begin{equation}
\bv{F}^{n}_{ijk}=\bv{F} \left( \bv{Q}^{\star} \left(
\bv{Q}^{n}_{ijk}, \bv{Q}^{n}_{(i-1)jk} \right) \right).
\label{F_one}
\end{equation}
Similarly we find $\bv{G}^{n}_{ijk}$ and $\bv{H}^{n}_{ijk}$.

In order to construct a second-order scheme we have to introduce a
structure inside a cell and to let quantities change during a time
interval $\Delta \mbox{t}$.

The second order in time is achieved by using a first order scheme
to get an intermediate solution $\bv{Q}^{n+1/2}_{ijk}$ at time
$\mbox{t}_n+\Delta \mbox{t}/2$ and by applying it to finding
fluxes.

In order to achieve the second order in space we approximate the
quantities inside a cell by linear functions. In doing so we
must be cautious not to break scheme's monotonicity. This is
important because non-monotonic schemes
lead to artificial oscillations in neighborhood of shock waves and
contact discontinuities. In order not to allow such effects an
averaging function is used to calculate cell gradients that
compares gradients in neighboring cells and chooses the smallest
one. Doing so we have
\begin{equation}
\left( \pdif{\bv{Q}}{\mbox{x}} \right)^{n+1/2}_{ijk}=\mbox{av}
\left(
\frac{\bv{Q}^{n+1/2}_{ijk}-\bv{Q}^{n+1/2}_{(i-1)jk}}{\mbox{h}_x},
\frac{\bv{Q}^{n+1/2}_{(i+1)jk}-\bv{Q}^{n+1/2}_{ijk}}{\mbox{h}_x}
\right), \label{grad_x}
\end{equation}
\begin{equation}
\left( \pdif{\bv{Q}}{\mbox{y}} \right)^{n+1/2}_{ijk}=\mbox{av}
\left(
\frac{\bv{Q}^{n+1/2}_{ijk}-\bv{Q}^{n+1/2}_{i(j-1)k}}{\mbox{h}_y},
\frac{\bv{Q}^{n+1/2}_{i(j+1)k}-\bv{Q}^{n+1/2}_{ijk}}{\mbox{h}_y}
\right), \label{grad_y}
\end{equation}
\begin{equation}
\left( \pdif{\bv{Q}}{\mbox{z}} \right)^{n+1/2}_{ijk}=\mbox{av}
\left(
\frac{\bv{Q}^{n+1/2}_{ijk}-\bv{Q}^{n+1/2}_{ij(k-1)}}{\mbox{h}_z},
\frac{\bv{Q}^{n+1/2}_{ij(k+1)}-\bv{Q}^{n+1/2}_{ijk}}{\mbox{h}_z}
\right), \label{grad_z},
\end{equation}
where the averaging function $\mbox{av(a,b)}$ must have the
following properties
\begin{equation}
\mbox{av(a,b)} \left\{ \begin{array}{ll} \rightarrow \frac{1}{2}
\left( \mbox{a}+\mbox{b} \right), & \mbox{if  a}\rightarrow \mbox{b} \\
=0, & \mbox{if  ab}<0 \\
\rightarrow \mbox{a}, & \mbox{if  } \left| \mbox{a} \right|/
\left| \mbox{b} \right|
\rightarrow \infty \\
\rightarrow \mbox{b}, & \mbox{if  } \left| \mbox{b} \right| /
\left| \mbox{a} \right| \rightarrow \infty
\end{array} \right.
\label{av}
\end{equation}
There exist an infinite number of such functions. We use one of
the simplest variants that has the following form
\begin{equation}
\mbox{av(a,b)}=\left\{ \begin{array}{ll} \frac{\displaystyle
\mbox{a}^2 \mbox{b}+\mbox{a} \mbox{b}^2}{\displaystyle
\mbox{a}^2+\mbox{b}^2},
& \mbox{if  } \mbox{a}^2+\mbox{b}^2>0 \\
=0, & \mbox{if  } \mbox{ab} \leq 0
\end{array} \right.
\label{av1}
\end{equation}
After we have calculated gradients
(\ref{grad_x})-(\ref{grad_z}) we can find second order left and
right states for a cell boundary Riemann problem. For example, when
estimating $\bv{F}^{n}_{ijk}$ we assume
\begin{equation}
\bv{Q}^{(l)}_{ijk}=\bv{Q}^{n+1/2}_{ijk}+\frac{\mbox{h}_x}{2}
\left( \pdif{\bv{Q}}{\mbox{x}} \right)^{n+1/2}_{ijk} \label{Ql}
\end{equation}
\begin{equation}
\bv{Q}^{(r)}_{ijk}=\bv{Q}^{n+1/2}_{(i+1)jk}-\frac{\mbox{h}_x}{2}
\left( \pdif{\bv{Q}}{\mbox{x}} \right)^{n+1/2}_{(i+1)jk}
\label{Qr}
\end{equation}
Correspondingly
\begin{equation}
\bv{F}^{n}_{ijk}=\bv{F} \left( \bv{Q}^{\star} \left(
\bv{Q}^{(l)}_{ijk}, \bv{Q}^{(r)}_{(i-1)jk} \right) \right).
\label{F_two}
\end{equation}
Similar relations are applied to find $\bv{G}^{n}_{ijk}$ and
$\bv{H}^{n}_{ijk}$. As we said above, application of quantities with
time index $n+1/2$ in order to find fluxes warrants the second
order of the scheme in time.

Note that in relativistic hydrodynamics the condition
\begin{equation}
\sqrt{\mbox{Q}_x^2+\mbox{Q}_y^2+\mbox{Q}_z^2}<\mbox{Q}_1
\label{q1}
\end{equation}
must be held in order to obey causality principle. In general,
however, applying relations (\ref{Ql})-(\ref{Qr}) may violate it.
In such cases we have to additionally adjust slopes in order to
hold (\ref{q1}).

Thus we have a complete numerical second order scheme. The
only missing constituent is the procedure of calculating solution
of a Riemann problem \ie{} a Riemann solver. Having it at hand we
can find $\bv{Q}^{\star} \left( \bv{Q}^{(l)}_{ijk},
\bv{Q}^{(r)}_{(i-1)jk} \right)$ and then find fluxes and then move
one step further in time.

\section{Exact one-dimensional Riemann solver}

In this section we formulate a procedure of finding solution of a
Riemann problem. We call our solver exact in a sense that
numerical methods are applied to exact relations that describe a
Riemann problem solution. This differs our method from numerous
approximate Riemann solvers that use approximate relations for
estimating intermediate state $\bv{Q}^{\star}$.

As was said above, we need to construct an algorithm for a
one-dimensional Riemann problem. For this we consider the
following system
\begin{equation}
\pdif{\bv{Q}}{\mbox{t}}+\pdif{\bv{F}\left( \bv{Q}
\right)}{\mbox{x}}=0, \label{one-dim}
\end{equation}
where
\begin{equation}
\bv{Q}=\left( \frac{\varepsilon+ \mbox{pv}^2}{1-\mbox{v}^2},
\frac{\left( \varepsilon+ \mbox{p} \right)
\mbox{v}}{1-\mbox{v}^2}, \frac{\mbox{n}}{\sqrt{1-\mbox{v}^2}}
\right), \label{Q_1}
\end{equation}
\begin{equation}
\bv{F}=\left( \frac{\left( \varepsilon +\mbox{p} \right)
\mbox{v}}{1-\mbox{v}^2},\frac{\mbox{p}+\varepsilon
\mbox{v}^2}{1-\mbox{v}^2}, \frac{\mbox{nv}}{1-\mbox{v}^2} \right).
\label{F_1}
\end{equation}
As we see, all the quantities here depend only on $\mbox{t}$ and
$\mbox{x}$ and velocity has only the $\mbox{x}$-th component
denoted as $\mbox{v}$. The influence on solution of non-zero
tangential velocities $\mbox{v}_y$ and $\mbox{v}_z$ will be
considered later.

For the system (\ref{one-dim}) we state the following initial
conditions at initial time $\mbox{t}=\mbox{t}_0$
\begin{equation}
\bv{Q}\left( 0,\mbox{x} \right) =\left\{ \begin{array}{ll}
\bv{Q}^{(l)}, & \mbox{if  } \mbox{x}<0 \\
\bv{Q}^{(r)}, & \mbox{if  } \mbox{x}>0
\end{array}, \right.
\label{rim}
\end{equation}
where $\bv{Q}^{(l)}$ and $\bv{Q}^{(r)}$ are some constant states.

\begin{figure*}
\begin{center}
\epsfysize=5cm \epsffile{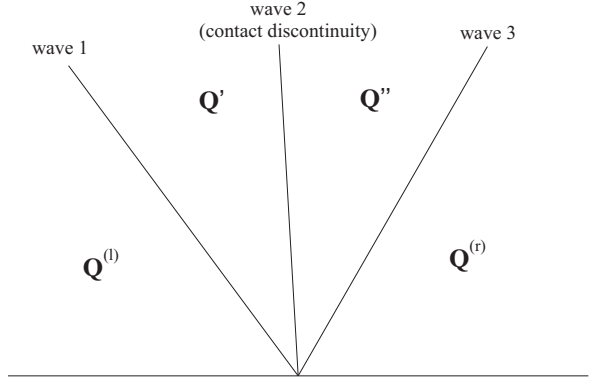}
\end{center}
\caption{\scriptsize Schematic depiction of break up of an arbitrary discontinuity. Initial left
$\bv{Q}^{(l)}$ and right $\bv{Q}^{(r)}$ states are separated by two intermediate states
$\bv{Q}'$ and $\bv{Q}''$. All the states are connected through the set of 3 waves. For linear systems these waves are discontinuities. For a non-linear system of relativistic hydrodynamics
with a thermodynamically normal equation of state wave 2 is a contact discontinuity and waves 1 and 3 can be either shock waves or simple rarefaction waves}
\label{f:waves}
\end{figure*}

Quite often numerical solution of (\ref{one-dim})-(\ref{rim}) is
based on a linear approximation. We too base our idea on solution
of a similar problem for a linear system. It is well known that
for linear systems, \ie{} systems with $\bv{F}\left( \bv{Q}
\right)=\hat{\mbox{A}}\bv{Q}$, $\hat{\mbox{A}}$ being a constant
matrix, solution of problem (\ref{one-dim})-(\ref{rim}) consists
of 3 discontinuities (or $\mbox{n}$ discontinuities for systems of
$\mbox{n}$ equations) divided by domains of constant flows 
(see Figure~\ref{f:waves}). But in
case of a nonlinear problem there arise continuous self-similar
solutions called simple waves.

At this point in order to be able to move further we need to say
that only so-called thermodynamically normal media are the
subject of present study. Medium is called thermodynamically
normal when the following condition holds
\begin{equation}
\left. \frac{\partial^2 \mbox{p}}{\partial \tau^2}
\right|_{\mbox{s}}>0, \label{Bethe}
\end{equation}
where $\displaystyle \tau=\frac{\varepsilon +\mbox{p}}{\mbox{n}}$
is so-called generalized specific volume and $\mbox{s}$ is
entropy. Condition (\ref{Bethe}) means that only compression shock
waves and rarefaction simple waves are physical. Rarefaction
shocks are nonphysical in the sense that entropy of a fluid
element decreases when crossing a shock front. Compression simple
waves are nonphysical in the sense that wave profile becomes
singular at a definite moment of time.

Media with violated condition (\ref{Bethe}) are called
thermodynamically anomalous. The consideration here is similar to
the thermodynamically normal case with only change that physical
are rarefaction shocks and compression simple waves. Cases when for
some parameters relation (\ref{Bethe}) holds and for some
parameters is violated are more complicated and are the subject of
further investigation. This is of importance because equations of
state where exists a phase transition represent both normal and
anomalous behavior at different values of thermodynamical
parameters.

Nevertheless from the point of view of pure mathematics both
compression and rarefaction shocks exist as generalized solutions
of equations (\ref{one-dim}). It means that we can search for pure
shock wave solution of problem (\ref{one-dim})-(\ref{rim}), and at
this point not take into account how physical are its
constituents.

Indeed, a discontinuity $\bv{Q}_1 \rightarrow \bv{Q}_2$ is a
generalized solution of system (\ref{one-dim}) if
\begin{equation}
\bv{F} \left( \bv{Q}_1 \right) - \bv{F} \left( \bv{Q}_2 \right) =
\mbox{S} \left( \bv{Q}_1 - \bv{Q}_2 \right), \label{shock}
\end{equation}
where $\mbox{S}$ is propagation speed of discontinuity and in
linear case $\bv{F}\left( \bv{Q} \right)=\hat{\mbox{A}}\bv{Q}$ is
just an eigenvalue of matrix $\hat{\mbox{A}}$.

Arbitrary states $\bv{Q}^{(l)}$ and $\bv{Q}^{(r)}$ in general does
not satisfy (\ref{shock}). But they can be sewn together through
two intermediate states $\bv{Q}'$ and $\bv{Q}''$ (see Figure~\ref{f:waves}). 
Then we have the following system of algebraical equations
\begin{equation}
\bv{F} \left( \bv{Q}^{(l)} \right) - \bv{F} \left( \bv{Q}' \right)
=\mbox{S}_1 \left( \bv{Q}^{(l)} - \bv{Q}' \right) \label{1}
\end{equation}
\begin{equation}
\bv{F} \left( \bv{Q}' \right) - \bv{F} \left( \bv{Q}'' \right)
=\mbox{S}_2 \left( \bv{Q}' - \bv{Q}'' \right) \label{2}
\end{equation}
\begin{equation}
\bv{F} \left( \bv{Q}'' \right) - \bv{F} \left( \bv{Q}^{(r)}
\right) =\mbox{S}_3 \left( \bv{Q}'' - \bv{Q}^{(r)} \right)
\label{3}
\end{equation}
This is system of 9 equations for 9 unknown variables - $\bv{Q}'$,
$\bv{Q}''$, $\mbox{S}_1$, $\mbox{S}_2$, $\mbox{S}_3$. Once have
solved it, we have a pure shock wave solution of a Riemann problem
(\ref{one-dim})-(\ref{rim}). Taking into account that wave 2 must
be a contact discontinuity for which $\mbox{p}'=\mbox{p}''$,
$\mbox{v}'=\mbox{v}''=\mbox{S}_2$, the number of equations in
(\ref{1})-(\ref{3}) can be reduced to 7. The system
(\ref{1})-(\ref{3}) is written in terms of conservative variables
$\bv{Q}$ but in our code is solved in terms of primitive variables
$\bv{P}=\left( \varepsilon, \mbox{v}, \mbox{n} \right)$. In
ultra-relativistic case we use instead of $\mbox{v}$
$\gamma$-factor $\displaystyle
\gamma=\frac{1}{\sqrt{1-\mbox{v}^2}}$. So, after we have solved
(\ref{1})-(\ref{3}), we have $\varepsilon'$, $\mbox{n}'$,
$\varepsilon''$, $\mbox{n}''$, $\mbox{S}_1$, $\mbox{S}_3$,
$\mbox{v}'$.

The numerical method we use to solve (\ref{1})-(\ref{3}) is the
Newton method. An obvious difficulty here is how to choose an initial
approximation. As we have a set of 7 equations its quite a
non-trivial problem. In some cases, when our initial approximation
appears to be bad, the method fails. In such cases we use a switch
to the alternative method of finding $\varepsilon'$, $\mbox{n}'$,
$\varepsilon''$, $\mbox{n}''$, $\mbox{S}_1$, $\mbox{S}_3$,
$\mbox{v}'$.

It is based on relations that connect variables on both sides of a
shock wave and on a well known fact that across a contact
discontinuity pressure and velocity are continuous. First, note,
that shock speeds $\mbox{S}_1$, $\mbox{S}_3$ and velocity
$\mbox{v}'$ can be explicitly expressed in terms of
thermodynamical variables. So, as a matter of fact we need 4
equations to find $\varepsilon'$, $\mbox{n}'$, $\varepsilon''$,
$\mbox{n}''$.

The first two equations we take from Hugoniot relations and obtain
\begin{equation}
\mbox{n}'=\mbox{n}^{(l)}\sqrt{ \frac{\left( \varepsilon' + \mbox{p}' \right)
\left( \varepsilon' + \mbox{p}^{(l)} \right)}{\left(
\varepsilon^{(l)} + \mbox{p}^{(l)} \right) \left(
\varepsilon^{(l)} + \mbox{p}' \right)}}, \label{hug1}
\end{equation}
\begin{equation}
\mbox{n}''=\mbox{n}^{(r)}\sqrt{ \frac{\left( \varepsilon'' + \mbox{p}' \right)
\left( \varepsilon'' + \mbox{p}^{(r)} \right)}{\left(
\varepsilon^{(r)} + \mbox{p}^{(r)} \right) \left(
\varepsilon^{(r)} + \mbox{p}' \right)}} \label{hug2}
\end{equation}
Here we used the fact that on contact discontinuity
$\mbox{p}'=\mbox{p}''$. This gives us the third equation
\begin{equation}
\mbox{p} \left( \varepsilon', \mbox{n}' \right)=\mbox{p} \left(
\varepsilon'', \mbox{n}'' \right) \label{pp}
\end{equation}
The fourth equation we obtain from relation
$\mbox{v}'=\mbox{v}''$. The velocity of fluid behind the shock
front (marked by index 2) in the rest frame of fluid ahead of the
shock front (marked by index 1) is
\begin{equation}
\mbox{v}^{\pm}_{sh}\left( \varepsilon_1, \mbox{n}_1, \varepsilon_2, \mbox{n}_2 \right)= 
\pm \sqrt{\frac{\left( \mbox{p}_2-\mbox{p}_1 \right) \left(\varepsilon_2-\varepsilon_1 \right)}
{\left( \varepsilon_1 + \mbox{p}_2 \right) \left( \varepsilon_2 + \mbox{p}_1 \right)}} \label{v12}
\end{equation}
where the sign is defined by two factors- whether the shock is
compression or rarefaction one and what direction does it go, right or left.

Then the fourth equation is written as
\begin{equation}
\frac{\mbox{v}^{(l)}+\mbox{v}^{\pm}_{12}\left( \varepsilon^{(l)}, \mbox{n}^{(l)}, \varepsilon',
\mbox{n}' \right)}{1+\mbox{v}^{(l)}\mbox{v}^{\pm}_{12}\left( \varepsilon^{(l)}, \mbox{n}^{(l)}, \varepsilon', \mbox{n}' \right)}=
\frac{\mbox{v}^{(r)}+\mbox{v}^{\pm}_{12}\left( \varepsilon^{(r)}, \mbox{n}^{(r)}, \varepsilon'',
\mbox{n}'' \right)}{1+\mbox{v}^{(r)}\mbox{v}^{\pm}_{12}\left( \varepsilon^{(r)}, \mbox{n}^{(r)}, \varepsilon'', \mbox{n}'' \right)}
\label{fourth}
\end{equation}
Depending on situation there can be 4 combinations of signs.

Shocks speeds $\mbox{S}_1$ and $\mbox{S}_3$ can be written as
\begin{equation}
\mbox{S}_1=\frac{\mbox{v}^{(l)}+\mbox{v}^{-}_{sh}\left( \varepsilon^{(l)}, \mbox{n}^{(l)}, \varepsilon', \mbox{n}' \right)}{1+\mbox{v}^{(l)}\mbox{v}^{-}_{sh}\left( \varepsilon^{(l)}, \mbox{n}^{(l)}, \varepsilon', \mbox{n}' \right)}
\label{s1}
\end{equation}
and
\begin{equation}
\mbox{S}_3=\frac{\mbox{v}^{(r)}+\mbox{v}^{+}_{sh}\left( \varepsilon^{(r)}, \mbox{n}^{(r)}, \varepsilon'', \mbox{n}'' \right)}{1+\mbox{v}^{(r)}\mbox{v}^{+}_{sh}\left( \varepsilon^{(r)}, \mbox{n}^{(r)}, \varepsilon'', \mbox{n}'' \right)},
\label{s3}
\end{equation}
where
$\displaystyle \mbox{v}^{\pm}_{12}\left(\varepsilon_1, \mbox{n}_1, \varepsilon_2,
\mbox{n}_2 \right) =\pm \sqrt{ \frac{\left( \mbox{p}_2 -
\mbox{p}_1 \right) \left( \varepsilon_2 + \mbox{p}_1
\right)}{\left( \varepsilon_2 - \varepsilon_1 \right) \left(
\varepsilon_1 + \mbox{p}_2 \right)}}$ is the speed of the shock front in the rest frame of fluid ahead of a shock (marked by 1) and the sign is "-" when a shock moves in the rest frame to the left (applies to shock wave 1) and is "+" when a shock moves to the right (applies to shock wave 3).

Thus, we have constructed a purely shock wave solution for a Riemann problem. But the real solution of a Riemann problem (\ref{one-dim})-(\ref{rim}) consists not only from shocks. In general, for a normal medium, waves 1 and 3 can be either compression shocks or rarefaction
simple waves. Thus, the next step is to find real pattern of waves.

\begin{figure*}
\begin{tabular}{ll}
(a) & (b) \\ \epsfysize=3.0cm \epsffile{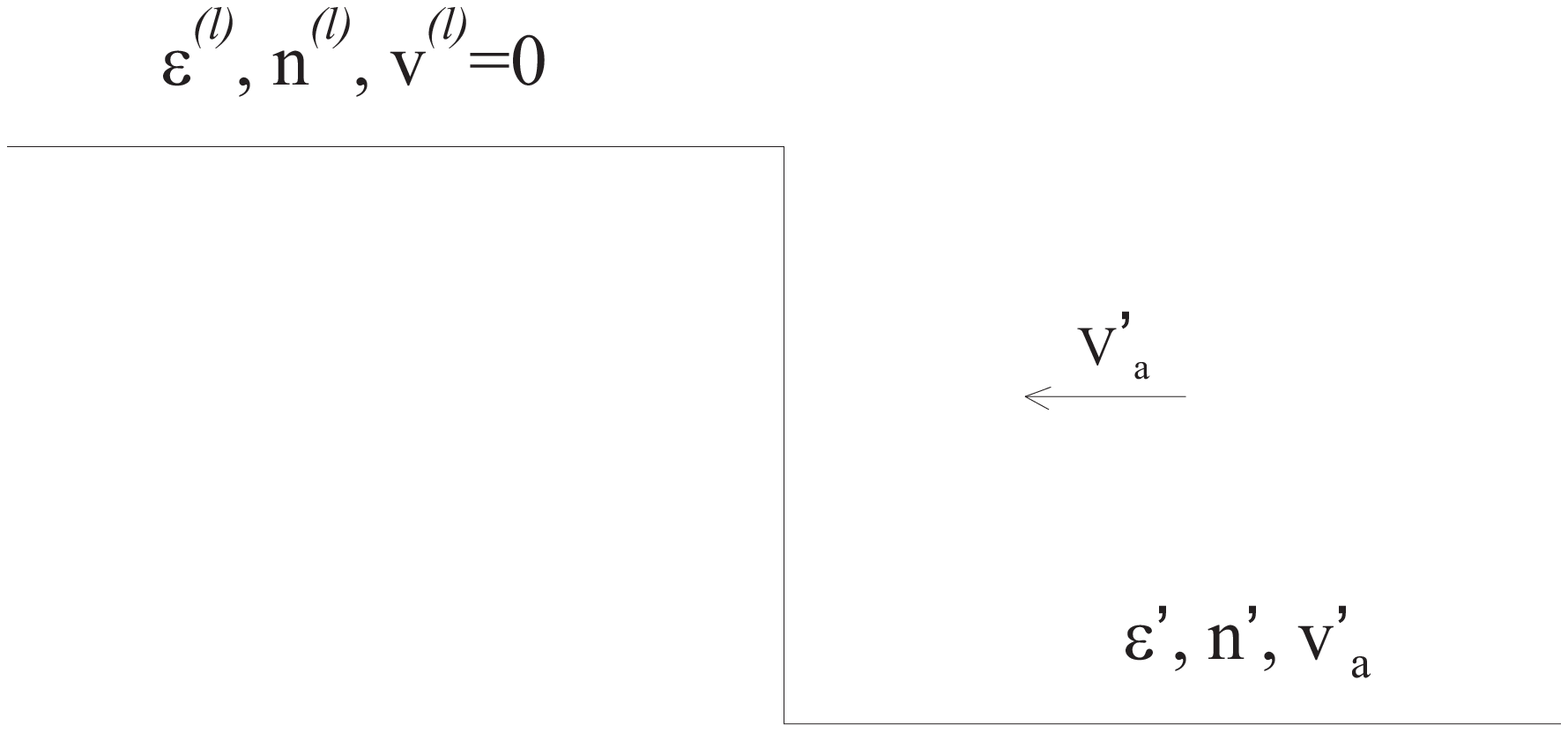} &
\epsfysize=3.0cm \epsffile{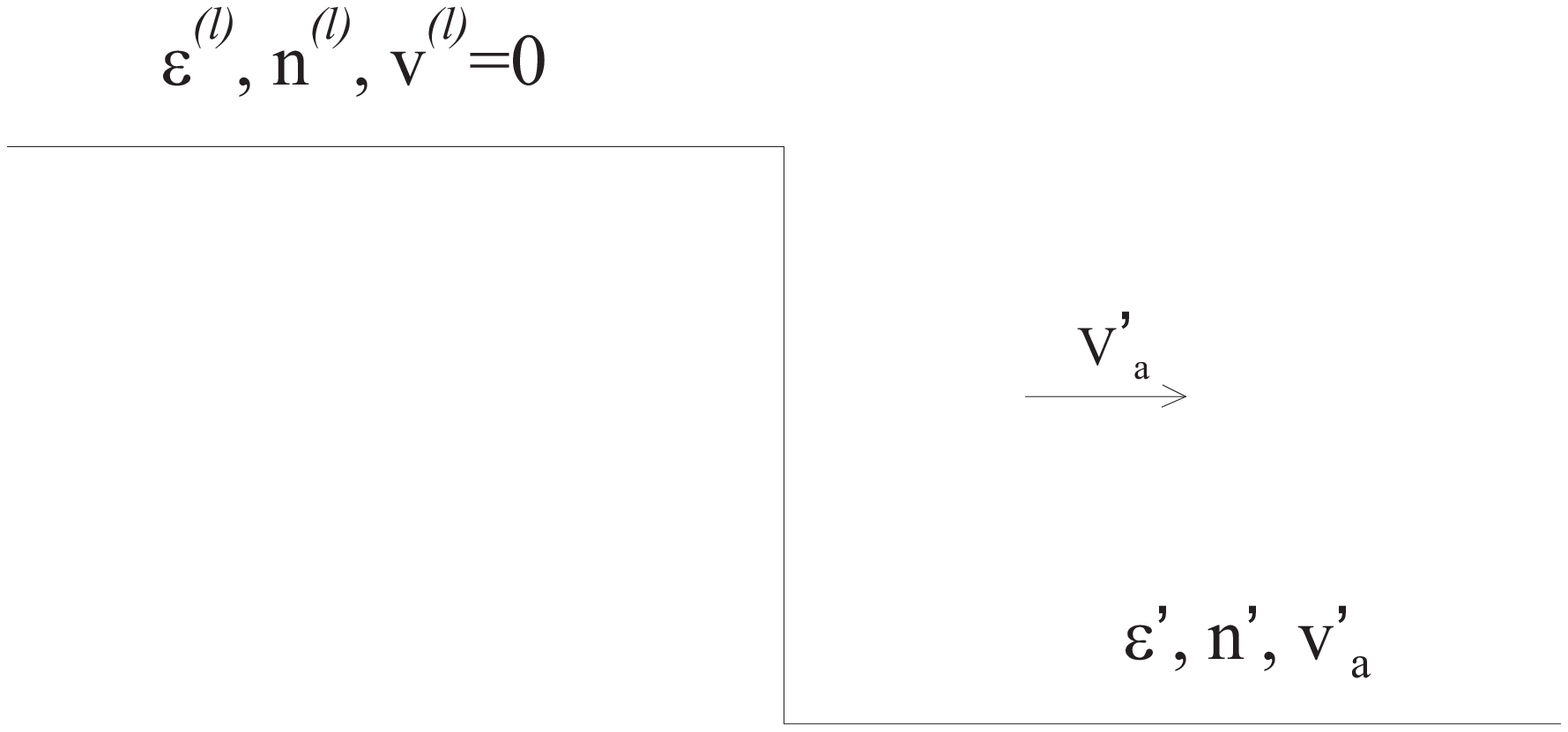} \\
\end{tabular}
\caption{\scriptsize A shock wave $\varepsilon^{(l)},\mbox{n}^{(l)},\mbox{v}^{(l)} \rightarrow
\varepsilon',\mbox{n}',\mbox{v}'$ in the rest frame of the left fluid. Here $\mbox{v}'_{a}=
\mbox{v}^{\pm}_{12}\left(\varepsilon_1, \mbox{n}_1, \varepsilon_2, \mbox{n}_2 \right)$ (see Eq.(\ref{v12})). If $\mbox{v}'_{a}<0$ then we have a compression shock. When $\mbox{v}'_{a}>0$ then there is a rarefaction wave.}
\label{f:pattern}
\end{figure*}

As an example consider a shock transition $\varepsilon^{(l)},\mbox{n}^{(l)},\mbox{v}^{(l)} \rightarrow \varepsilon',\mbox{n}',\mbox{v}'$ with $\mbox{n}^{(l)}>\mbox{n}'$ (see Figure~\ref{f:pattern}). In order to find out whether such a shock is compression or rarefaction shock we have to go to the rest frame of the left fluid. If velocity to the right $\mbox{v}'_{a}>0$
(it means that $\mbox{v}^{(l)}<\mbox{v}'$ in laboratory frame) then fluid crosses the shock front from the left to the right. As we supposed that $\mbox{n}^{(l)}>\mbox{n}'$ it means that the shock is a rarefaction one. If $\mbox{v}'_{a}<0$ ($\mbox{v}^{(l)}>\mbox{v}'$ in laboratory frame) than fluid crosses shock front from the right to the left and we have a compression shock. Similar considerations should be applied to case $\mbox{n}^{(l)}<\mbox{n}'$ and to a shock transition $\varepsilon^{(r)},\mbox{n}^{(r)},\mbox{v}^{(r)} \rightarrow \varepsilon'',\mbox{n}'',\mbox{v}''$. 

Once we have found types of shocks, our considerations are as follows. 
If both shock waves, 1 and 3, are compression shocks, then solution of a Riemann problem is over. If one of waves appears to be a rarefaction one, we should replace it by a simple rarefaction wave. In such a case we have to sew together a simple and a shock wave. As an example consider the case when wave 1 is a simple rarefaction wave and wave 3 is a shock wave.

Profile in a simple rarefaction wave is described by the following system of ordinary differential equations
\begin{equation}
\frac{\mbox{dv}}{\mbox{dn}}=\pm \frac{\mbox{c}_s\left( \varepsilon,\mbox{n}\right) \left( 1-\mbox{v}^2 \right)}{\mbox{n}},
\label{vrar}
\end{equation}
\begin{equation}
\frac{\mbox{d} \varepsilon}{\mbox{dn}}=\frac{\varepsilon +\mbox{p} \left( \varepsilon,\mbox{n}\right)}{\mbox{n}},
\label{erar}
\end{equation}
where $\displaystyle \mbox{c}_s\left( \varepsilon,\mbox{n}\right)=\sqrt{\left. \frac{\partial \mbox{p}}{\partial \varepsilon} \right|_{\mbox{s}}}$ is the speed of sound. Sign in Eq.(\ref{vrar}) depends on direction of wave propagation. If wave propagates into the left fluid than there is "-", if it goes to the right fluid, there is "+". In our case of wave 1 we have "-". Equations (\ref{vrar}),(\ref{erar}) define functions $\mbox{v}^{\pm}_{rar}\left( \mbox{n} \right)$ and $\varepsilon_{rar}\left( \mbox{n} \right)$. Thus, sewing of a simple rarefaction wave 1 and a shock wave 3 means solution of the following system of equations.
\begin{equation}
\mbox{v}^{-}_{rar}\left( \mbox{n}' \right)=\frac{\mbox{v}^{+}_{12} \left( \varepsilon^{(r)}, \mbox{n}^{(r)}, \varepsilon'', \mbox{n}'' \right)+\mbox{v}^{(r)}}{1+\mbox{v}^{(r)}\mbox{v}^{+}_{12} \left( \varepsilon^{(r)}, \mbox{n}^{(r)}, \varepsilon'', \mbox{n}'' \right)}
\label{shrar1}
\end{equation}
\begin{equation}
\mbox{p}\left( \varepsilon',\mbox{n}' \right)=\mbox{p}\left( \varepsilon'',\mbox{n}'' \right)
\label{shrar2}
\end{equation}
\begin{equation}
\varepsilon'=\varepsilon_{rar}\left( \mbox{n}' \right)
\label{shrar3}
\end{equation}
\begin{equation}
\mbox{n}''=\sqrt{ \frac{\left( \varepsilon'' + \mbox{p}\left( \varepsilon', \mbox{n}' \right) \right) \left( \varepsilon'' + \mbox{p}^{(r)} \right)}{\left(
\varepsilon^{(r)} + \mbox{p}^{(r)} \right) \left(
\varepsilon^{(r)} + \mbox{p} \left( \varepsilon', \mbox{n}' \right) \right)}},
\label{shrar4}
\end{equation}
where $\mbox{v}_{12}$ is defined by Eq.(\ref{v12}). Thus we find $\varepsilon', \mbox{n}', \varepsilon'', \mbox{n}''$. Corresponding velocity is $\mbox{v}'=\mbox{v}''=\mbox{v}^{-}_{rar}\left( \mbox{n}' \right)$. System (\ref{shrar1})-(\ref{shrar4}) is solved by Newton method. Integration of system (\ref{vrar})-(\ref{erar}) to find $\mbox{v}^{\pm}_{rar} \left( \mbox{n} \right)$ and $\varepsilon^{\pm}_{rar} \left( \mbox{n} \right)$ is conducted by using a Runge-Kutta method.

The velocity of wave 1 is defined by relation
\begin{equation}
\mbox{S}_1=\frac{\mbox{v}^{(l)}-\mbox{c}_s}{1-\mbox{v}^{(l)}\mbox{c}_s}
\label{s_rar}
\end{equation}

Similar considerations are applied to the case when we have a simple rarefaction wave moving to the right state and a shock wave moving to the left state.

In case when in all-shock solution waves 1 and 3 are rarefaction shocks, we have to replace both by simple rarefaction waves. In this case we solve the following system of equations
\begin{equation}
\mbox{v}^{-}_{rar} \left( \mbox{n}' \right)=\mbox{v}^{+}_{rar} \left( \mbox{n}'' \right)
\label{2rar1}
\end{equation}
\begin{equation}
\mbox{p}\left( \varepsilon',\mbox{n}' \right)=\mbox{p}\left( \varepsilon'',\mbox{n}'' \right)
\label{2rar2}
\end{equation}
\begin{equation}
\varepsilon'=\varepsilon_{rar} \left( \mbox{n}' \right)
\label{2rar3}
\end{equation}
\begin{equation}
\varepsilon''=\varepsilon_{rar} \left( \mbox{n}'' \right)
\label{2rar4}
\end{equation}

Now, after we have solved a Riemann problem on a cell interface, we have to find state $\bv{Q}^{\star}$ (see Eq(\ref{F_one}) and (\ref{F_two})). In case when wave 1 and 3 are shock waves $\bv{Q}^{\star}=\bv{Q}^{(l)}$ when $\mbox{S}_1>0$, $\bv{Q}^{\star}=\bv{Q}^{(r)}$ when $\mbox{S}_3<0$, $\bv{Q}^{\star}=\bv{Q}'$ when $\mbox{S}_1<0$ and $\mbox{S}_2>0$, $\bv{Q}^{\star}=\bv{Q}''$ when $\mbox{S}_3>0$ and $\mbox{S}_2<0$.

In case when there are simple rarefaction waves situation is more complicated. Take again the example with a rarefaction propagating to the left side and a shock propagating to the right side.
Then $\mbox{S}_1$ is defined by Eq.(\ref{s_rar}). If $\mbox{S}_1>0$ then $\bv{Q}^{\star}=\bv{Q}^{(l)}$. If $\mbox{S}_3<0$ then $\bv{Q}^{\star}=\bv{Q}^{(r)}$. If $\mbox{S}_3>0$ and $\mbox{S}_2<0$ then $\bv{Q}^{\star}=\bv{Q}''$. But if $\mbox{S}_1<0$ and $\mbox{S}_2>0$ situation depends on how strong rarefaction is. If $\mbox{v}'<\mbox{c}_s$ then rarefaction is subsonic and $\bv{Q}^{\star}=\bv{Q}'$. If $\mbox{v}'>\mbox{c}_s$ then rarefaction is supersonic and we have to integrate system (\ref{vrar})-(\ref{erar}) starting from the left-side state to the sonic point $\bv{Q}_{sn}=\left(\varepsilon_{sn}, \mbox{n}_{sn}, \mbox{v}_{sn} \right)$, \ie{} to the state where $\mbox{v}^{-}_{rar} \left( \mbox{n}_{sn} \right)= \mbox{c}_s$. Then $\bv{Q}^{\star}=\bv{Q}_{sn}$.

Similar considerations are made to find $\bv{Q}^{\star}$ in case of a shock wave propagating to the left and a rarefaction propagating to the right and in case of two rarefactions.

Now, after we have defined state $\bv{Q}^{\star}$, construction of a one-dimensional numerical scheme is complete. In three dimensions we have to take into account presence of  non-zero tangential velocities. This problem is considered in Section 4.

\section{Special equation of state $\mbox{p} \left( \varepsilon, \mbox{n} \right)=0$}

In case of a pressureless medium construction made in Section 2 does not work. Here we have to apply a special algorithm. Flux we want to find is
\begin{equation}
\bv{F}^{n}_{i}=\frac{1}{\Delta \mbox{t}}
\int^{\mbox{t}_{n+1}}_{\mbox{t}_n} \bv{F} \left( \bv{Q} \left(
i\mbox{h}_x, \mbox{t} \right) \right) \mbox{dt}, \label{F_nump0}
\end{equation}

In predictor step distribution of quantities over a cell is uniform and obviously
\begin{equation}
\bv{F}^{n}_{i}=\left\{ \begin{array}{ll}
\bv{F} \left( \bv{Q}^{n}_{i} \right), & \mbox{if  } \mbox{v}^{n}_{i}>0 \\
\bv{F} \left( \bv{Q}^{n}_{i+1} \right), & \mbox{if  } \mbox{v}^{n}_{i+1}<0
\end{array}, \right.
\label{fluxpp0}
\end{equation}

In corrector step we have to take into account linear distribution of quantities over a cell. As an example take the case with $\mbox{v}^{n}_{i}>0$. Case $\mbox{v}^{n}_{i+1}<0$ can be taken similarly.

In this case we take approximately
\begin{equation}
\bv{F}^{n}_{i}=\frac{1}{2} \left( \bv{F} \left( \bv{Q}^{n}_{i} \right)+ \bv{F} \left( \bv{Q}^{n+1}_{i} \right) \right),
\label{fluxcp0}
\end{equation}
where the second term is
\begin{equation}
\bv{F}\left( \bv{Q}^{n+1}_{i} \right)=\bv{F} \left( \bv{Q} \left( i\mbox{h}_x, \mbox{t}_n+\Delta \mbox{t} \right) \right)=\bv{F} \left( \bv{Q} \left( i\mbox{h}_x-\Delta \mbox{x}, \mbox{t}_n \right) \right)
\label{fp0}
\end{equation}

Here $\Delta \mbox{x}$ is defined by algebraical equation
\begin{equation}
\mbox{v} \left( i\mbox{h}_x-\Delta \mbox{x}, \mbox{t}_n \right) \Delta \mbox{t}-\Delta \mbox{x}=0
\label{vp0}
\end{equation}

It is solved by Newton method. $\mbox{v} \left( i\mbox{h}_x-\Delta \mbox{x}, \mbox{t}_n \right)$ is defined using linear structure of quantities inside a cell. This gives us the system
\begin{equation}
\mbox{Q}_{1i}^{n}+\alpha_1 \left( \frac{\mbox{h}_x}{2}-\Delta \mbox{x} \right)=\frac{\varepsilon \left( i\mbox{h}_x-\Delta \mbox{x}, \mbox{t}_n \right)}{1-\mbox{v}^2 \left( i\mbox{h}_x-\Delta \mbox{x}, \mbox{t}_n \right)}
\label{p01}
\end{equation}
\begin{equation}
\mbox{Q}_{xi}^{n}+\alpha_x \left( \frac{\mbox{h}_x}{2}-\Delta \mbox{x} \right)=\frac{\varepsilon \left( i\mbox{h}_x-\Delta \mbox{x}, \mbox{t}_n \right) \mbox{v} \left( i\mbox{h}_x-\Delta \mbox{x}, \mbox{t}_n \right)}{1-\mbox{v}^2 \left( i\mbox{h}_x-\Delta \mbox{x}, \mbox{t}_n \right)},
\label{p011}
\end{equation}
where $\alpha_1$ and $\alpha_x$ are cell slopes of corresponding conservative variables.

Solving Eq.(\ref{vp0}) and using Eq.(\ref{fp0}) we complete construction of a scheme for special case of equation of state $\mbox{p} \left( \varepsilon, \mbox{n} \right)=0$.

\section{Boundary with vacuum}

Another difficult problem is defining fluxes at boundary of medium and vacuum. Let $i$-th cell at time $\mbox{t}_n$ be vacuum and $\left( i-1 \right)$-th cell be medium. Consider it for predictor step when distribution of quantities over $\left( i-1 \right)$-th cell is uniform and equal $\bv{Q}^{n}_{i-1}$. Having this, we want to find $\bv{Q}_{i}^{n+1}$.
Then
\begin{equation}
\bv{Q}^{n+1}_{i}=\bv{Q}^{n}_{i}+\frac{\Delta
\mbox{t}}{\mbox{h}_x} \bv{F}^{n}_{i-1}=\frac{\Delta \mbox{t}}{\mbox{h}_x} \bv{F}^{n}_{i-1},
\label{vac1}
\end{equation}
where
\begin{equation}
\bv{F}^{n}_{i-1}=\frac{1}{\Delta \mbox{t}}
\int^{\mbox{t}_{n+1}}_{\mbox{t}_n} \bv{F} \left( \bv{Q} \left(
(i-1)\mbox{h}_x, \mbox{t} \right) \right) \mbox{dt}
\label{fvac}
\end{equation}
is the flux between $(i-1)$-th and $i$-th cells. In order to estimate it we use the following approximation. Knowing that under expansion into vacuum velocity must increase and all other quantities decrease and supposing that $\Delta \mbox{t}$ is small enough, we use a linear approximation for quantities on cell boundary
\begin{equation}
\mbox{v}^{b}\left( \mbox{t} \right)=\mbox{v}^{n}_{i-1}+\alpha \left( \mbox{t}-\mbox{t}_n \right)
\label{vac2}
\end{equation}
\begin{equation}
\varepsilon^{b}\left( \mbox{t} \right)=\varepsilon^{n}_{i-1} -\beta \left( \mbox{t}-\mbox{t}_n \right)
\label{vac3}
\end{equation}
\begin{equation}
\mbox{n}^{b}\left( \mbox{t} \right)=\mbox{n}^{n}_{i-1}-\kappa \left( \mbox{t}-\mbox{t}_n \right)
\label{vac4}
\end{equation}
\begin{equation}
\mbox{p}^{b}\left( \mbox{t} \right)=\mbox{p}^{n}_{i-1} -\gamma \left( \mbox{t}-\mbox{t}_n \right)
\label{vac5}
\end{equation}

Substituting Eqs.(\ref{vac2})-(\ref{vac5}) into Eq.(\ref{fvac}) we obtain
\begin{eqnarray}
\mbox{F}^{n}_{1(i-1)}=\frac{\left( \varepsilon^{n}_{i-1}+\mbox{p}^{n}_{i-1} \right) \mbox{v}^{n}_{i-1}}{1-\left( \mbox{v}^{n}_{i-1} \right)^2}+ \nonumber & \\ \frac{1}{2} \left[ \frac{ \left( \varepsilon^{n}_{i-1}+\mbox{p}^{n}_{i-1} \right) \left( 1+\left( \mbox{v}^{n}_{i-1} \right)^2 \right)}{\left( 1-\left( \mbox{v}^{n}_{i-1} \right)^2 \right)^2} \Delta \mbox{v}+
\frac{\left( \Delta \varepsilon + \Delta \mbox{p} \right) \mbox{v}^{n}_{i-1}}{1-\left( \mbox{v}^{n}_{i-1} \right)^2} \right]
\label{fvac1} &
\end{eqnarray}
\begin{eqnarray}
\mbox{F}^{n}_{x(i-1)}=\frac{\mbox{p}^{n}_{i-1}+\varepsilon^{n}_{i-1} \left( \mbox{v}^{n}_{i-1} \right)^2}{1-\left( \mbox{v}^{n}_{i-1} \right)^2}+\frac{1}{2} \left[ \frac{2\mbox{v}^{n}_{i-1} \left( \varepsilon^{n}_{i-1}+\mbox{p}^{n}_{i-1} \right)}{\left( 1-\left( \mbox{v}^{n}_{i-1} \right)^2 \right)^2} \Delta \mbox{v}+ \right. \nonumber & \\ \left. \frac{\Delta \mbox{p}+ \Delta \varepsilon \left( \mbox{v}^{n}_{i-1} \right)^2}{1-\left( \mbox{v}^{n}_{i-1} \right)^2} \right]
\label{fvac2} &
\end{eqnarray}
\begin{equation}
\mbox{F}^{n}_{n(i-1)}=\frac{\mbox{n}^{n}_{i-1} \mbox{v}^{n}_{i-1}}{\sqrt{1-\left( \mbox{v}^{n}_{i-1} \right)^2}} +\frac{1}{2} \left[ \frac{\mbox{n}^{n}_{i-1} \Delta \mbox{v}}{\left( 1-\left( \mbox{v}^{n}_{i-1} \right)^2 \right)^{3/2}}+\frac{\mbox{v}^{n}_{i-1} \Delta \mbox{n}}{\sqrt{1-\left( \mbox{v}^{n}_{i-1} \right)^2}} \right],
\label{fvac3}
\end{equation}
where $\Delta \mbox{A}= \mbox{A}^{b} \left( \mbox{t}_{n+1} \right)-\mbox{A}^{n}_{i-1}$. Take into account the following
\begin{eqnarray}
\bv{Q}^{n+1}_{i}=\frac{1}{\mbox{h}_x} \int^{i\mbox{h}_x}_{(i-1)\mbox{h}_x}
 \bv{Q} \left( \mbox{x}, \mbox{t}_{n+1} \right) \mbox{dx}=\frac{1}{\mbox{h}_x} \int^{(i-1)\mbox{h}_x+\mbox{v}^{vac}\Delta \mbox{t}}_{(i-1)\mbox{h}_x} \bv{Q} \left( \mbox{x}, \mbox{t}_{n+1} \right) \mbox{dx} \nonumber & \\ \approx \frac{\Delta \mbox{t}}{\mbox{h}_x} \mbox{v}^{vac} \bv{Q}^{b} \left( \mbox{t}_{n+1} \right),
\label{qvac} &
\end{eqnarray}
where $\mbox{v}^{vac}$ is the velocity of boundary with vacuum that at time $\mbox{t}_{n+1}$ is somewhere inside the cell. In order to close the system we approximately put $\mbox{v}^{vac} \approx \mbox{v}^{b} \left( \mbox{t}_{n+1} \right)$. After this we substitute (\ref{fvac1})-(\ref{qvac}) into (\ref{vac1}) and obtain algebraical system of 3 equations for $\varepsilon^{b} \left( \mbox{t}_{n+1} \right), \mbox{n}^{b} \left( \mbox{t}_{n+1} \right),\mbox{v}^{b} \left( \mbox{t}_{n+1} \right)$. After solving it by Newton method and using relation (\ref{qvac}) we obtain $\bv{Q}^{n+1}_{i}$. In a corrector step we use $\bv{Q}^{(l)}_{i-1}$ instead of $\bv{Q}^{n}_{i-1}$. 

The case when vacuum is located on the right side from the medium is considered similarly.

Note, that after we have found $\bv{Q}^{n+1}_i$, there will be non-zero $\bv{Q}^{n+2}_{i+1}$. But in reality boundary with vacuum crosses the boundary of between the $i$-th and $\left( i+1 \right)$-th cells in several time steps. In order to avoid such artificial superlight expansion of matter, we should allow matter to expand into $(i+1)$-th cell only after boundary with vacuum reaches the right boundary of $i$-th cell.

\section{Three-dimensional scheme: the role of tangential velocities}

In three-dimensional case for one cell we have to solve three Riemann problems - in $\mbox{x}$-, $\mbox{y}$- and $\mbox{z}$- directions. As an example consider a Riemann problem in $x$- direction. Then in Fig.~\ref{f:waves} on the left and right states there are tangential components of velocity- $\mbox{v}_y^{(l)}$, $\mbox{v}_z^{(l)}$ and $\mbox{v}_y^{(r)}$, $\mbox{v}_z^{(r)}$ respectively. These components constitute vectors $\bv{v}_t^{(l)}$ and $\bv{v}_t^{(r)}$. As a result, the system (\ref{1})-(\ref{3}) has 13, not 9, equations. The structure of wave pattern is the same as in one-dimensional case (see Fig.\ref{f:waves}). It is known that due to Lorentz factor coupling tangential velocities are incorporated into solution of the Riemann problem \cite{martimuller94,ponsetal00}. This makes solution of exact equations rather numerically expensive. In order to simplify things and reduce to one-dimensional Riemann problem we make an approximation that tangential velocities are continuous over shock waves and simple rarefactions but undergo a jump over a contact discontinuity. It means that $\bv{v}_{t}'=\bv{v}_t^{(l)}$ and $\bv{v}_{t}''=\bv{v}_t^{(r)}$. Having this we can again reduce in (\ref{1})-(\ref{3}) to 9 equations and taking into account that Eq. (\ref{2}) is an identity to 7. To make things completely similar to one-dimensional case we must  use Eq. (\ref{1}) in the reference frame where $\bv{v}_t^{(l)}=0$. In this case we have to substitute $\displaystyle \mbox{v}^{(l)} \rightarrow \mbox{v}^{(l)}_{m}=\mbox{v}^{(l)}_x/\sqrt{1-\left( \bv{v}_t^{(l)} \right)^2}$ and $\displaystyle \mbox{v}' \rightarrow \mbox{v}'_{m}=\mbox{v}'_x/\sqrt{1-\left( \bv{v}_t^{(l)} \right)^2}$. Eq. (\ref{3}) thus must be used in reference frame where $\bv{v}_t^{(r)}=0$ with substitutions $\displaystyle \mbox{v}^{(r)} \rightarrow \mbox{v}^{(r)}_{m}=\mbox{v}^{(r)}_x/\sqrt{1-\left( \bv{v}_t^{(r)} \right)^2}$, $\displaystyle \mbox{v}'' \rightarrow \mbox{v}''_{m}=\mbox{v}''_x/\sqrt{1-\left( \bv{v}_t^{(r)} \right)^2}$. After this we have again one-dimensional system of equations. Correspondingly, condition of normal velocity continuity over contact discontinuity $\mbox{v}'_x=\mbox{v}''_x$ takes form $\displaystyle \mbox{v}'_m\sqrt{1-\left( \bv{v}_t^{(l)} \right)^2}=\mbox{v}''_m\sqrt{1-\left( \bv{v}_t^{(r)} \right)^2}$.

All the rest of the scheme can be reduced to one-dimensional case described in Section 2 in similar way. For example Eq. (\ref{fourth}) can be rewritten as
\begin{eqnarray}
\frac{\mbox{v}^{(l)}_m+\mbox{v}^{\pm}_{12}\left( \varepsilon^{(l)}, \mbox{n}^{(l)}, \varepsilon',
\mbox{n}' \right)}{1+\mbox{v}^{(l)}_m\mbox{v}^{\pm}_{12}\left( \varepsilon^{(l)}, \mbox{n}^{(l)}, \varepsilon', \mbox{n}' \right)}\sqrt{1-\left( \bv{v}_t^{(l)} \right)^2}= \frac{\mbox{v}^{(r)}_m+\mbox{v}^{\pm}_{12}\left( \varepsilon^{(r)}, \mbox{n}^{(r)}, \varepsilon'',
\mbox{n}'' \right)}{1+\mbox{v}^{(r)}_m\mbox{v}^{\pm}_{12}\left( \varepsilon^{(r)}, \mbox{n}^{(r)}, \varepsilon'', \mbox{n}'' \right)}\sqrt{1-\left( \bv{v}_t^{(r)} \right)^2} \nonumber & \\
&
\label{fourth1} &
\end{eqnarray}
Similarly we find fluxes in $\mbox{y}$- and $\mbox{z}$-directions and complete a time step for fully three-dimensional hydrodynamics equations.

\section{Conservative and primitive variables}

In Eq.(\ref{exact}) conservative variables $\bv{Q}$ are updated. But in the Riemann solver we use primitive variables $\bv{P}=\left( \varepsilon, \mbox{n}, \bv{v} \right)$. It means that at every time step we have to convert conservative variables into primitive ones.

Conservative variables are defined by Eq.(\ref{Q}) and one can express thermodynamical parameters $\varepsilon$ and $\mbox{n}$ via velocities
\begin{equation}
\mbox{n}=\mbox{Q}_n \sqrt{1-\bv{v}^2}
\label{pn}
\end{equation}
\begin{equation}
\varepsilon=\mbox{Q}_1-\mbox{Q}_x\mbox{v}_x-\mbox{Q}_y\mbox{v}_y-\mbox{Q}_z\mbox{v}_z
\label{pe}
\end{equation}
Consider as an example case when $\mbox{Q}_x \neq 0$. Then it is obvious that
\begin{equation}
\mbox{v}_y=\frac{\mbox{Q}_y \mbox{v}_x}{\mbox{Q}_x}, \mbox{   } \mbox{v}_z=\frac{\mbox{Q}_z \mbox{v}_x}{\mbox{Q}_x}
\label{vyz}
\end{equation}
One can check that the following relation holds
\begin{equation}
\mbox{Q}_x-\mbox{Q}_1 \mbox{v}_x -\mbox{p} \left( \varepsilon, \mbox{n} \right) \mbox{v}_x=0
\label{pq}
\end{equation}
Substituting in it relations (\ref{pn})-(\ref{vyz}) we have just one single algebraical equation for $\mbox{v}_x$. It can be easily solved by Newton method.

\section{Source term}

In our numerical scheme we include the possibility for presence of a source term. It means that we solve the following system of equations
\begin{equation}
\pdif{\bv{Q}}{\mbox{t}}+\pdif{\bv{F(Q)}}{\mbox{x}}+
\pdif{\bv{G(Q)}}{\mbox{y}}+\pdif{\bv{H(Q)}}{\mbox{z}}=\bv{S} \left( \bv{Q},\bv{r},t \right),
\label{cons_t}
\end{equation}
where $\bv{S} \left( \bv{Q},\bv{r},t \right)$ is a source term.

In order to solve such a system we use the standard fractional step method. It means that as a corrector step we solve equations without source term
\begin{equation}
\pdif{\bv{Q}^{p}}{\mbox{t}}+\pdif{\bv{F}(\bv{Q}^{p})}{\mbox{x}}+
\pdif{\bv{G}(\bv{Q}^{p})}{\mbox{y}}+\pdif{\bv{H}(\bv{Q}^{p})}{\mbox{z}}=0.
\label{sq}
\end{equation}
We just have constructed the procedure how to solve this system. As a result we obtain $\left(\bv{Q}^p \right)^{n+1}_i$. Then we use these values as initial conditions for the following system of ordinary differential equations
\begin{equation}
\frac{\mbox{d} \bv{Q}}{\mbox{dt}}=\bv{S} \left( \bv{Q},\bv{r},t \right)
\label{sc}
\end{equation}
To solve the system (\ref{sc}) we use a Runge-Kutta method. Time step is $\Delta \mbox{t}$. This solution gives us numerical solution of system (\ref{cons_t}).

\section{Test simulations: One dimension}

\subsection{Shock tube problems}

Here we consider two shock tube problems. Equation of state is $\mbox{p}=\mbox{c}_c^2 \varepsilon$ with $\mbox{c}_s^2=1/3$.

\begin{figure*}
\begin{center}
\begin{tabular}{ll}
(a) & (b) \\ \epsfysize=4.5cm \epsffile{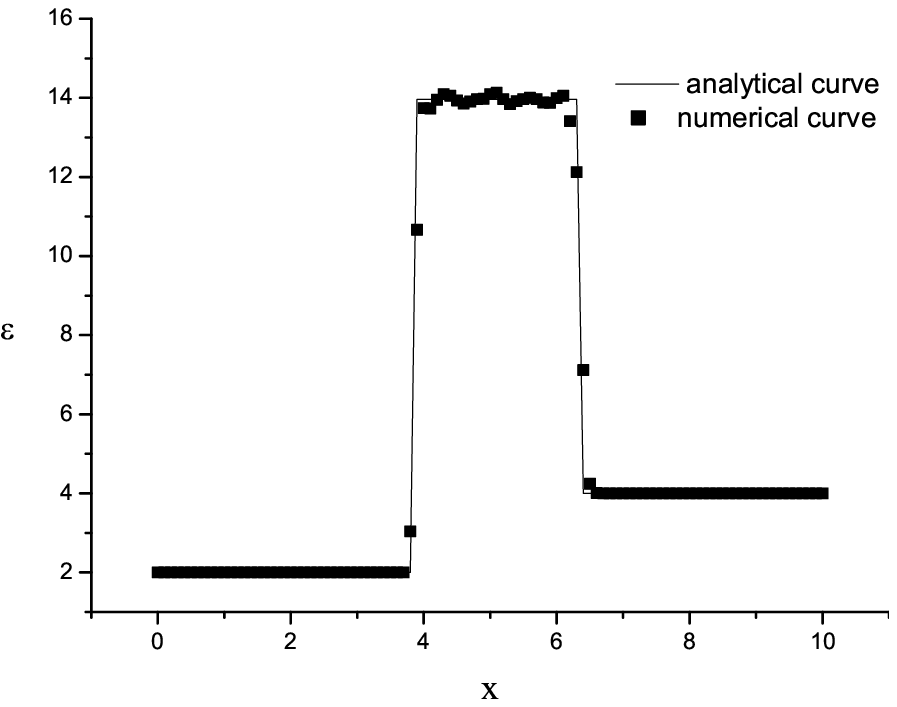} &
\epsfysize=4.5cm \epsffile{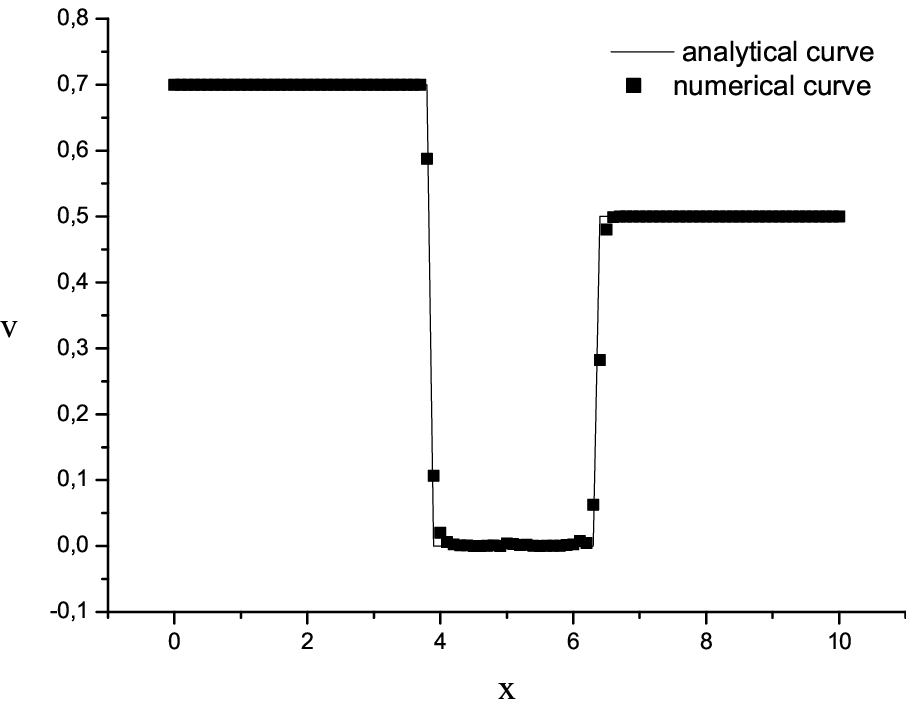} \\
(c) & (d) \\
\epsfysize=4.5cm \epsffile{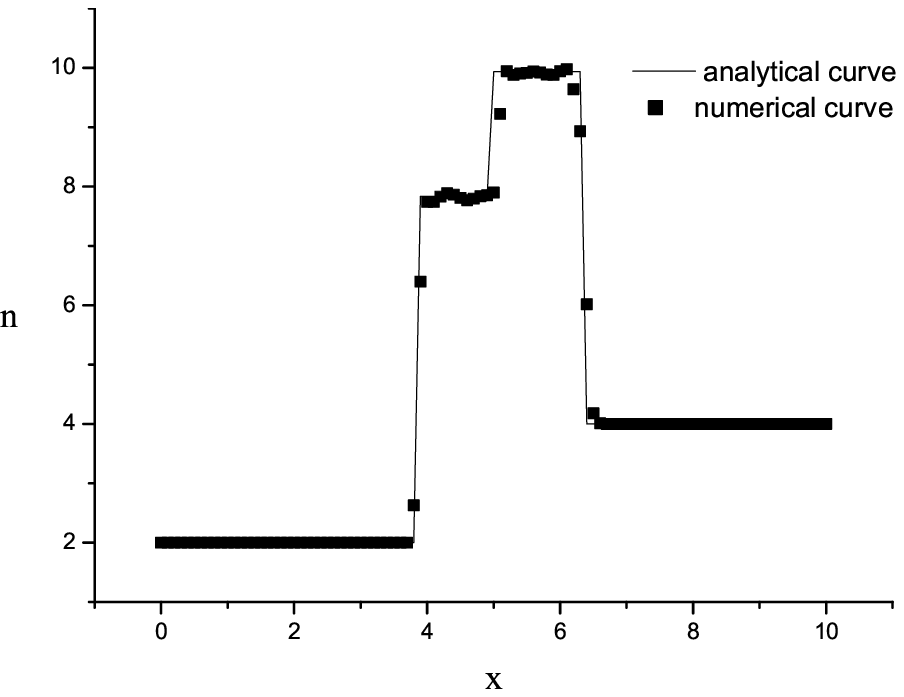} & \epsfysize=4.5cm \epsffile{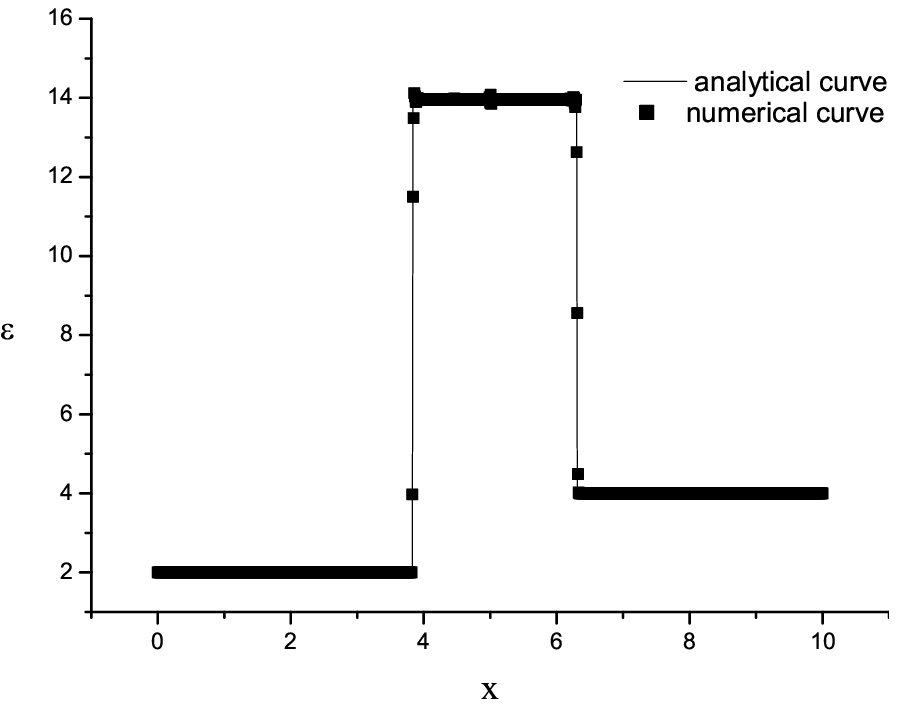} \\
(e) & (f) \\
\epsfysize=4.5cm \epsffile{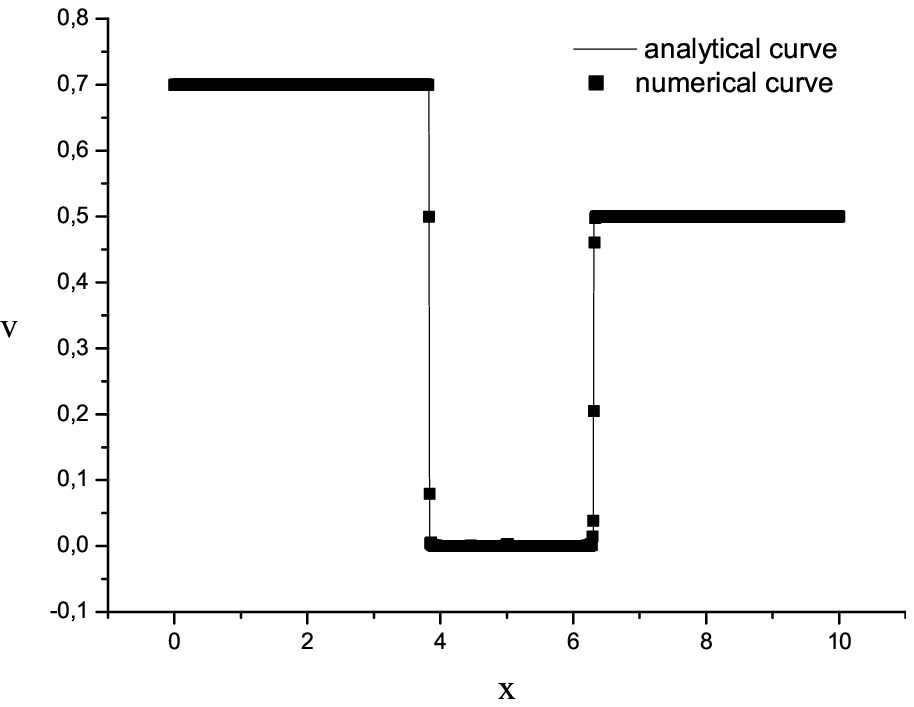} & \epsfysize=4.5cm \epsffile{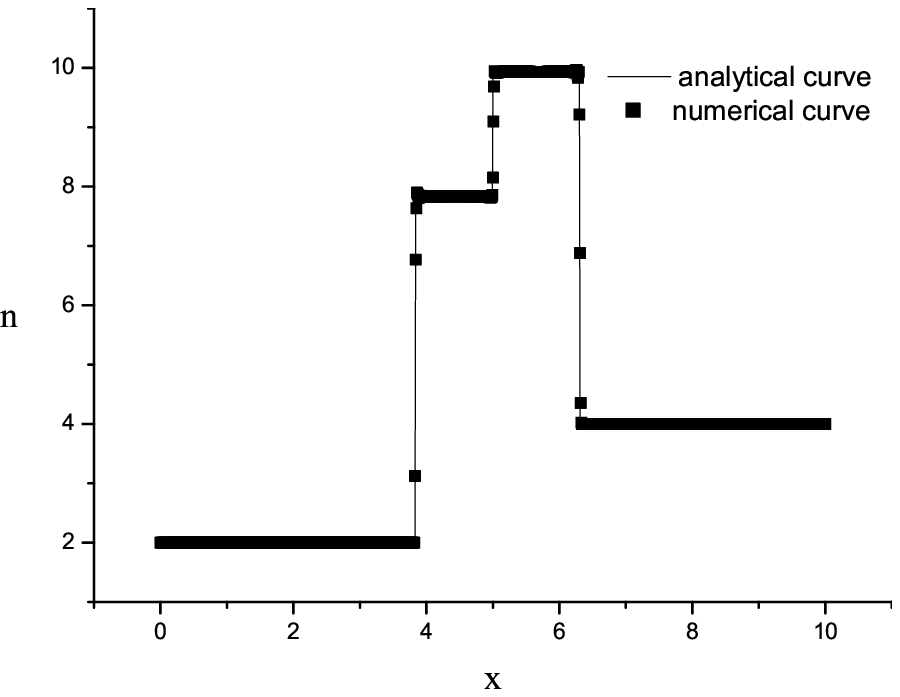} \\
\end{tabular}
\end{center}
\caption{\scriptsize Results of numerical simulations for a shock tube problem with following parameters: $\varepsilon^{(l)}=2$, $\mbox{v}^{(l)}=0.7$, $\mbox{n}^{(l)}=2$, $\varepsilon^{(r)}=4$, $\mbox{v}^{(r)}=-0.5$, $\mbox{n}^{(r)}=4$. Initial moment of time is $\mbox{t}_0=0$, final point of time is $\mbox{t}=3$. All quantities are taken at time $\mbox{t}$. (a)-(c) are energy density, velocity and baryonic charge density correspondingly. (d)-(f) are the same quantities but calculated with better spatial resolution. (a)-(c) have 100 grid points and (d)-(f) have 1000 grid points in space.}
\label{sh-sh}
\end{figure*}

The first problem is collision of two gas masses that gives two shock waves propagating to opposite sides. To model this situation we take the following parameters: $\varepsilon^{(l)}=2$, $\mbox{v}^{(l)}=0.7$, $\mbox{n}^{(l)}=2$ to the left and $\varepsilon^{(r)}=4$, $\mbox{v}^{(r)}=-0.5$, $\mbox{n}^{(r)}=4$ to the right. Results of numerical simulations are depicted on Fig~\ref{sh-sh}.

\begin{figure*}
\begin{center}
\begin{tabular}{ll}
(a) & (b) \\ \epsfysize=4.5cm \epsffile{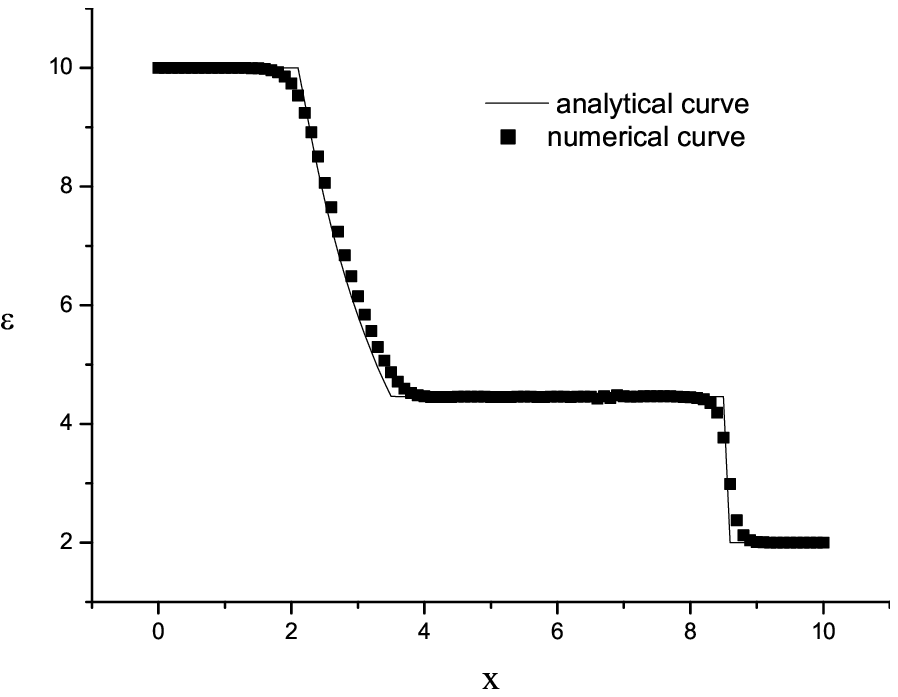} &
\epsfysize=4.5cm \epsffile{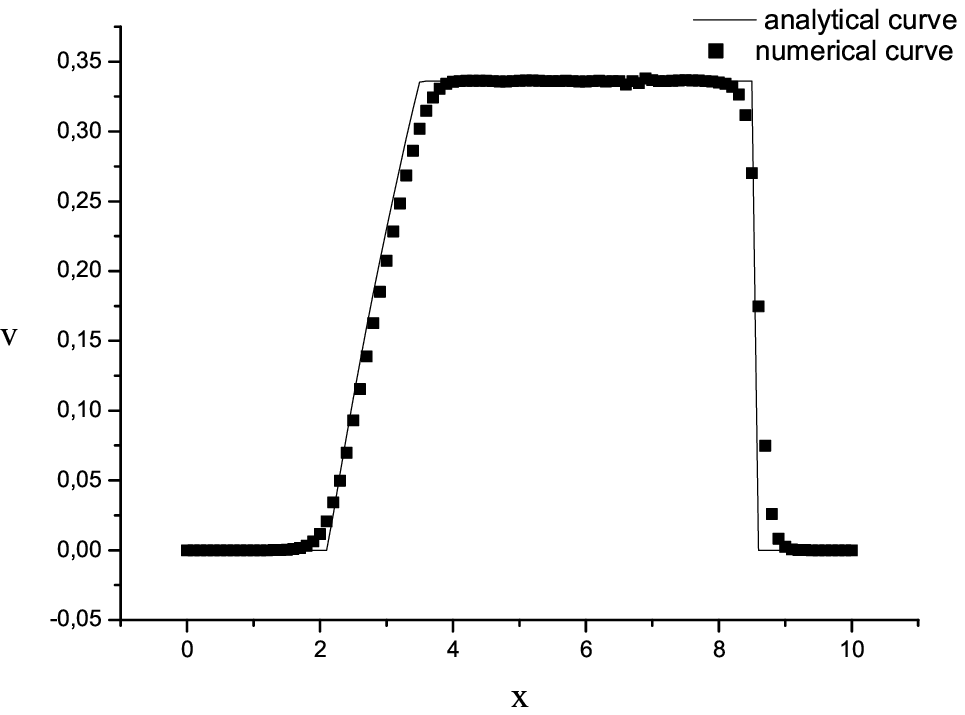} \\
(c) & (d) \\
\epsfysize=4.5cm \epsffile{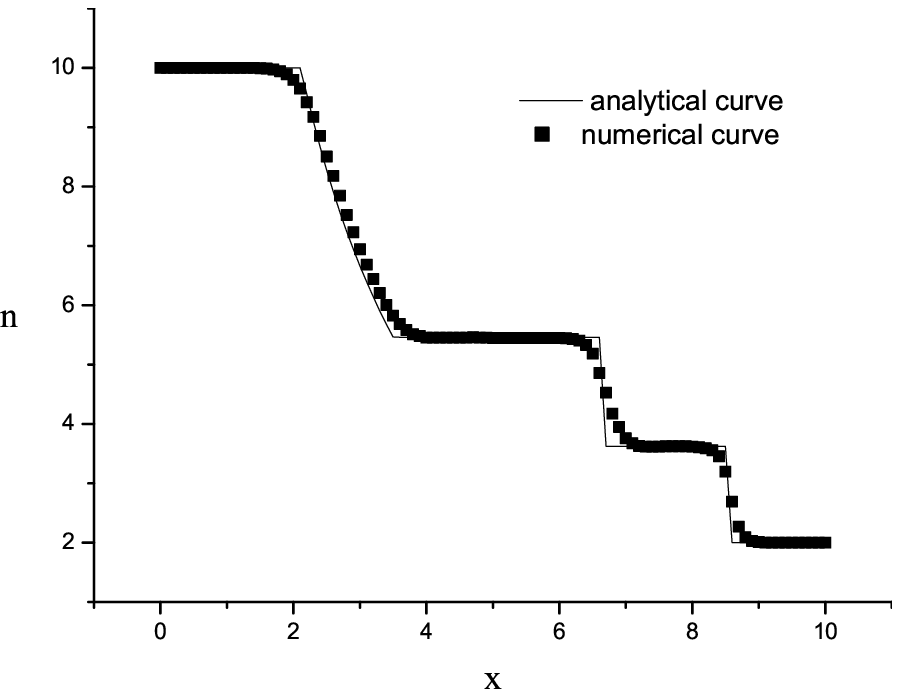} & \epsfysize=4.5cm \epsffile{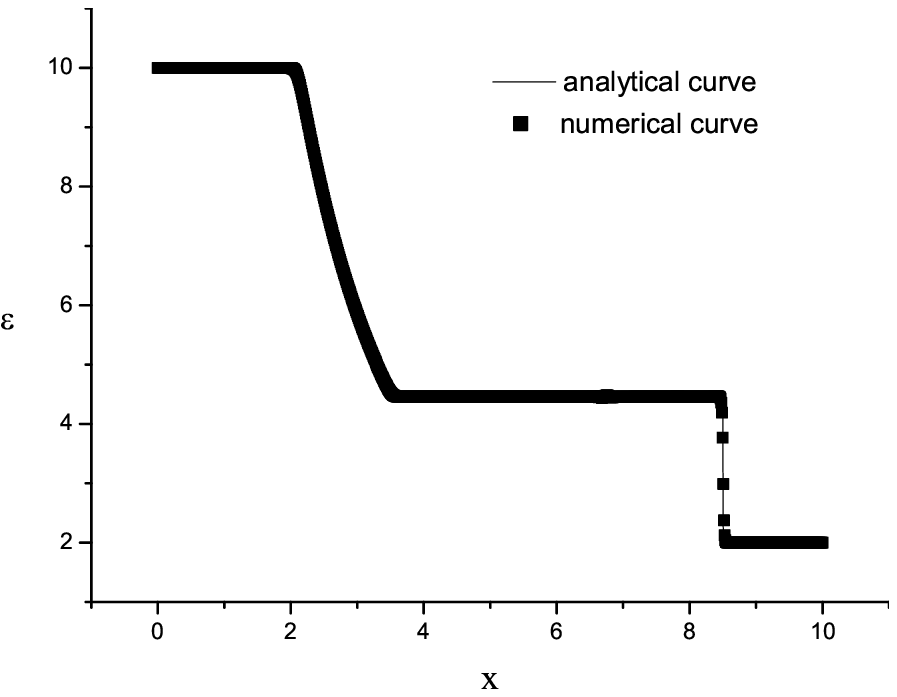} \\
(e) & (f) \\
\epsfysize=4.5cm \epsffile{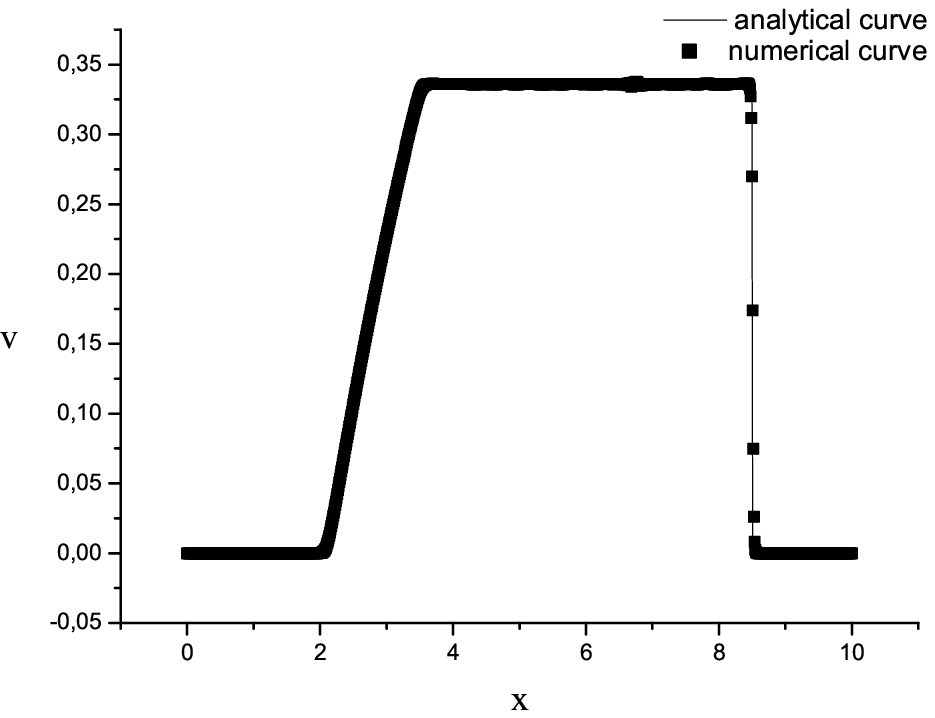} & \epsfysize=4.5cm \epsffile{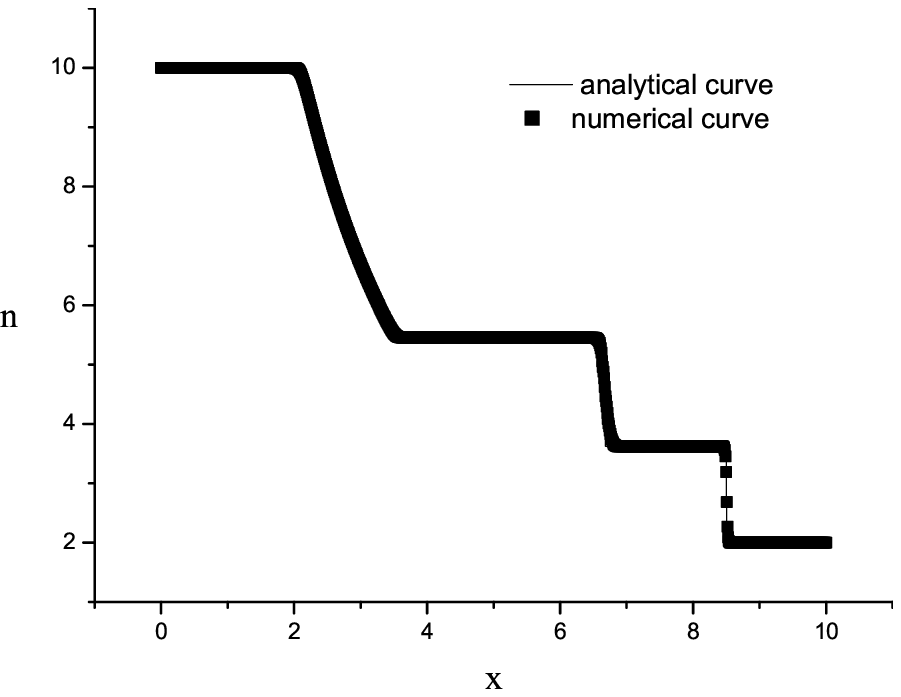} \\
\end{tabular}
\end{center}
\caption{\scriptsize Results of numerical simulations for a shock tube problem with following parameters: $\varepsilon^{(l)}=10$, $\mbox{v}^{(l)}=0$, $\mbox{n}^{(l)}=10$, $\varepsilon^{(r)}=2$, $\mbox{v}^{(r)}=0$, $\mbox{n}^{(r)}=2$. Initial moment of time is $\mbox{t}_0=0$, final point of time is $\mbox{t}=5$. All quantities are taken at time $\mbox{t}$. (a)-(c) are energy density, velocity and baryonic charge density correspondingly. (d)-(f) are the same quantities but calculated with better spatial resolution. (a)-(c) have 100 grid points and (d)-(f) have 1000 grid points in space.}
\label{rar-sh}
\end{figure*}

The second problem is the contact of two motionless masses of gas with different densities. This can give a rarefaction wave and a shock wave. To model this situation we take the following parameters: $\varepsilon^{(l)}=10$, $\mbox{v}^{(l)}=0$, $\mbox{n}^{(l)}=10$ to the left and $\varepsilon^{(r)}=2$, $\mbox{v}^{(r)}=0$, $\mbox{n}^{(r)}=2$ to the right. Results of numerical simulations are depicted on Fig~\ref{rar-sh}.

\subsection{Bjorken-like flow}

In this subsection we numerically simulate Bjorken-like expansion solution. This solution was introduced and analyzed in Refs. \cite{hwa,gorensteinetal,bjorken}. Equation of state is $\mbox{p}=\mbox{c}_s^2 \varepsilon$. Solution has the following form
\begin{equation}
\varepsilon=\varepsilon_0 \left( \frac{\tau_0}{\tau} \right)^{1+\mbox{c}_s^2}
\label{be}
\end{equation}
\begin{equation}
\mbox{n}=\mbox{n}_0 \frac{\tau_0}{\tau}
\label{bn}
\end{equation}
\begin{equation}
\mbox{v}=\frac{\mbox{x}}{\mbox{t}},
\label{bv}
\end{equation}
where $\displaystyle \tau=\sqrt{\mbox{t}^2-\mbox{x}^2}$.

\begin{figure*}
\begin{center}
\begin{tabular}{ll}
(a) & (b) \\ \epsfysize=4.5cm \epsffile{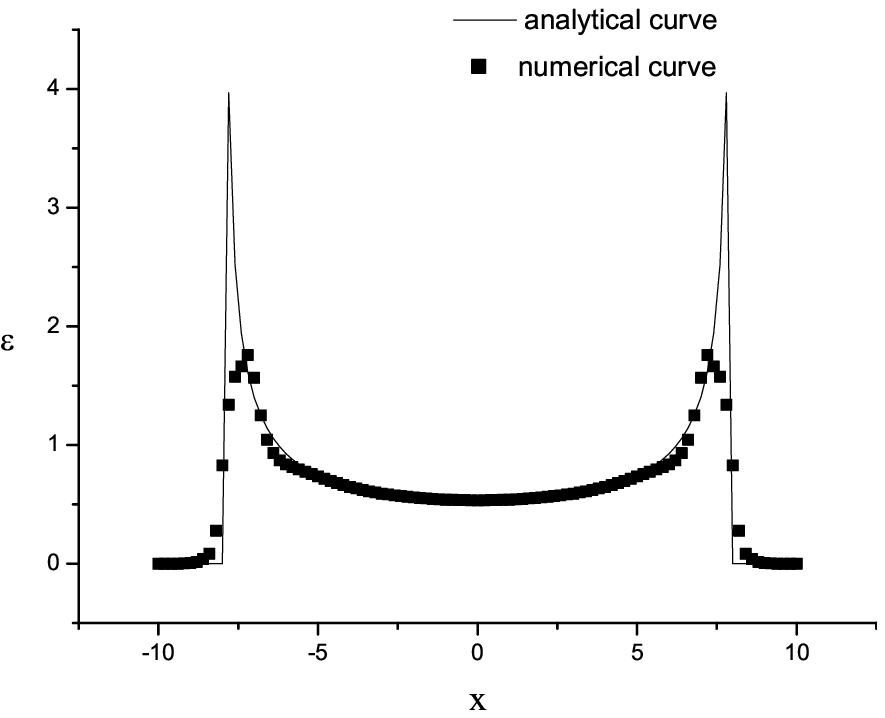} &
\epsfysize=4.5cm \epsffile{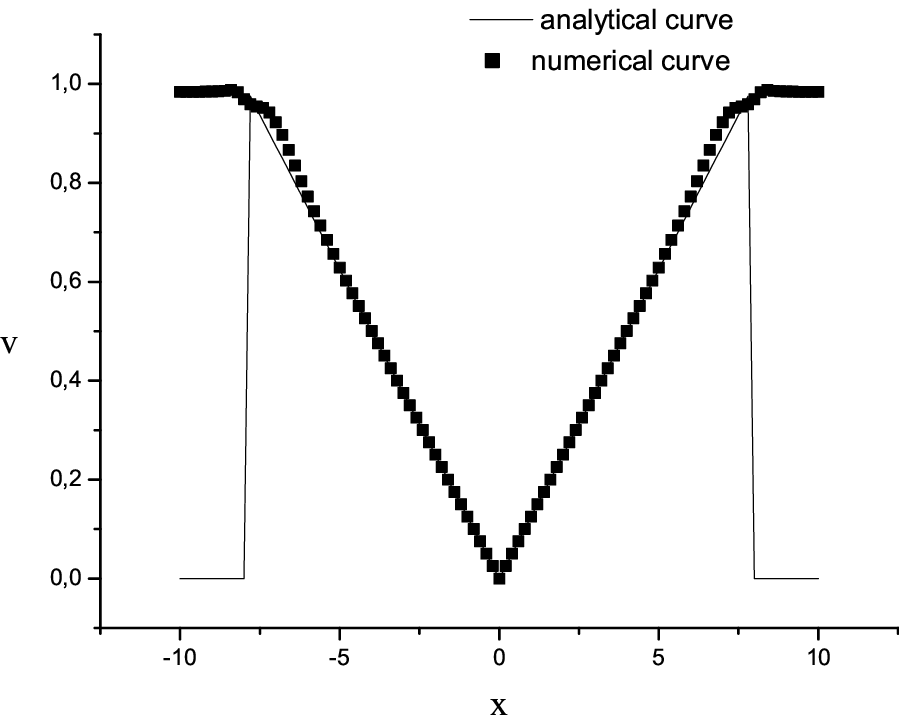} \\
\multicolumn{2}{l}{(c)} \\
\multicolumn{2}{c}{\epsfysize=4.5cm \epsffile{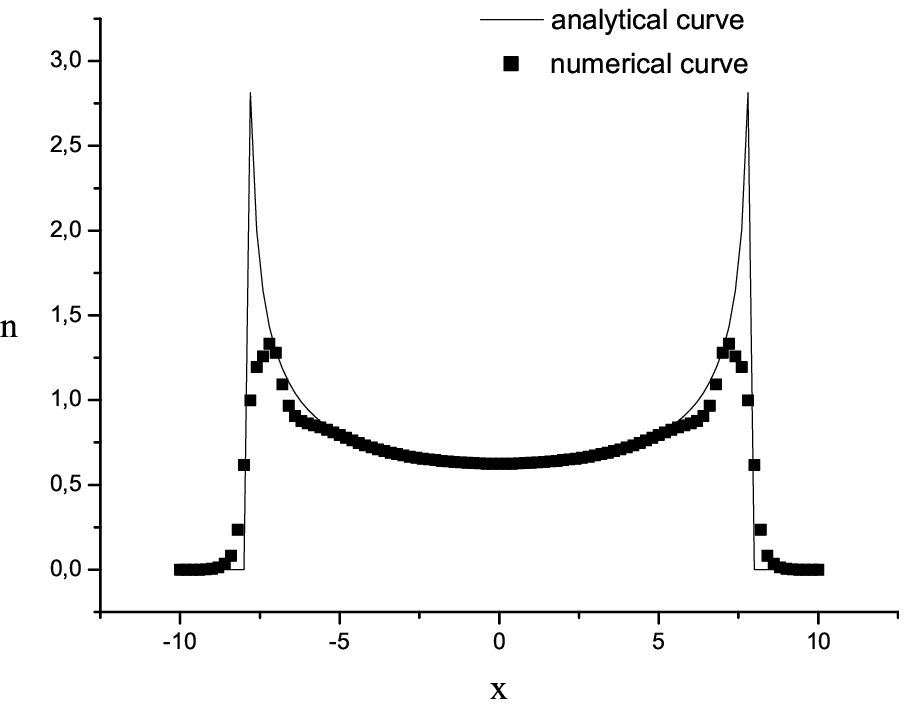}} \\
\end{tabular}
\end{center}
\caption{\scriptsize Results of numerical simulations for a Bjorken-like flow. (a) is energy density, (b) is velocity and (c) is baryonic charge density. Evolution starts  at time $\mbox{t}_0=5$ and is finished at time $\mbox{t}=8$. All the graphs are taken at time $\mbox{t}$. One can see a rarefaction wave expanding into vacuum and distorting analytical solution on the edges. In the middle of the region numerical and analytical solutions coincide.}
\label{f:bjorken}
\end{figure*}

We take the following parameters: $\mbox{c}_s^2=1/3$, $\tau_0=5$, $\mbox{n}_0=1$ and $\varepsilon_0=1$. Initial time is $\mbox{t}_0=5$ and final time is $\mbox{t}=8$. At moment of time $\mbox{t}_0$ we take all the quantities in region $\left| \mbox{x} \right| < \mbox{t}_0-\alpha$, where $\alpha$ is small enough, being defined by relations (\ref{be})-(\ref{bv}). Outside this region all the quantities are put to zero. Results of numerical simulations are depicted on Fig.~\ref{f:bjorken}.

\subsection{Hubble-like flow}

Another example is Hubble-like flow. It is constructed similarly to Bjorken-like solution, see Ref. \cite{chiuetal}, except that Hubble-like solution is spherically symmetric. It means that $\mbox{x}$ in system (\ref{one-dim}) means radial coordinate and here arises geometrical source term $\bv{S} \left( \bv{Q},\mbox{x},\mbox{t} \right)$ that has the following form
\begin{equation}
\bv{S} \left( \bv{Q},\mbox{x},\mbox{t} \right)=\left( -\frac{2\left( \varepsilon +\mbox{p} \right) \mbox{v}}{\mbox{x}\left( 1-\mbox{v}^2 \right)}, -\frac{2\left( \varepsilon +\mbox{p} \right) \mbox{v}^2}{\mbox{x}\left( 1-\mbox{v}^2 \right)}, -\frac{2\mbox{nv}}{\mbox{x} \sqrt{1-\mbox{v}^2}} \right).
\label{sr}
\end{equation}

Solution itself has the following form
\begin{equation}
\varepsilon=\varepsilon_0 \left( \frac{\tau_0}{\tau} \right)^{3\left( 1+\mbox{c}_s^2 \right)}
\label{he}
\end{equation}
\begin{equation}
\mbox{n}=\mbox{n}_0 \left( \frac{\tau_0}{\tau} \right)^3
\label{hn}
\end{equation}
\begin{equation}
\mbox{v}=\frac{\mbox{x}}{\mbox{t}},
\label{hv}
\end{equation}
where $\displaystyle \tau=\sqrt{\mbox{t}^2-\mbox{x}^2}$.

\begin{figure*}
\begin{center}
\begin{tabular}{ll}
(a) & (b) \\ \epsfysize=4.5cm \epsffile{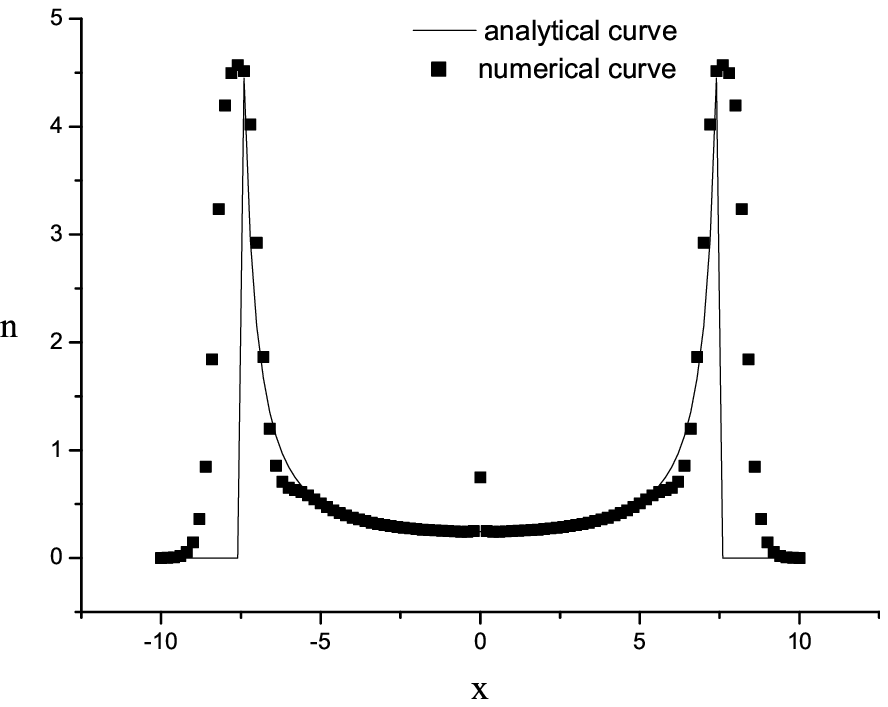} &
\epsfysize=4.5cm \epsffile{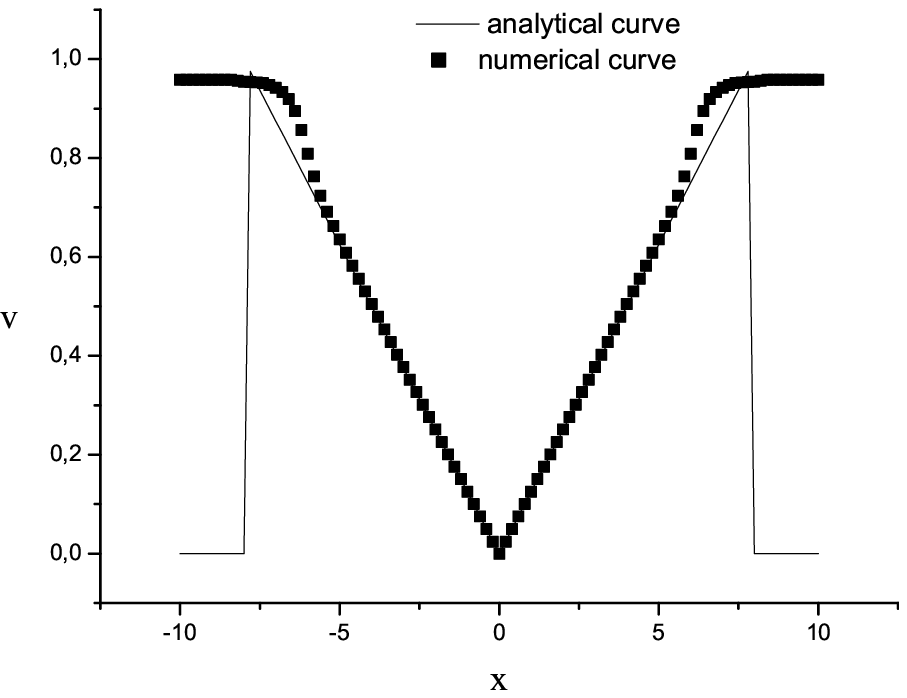} \\
\multicolumn{2}{l}{(c)} \\
\multicolumn{2}{c}{\epsfysize=4.5cm \epsffile{n_hubble1}} \\
\end{tabular}
\end{center}
\caption{\scriptsize Results of numerical simulations for a Hubble-like flow. (a) is energy density, (b) is velocity and (c) is baryonic charge density. Evolution starts  at time $\mbox{t}_0=5$ and is finished at time $\mbox{t}=8$. All the graphs are taken at time $\mbox{t}$. One can see a rarefaction wave expanding into vacuum and distorting analytical solution on the edges. In the middle of the region numerical and analytical solutions coincide. The difference is at $\mbox{x}=0$ because source term (\ref{sr}) has there singularity.}
\label{f:hubble}
\end{figure*}

We take the following parameters: $\mbox{c}_s^2=1/3$, $\tau_0=5$, $\mbox{n}_0=1$ and $\varepsilon_0=1$. Initial time is $\mbox{t}_0=5$ and final time is $\mbox{t}=8$. At moment of time $\mbox{t}_0$ we take all the quantities in region $\left| \mbox{x} \right| < \mbox{t}_0-\alpha$, where $\alpha$ is small enough, being defined by relations (\ref{he})-(\ref{hv}). Outside this region all the quantities are put to zero. An artificial extension of solution is made into region $\mbox{x}<0$ in order to avoid setting boundary conditions at $\mbox{x}=0$. Results of numerical simulations are depicted on Fig.~\ref{f:hubble}.

\subsection{Spherically symmetric solution with $\mbox{p}=0$}

In the paper by Sinyukov \& Karpenko \cite{sinyukovkarpenko} a new class of 3D anisotropic analytical solutions with quasi-inertial flows has been found. Here we consider a spherically symmetric solution for expanding finite
system with equation of state $\mbox{p}=0$ that  belongs to this new class. For testing of our code we choose the following specific form of the solution
\begin{equation}
\varepsilon=\frac{\mbox{c}_e}{\left( \mbox{t}+\mbox{t}_x \right)^3} e^{-\frac{\mbox{b}_e^2\left( \mbox{t}+\mbox{t}_x \right)^2}{\left( \mbox{t}+\mbox{t}_x \right)^2-\mbox{x}^2}}
\label{se}
\end{equation}
\begin{equation}
\mbox{v}=\frac{\mbox{x}}{\mbox{t}+\mbox{t}_x}
\label{sv}
\end{equation}
\begin{equation}
\mbox{n}=\frac{\mbox{c}_n}{\left( \mbox{t}+\mbox{t}_x \right)^3} e^{-\frac{\mbox{b}_n^2\left( \mbox{t}+\mbox{t}_x \right)^2}{\left( \mbox{t}+\mbox{t}_x \right)^2-\mbox{x}^2}}
\label{cn}
\end{equation}

\begin{figure*}
\begin{center}
\begin{tabular}{ll}
(a) & (b) \\ \epsfysize=4.5cm \epsffile{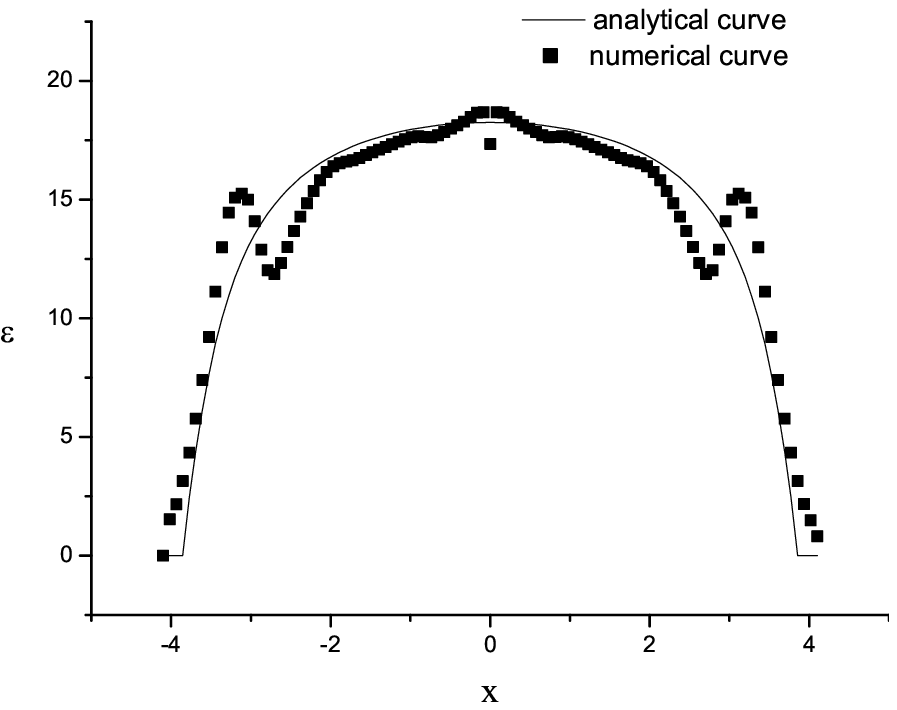} &
\epsfysize=4.5cm \epsffile{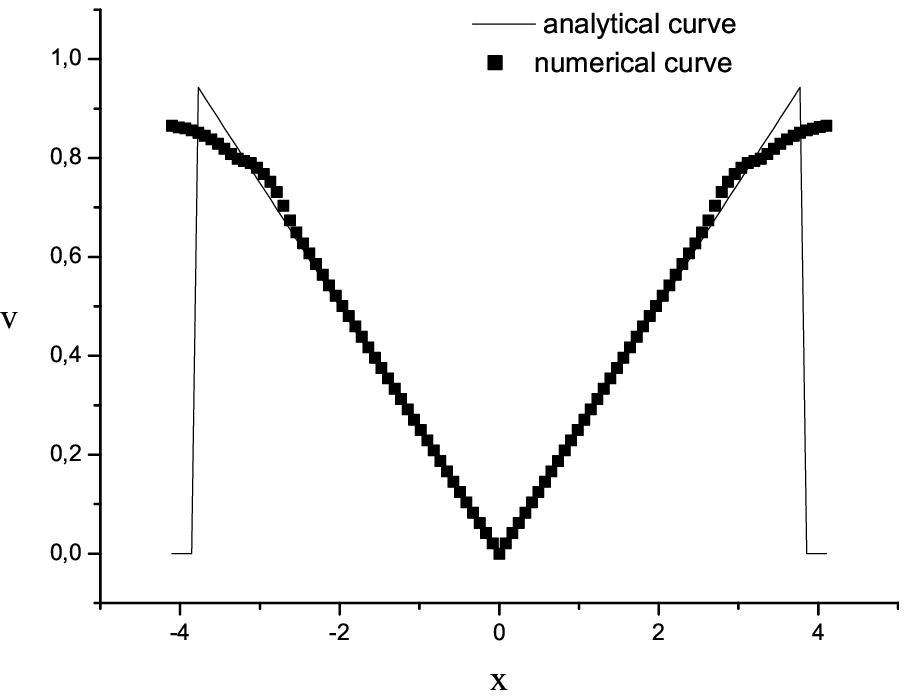} \\
(c) & (d) \\
\epsfysize=4.5cm \epsffile{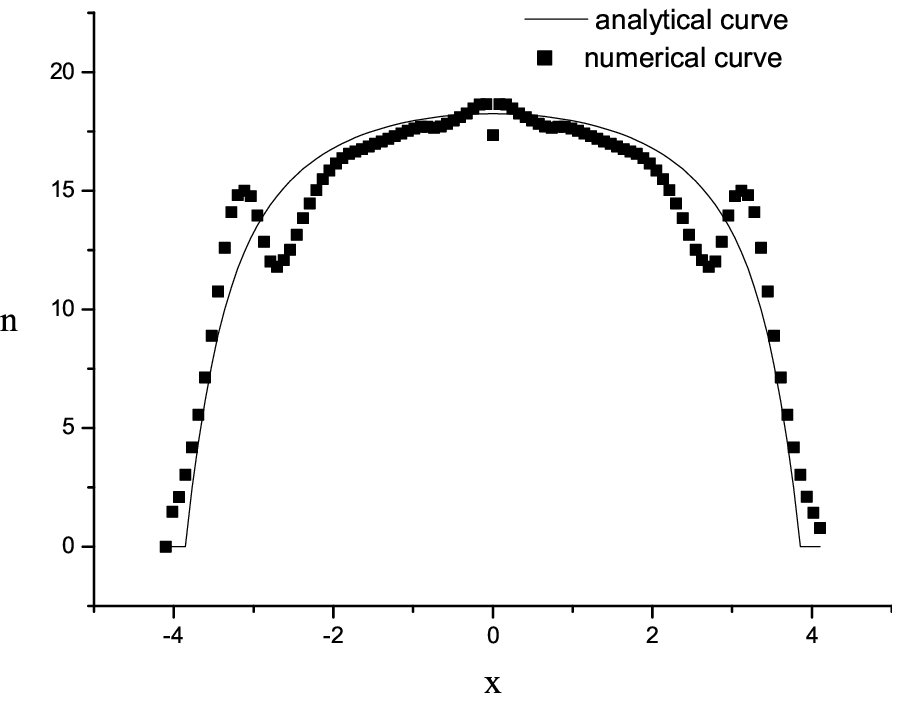} & \epsfysize=4.5cm \epsffile{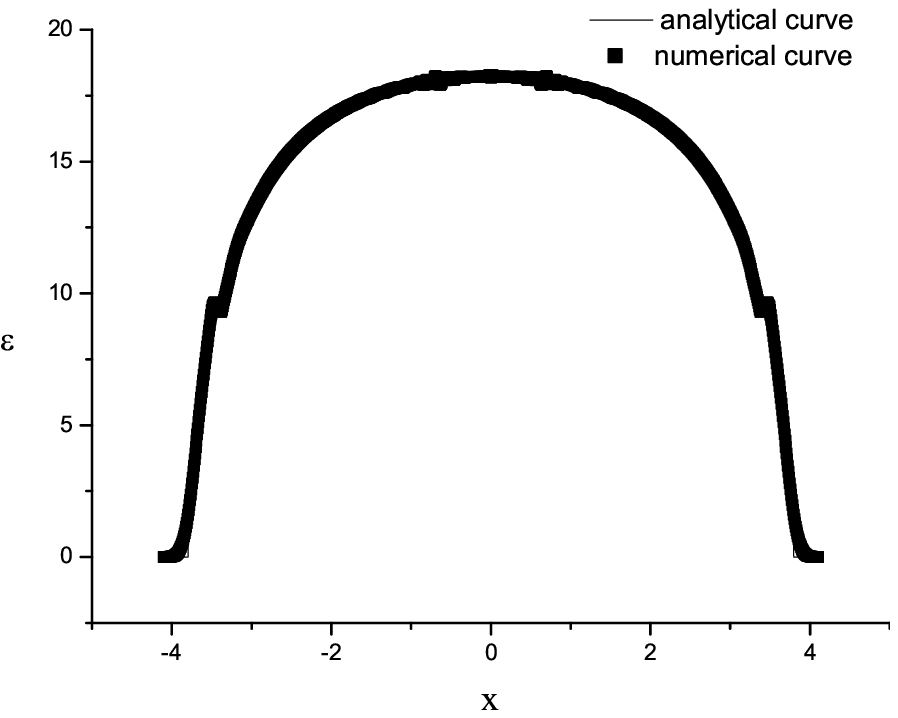} \\
\end{tabular}
\end{center}
\caption{\scriptsize Results of numerical simulations for a spherically symmetric flow with $\mbox{p}=0$. (a) is energy density, (b) is velocity and (c) is baryonic charge density. Evolution starts  at time $\mbox{t}_0=0$ and is finished at time $\mbox{t}=3$. All the graphs are taken at time $\mbox{t}$. In graphs (a)-(c) number of cells in space grid is 100. In graph (d) is depicted energy density simulated with number of cells in space grid being 1000. It is seen that irregularities that exist for 100 grid cells almost disappear if take 1000 grid points}
\label{f:symmetric}
\end{figure*}

Here as in previous subsection exists geometrical source term. We take the following parameters: $\mbox{c}_e=1500$, $\mbox{c}_n=1500$, $\mbox{b}_e^2=0.5$, $\mbox{b}_n^2=0.5$ and $\mbox{t}_x=1$. Initial time is $\mbox{t}_0=0$ and final time is $\mbox{t}=3$. At moment of time $\mbox{t}_0$ we take all the quantities in region $\left| \mbox{x} \right| < \mbox{t}_0+\mbox{t}_x-\alpha$, where $\alpha$ is small enough, being defined by relations (\ref{se})-(\ref{sv}). Outside this region all the quantities are put to zero. An artificial extension of solution is made into region $\mbox{x}<0$ in order to avoid setting boundary conditions at $\mbox{x}=0$. Results of numerical simulations are depicted on Fig.~\ref{f:symmetric}.

\section{Test simulations: three dimensions}

\subsection{Hubble-like flow}

\begin{figure*}
\begin{center}
\begin{tabular}{ll}
(a) & (b) \\ \epsfysize=4.0cm \epsffile{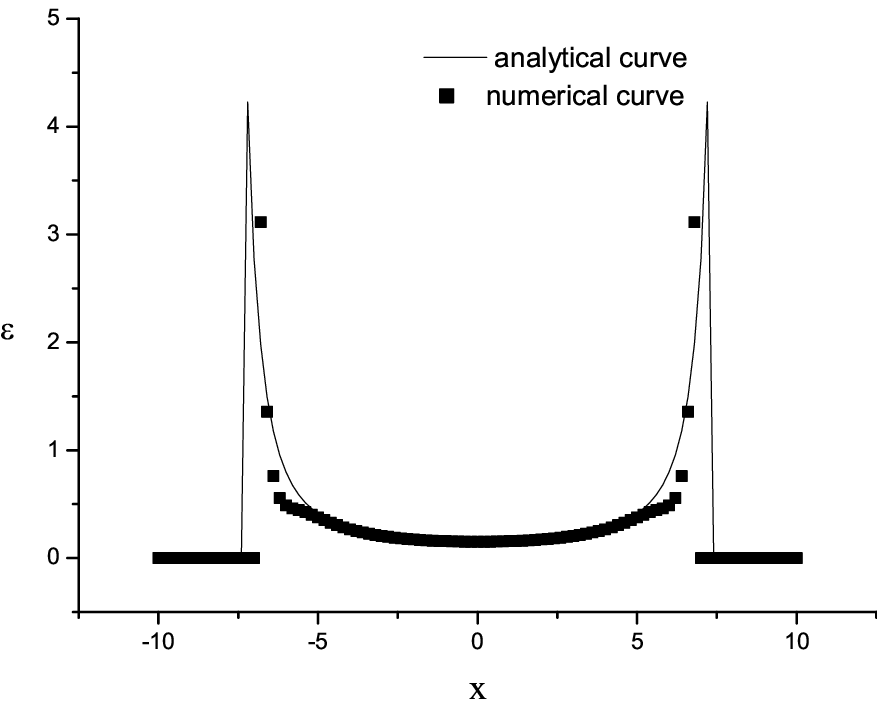} &
\epsfysize=4.0cm \epsffile{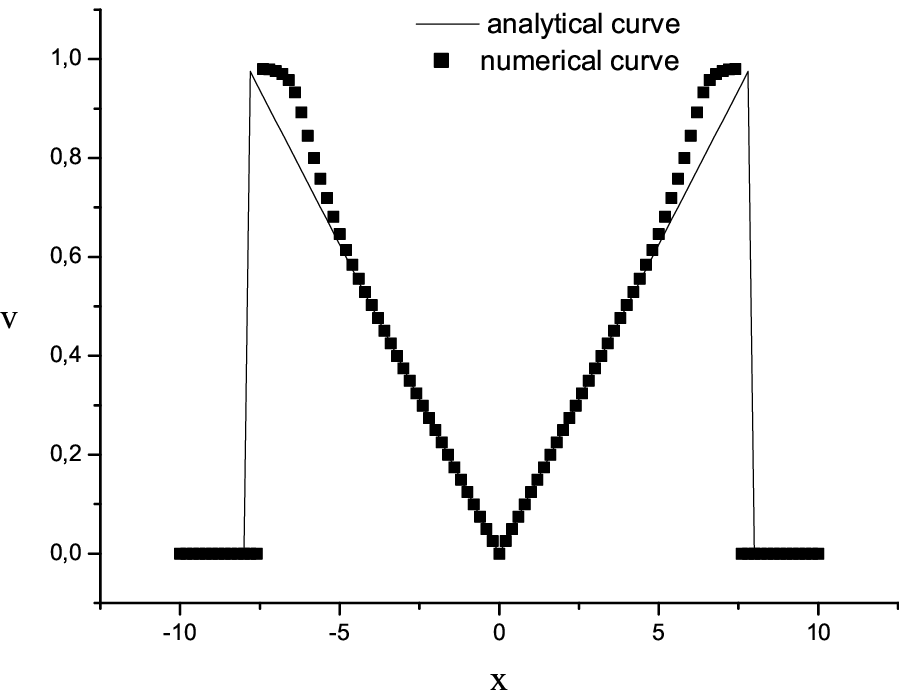} \\
(c) & (d) \\
\epsfysize=4.0cm \epsffile{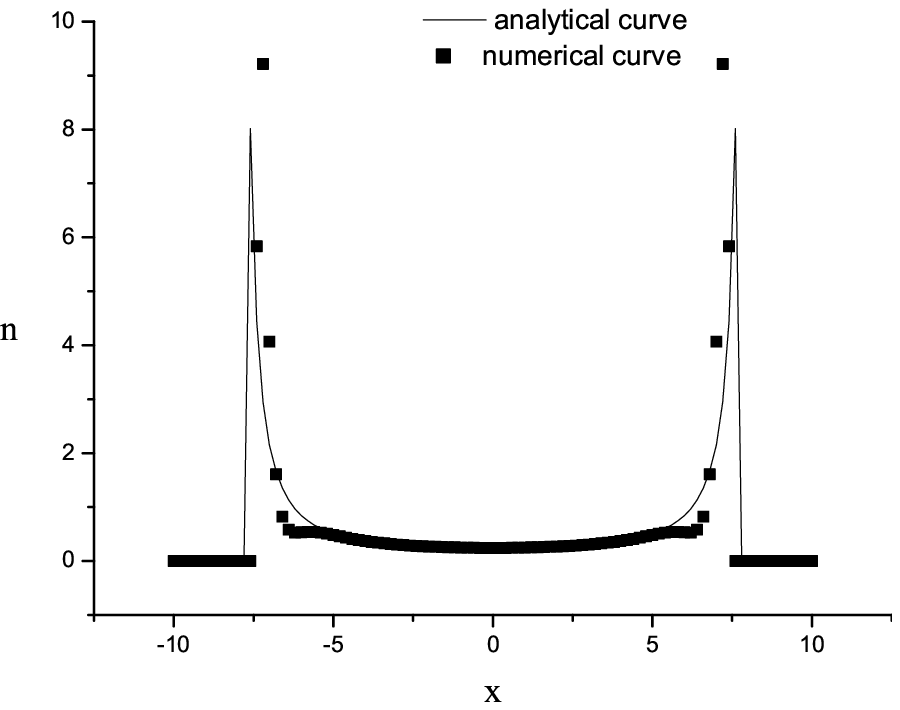} & \epsfysize=4.0cm \epsffile{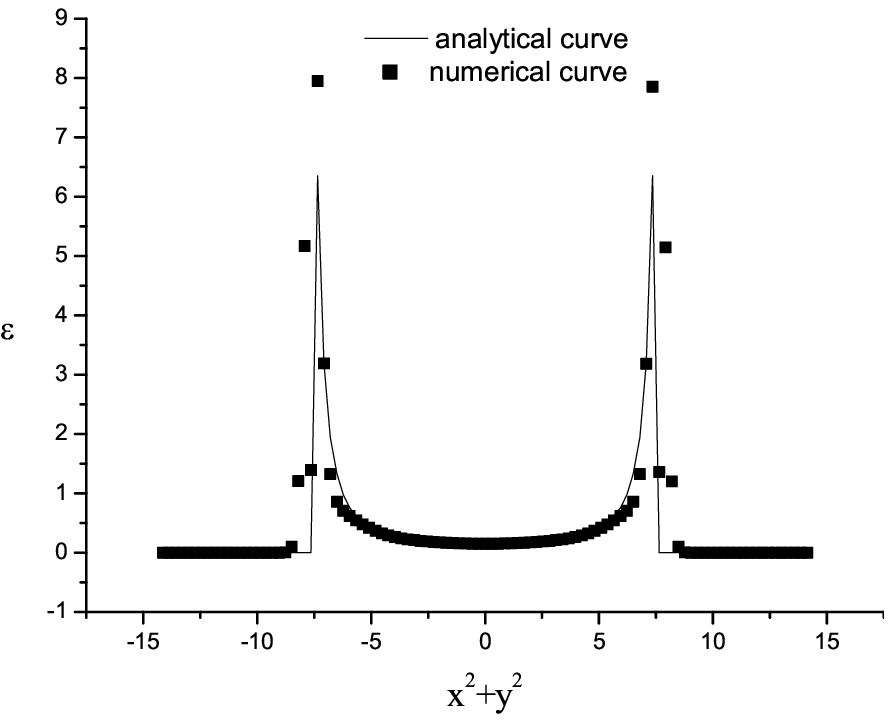} \\
(e) & (f) \\
\epsfysize=4.0cm \epsffile{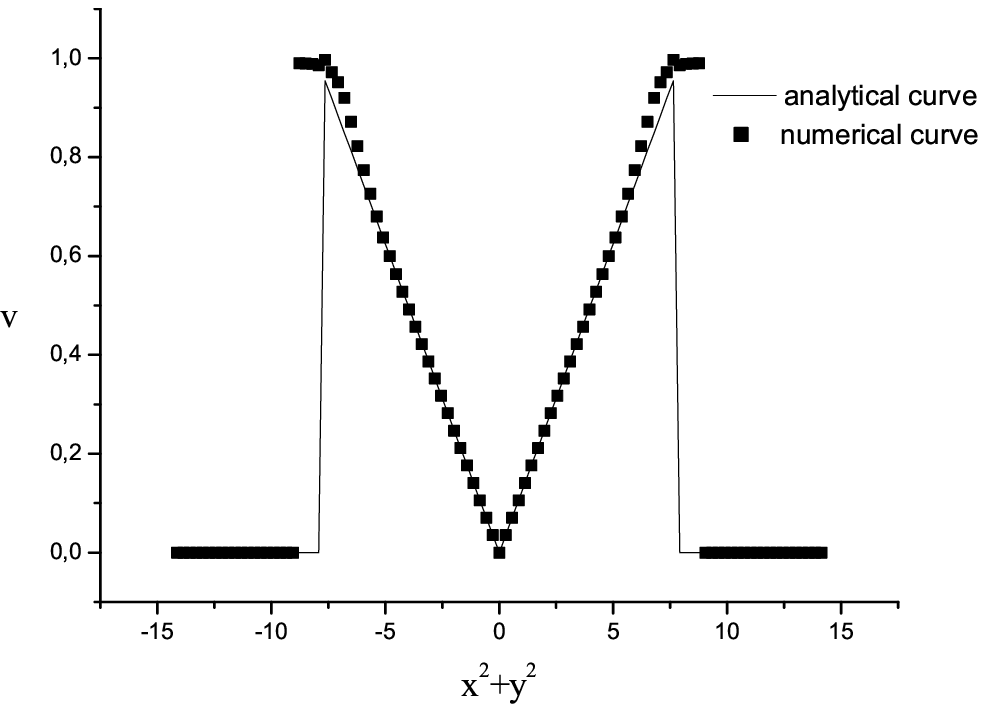} & \epsfysize=4.0cm \epsffile{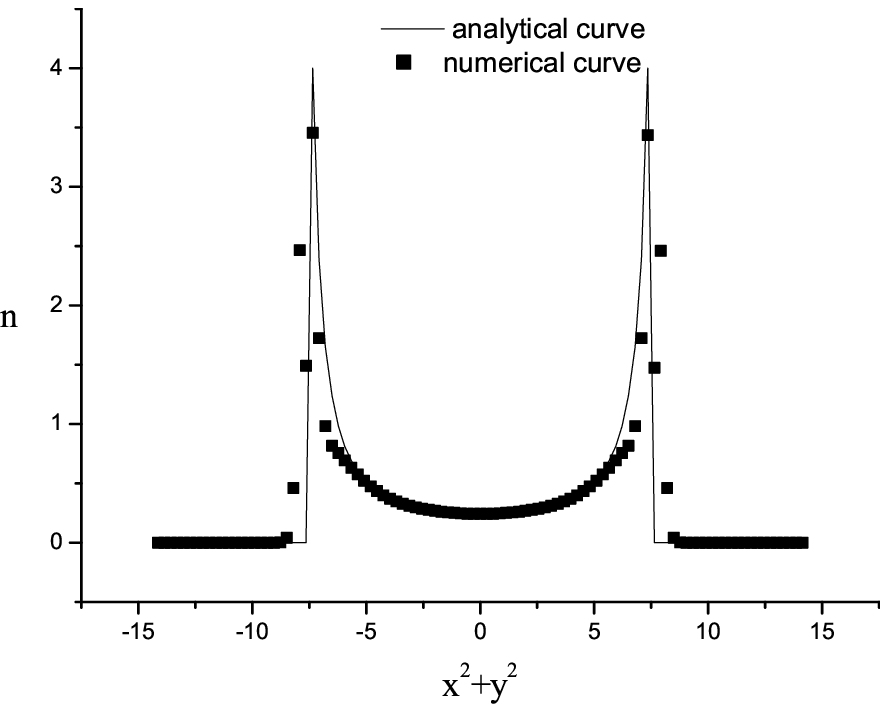} \\
(g) & (h) \\
\epsfysize=4.0cm \epsffile{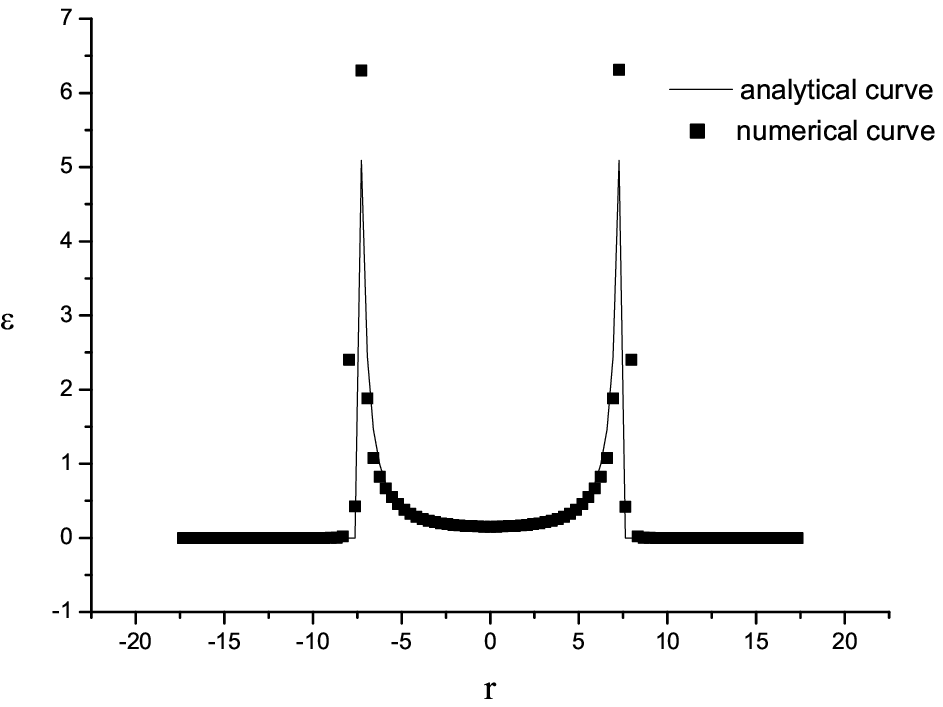} & \epsfysize=4.0cm \epsffile{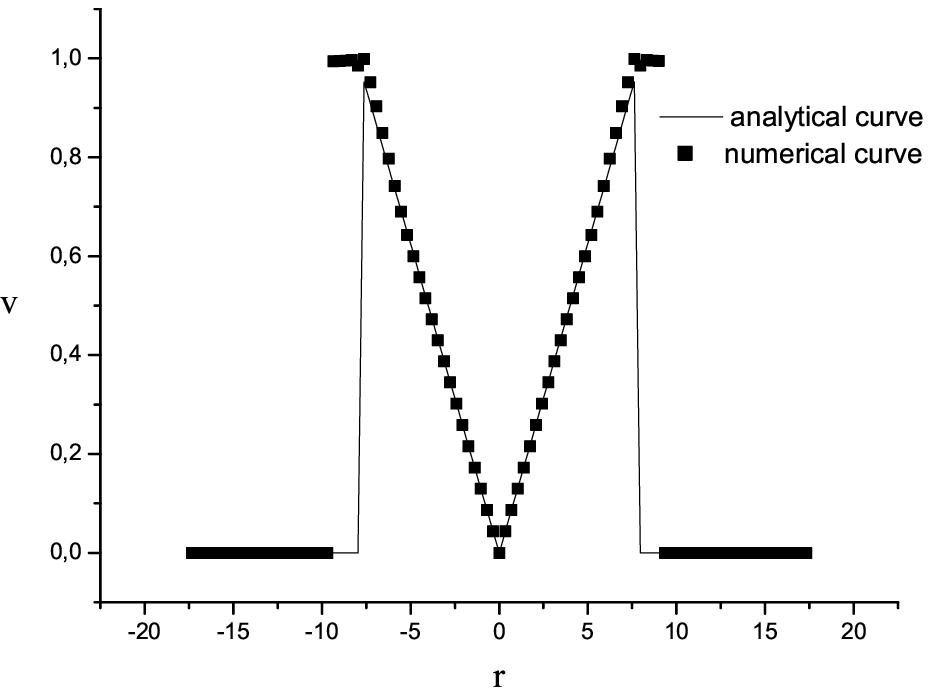} \\
\end{tabular}
\end{center}
\caption{\scriptsize Results of numerical simulations for a Hubble-like flow in three dimensions. (a)-(c) are energy density, velocity and baryonic charge density respectively along $\mbox{x}$-axis. (d)-(f) are the same quantities in $\mbox{xy}$-plane for $\mbox{x}=\mbox{y}$.  (g)-(h) are energy density and velocity in a radial direction for which $\mbox{x}=\mbox{y}=\mbox{z}$. Baryonic charge density is depicted on Fig~\ref{f:hubble32}. Evolution starts  at time $\mbox{t}_0=5$ and is finished at time $\mbox{t}=8$. All the graphs are taken at time $\mbox{t}$. One can see a rarefaction wave expanding into vacuum and distorting analytical solution on the edges. In the middle of the region numerical and analytical solutions coincide.}
\label{f:hubble31}
\end{figure*}

\begin{figure*}
\begin{center}
\begin{tabular}{ll}
\multicolumn{2}{l}{(i)} \\
\multicolumn{2}{c}{\epsfysize=4.5cm \epsffile{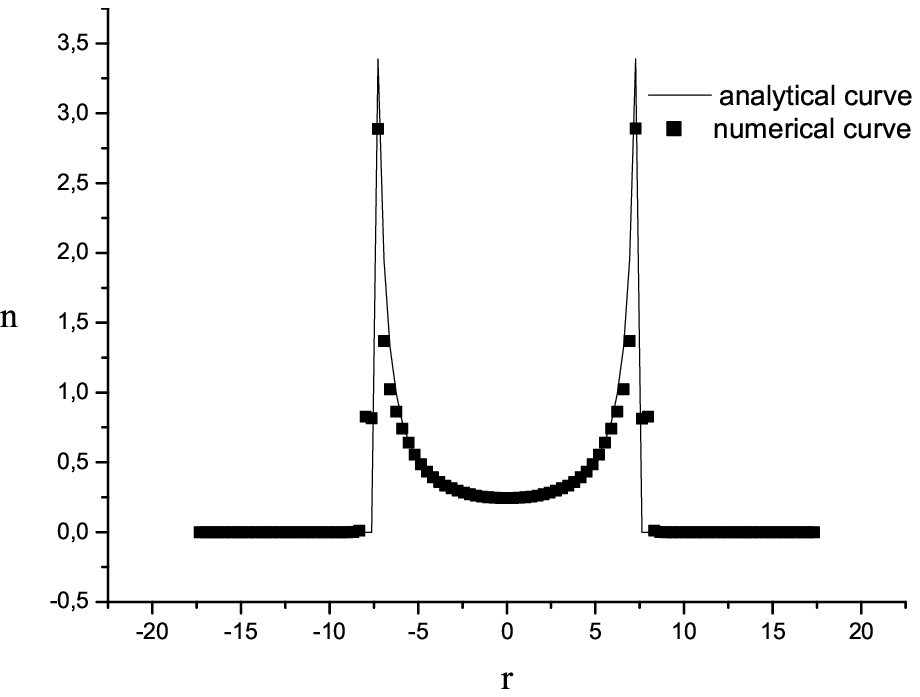}} \\
\end{tabular}
\end{center}
\caption{\scriptsize Baryonic charge density in a radial direction for which $\mbox{x}=\mbox{y}=\mbox{z}$ for a Hubble-like flow. Three-dimensional simulations.}
\label{f:hubble32}
\end{figure*}

Here we again use as a Hubble-like flow (\ref{he})-(\ref{hv}) as a test. But, unlike the previous section, we use fully three-dimensional code to simulate it. Of course, for three-dimensional calculations there is no source term. The results of numerical calculations are depicted on Figs.~\ref{f:hubble31}-\ref{f:hubble32}. Parameters of flow are taken the same as in subsection 9.2. We show dependences on distance from center along $\mbox{x}$-axis, in $\mbox{x} \mbox{y}$-plane for $\mbox{x}=\mbox{y}$ and in a radial direction for which $\mbox{x}=\mbox{y}=\mbox{z}$.

\subsection{Elliptical solution for $\mbox{p}=0$}

Here we take a solution similar to that taken in subsection 9.4. The difference is that now in order to test three-dimensional code solution is not spherically symmetric. In accordance with results of \cite{sinyukovkarpenko} we choose the solution in the the following form
\begin{equation}
\varepsilon=\frac{\mbox{c}_e}{\left( \mbox{t}+\mbox{t}_x \right) \left( \mbox{t}+\mbox{t}_y \right) \left( \mbox{t}+\mbox{t}_z \right)} e^{-\frac{\mbox{b}_e^2}{1-\mbox{s}}}
\label{3se}
\end{equation}
\begin{equation}
\bv{v}=\left( \frac{\mbox{x}}{\mbox{t}+\mbox{t}_x}, \frac{\mbox{y}}{\mbox{t}+\mbox{t}_y}, \frac{\mbox{z}}{\mbox{t}+\mbox{t}_z} \right)
\label{3sv}
\end{equation}
\begin{equation}
\mbox{n}=\frac{\mbox{c}_n}{\left( \mbox{t}+\mbox{t}_x \right) \left( \mbox{t}+\mbox{t}_y \right) \left( \mbox{t}+\mbox{t}_z \right)} e^{-\frac{\mbox{b}_n^2}{1-\mbox{s}}},
\label{3sn}
\end{equation}
where $\displaystyle \mbox{s}=\left( \frac{\mbox{x}}{\mbox{t}+\mbox{t}_x} \right)^2+ \left( \frac{\mbox{y}}{\mbox{t}+\mbox{t}_y} \right)^2+ \left( \frac{\mbox{z}}{\mbox{t}+\mbox{t}_z} \right)^2$.
This is a generalization of spherically symmetric flow considered in subsection 9.2.

\begin{figure*}
\begin{center}
\begin{tabular}{ll}
(a) & (b) \\ \epsfysize=4.0cm \epsffile{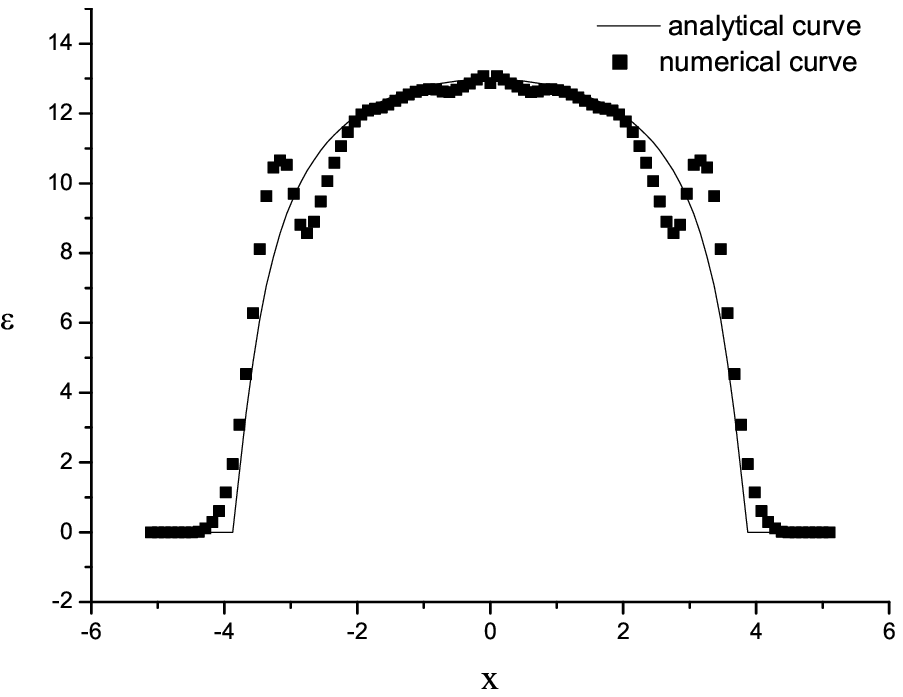} &
\epsfysize=4.0cm \epsffile{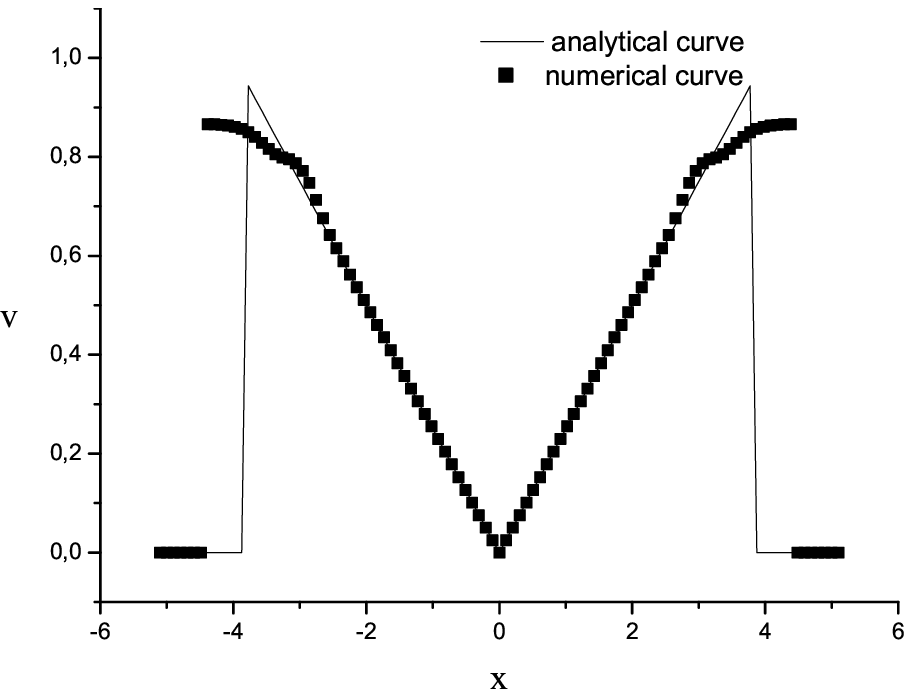} \\
(c) & (d) \\
\epsfysize=4.0cm \epsffile{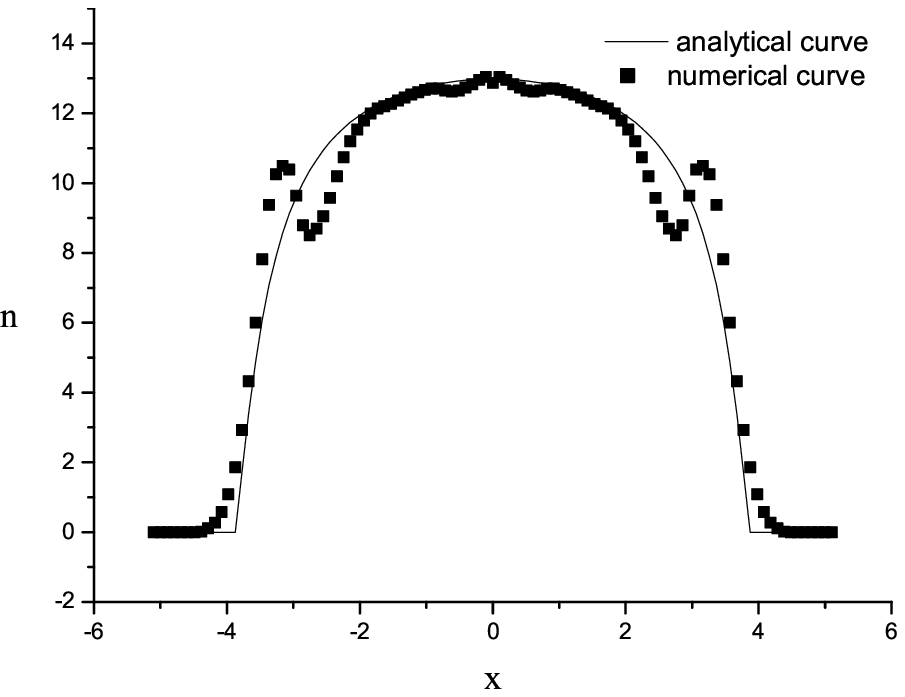} & \epsfysize=4.0cm \epsffile{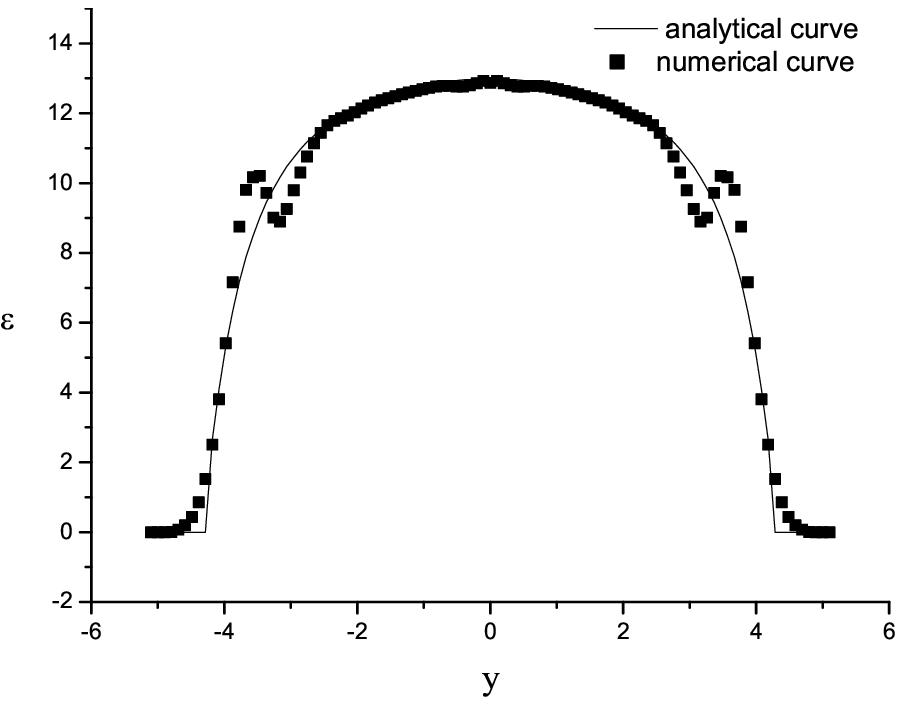} \\
(e) & (f) \\
\epsfysize=4.0cm \epsffile{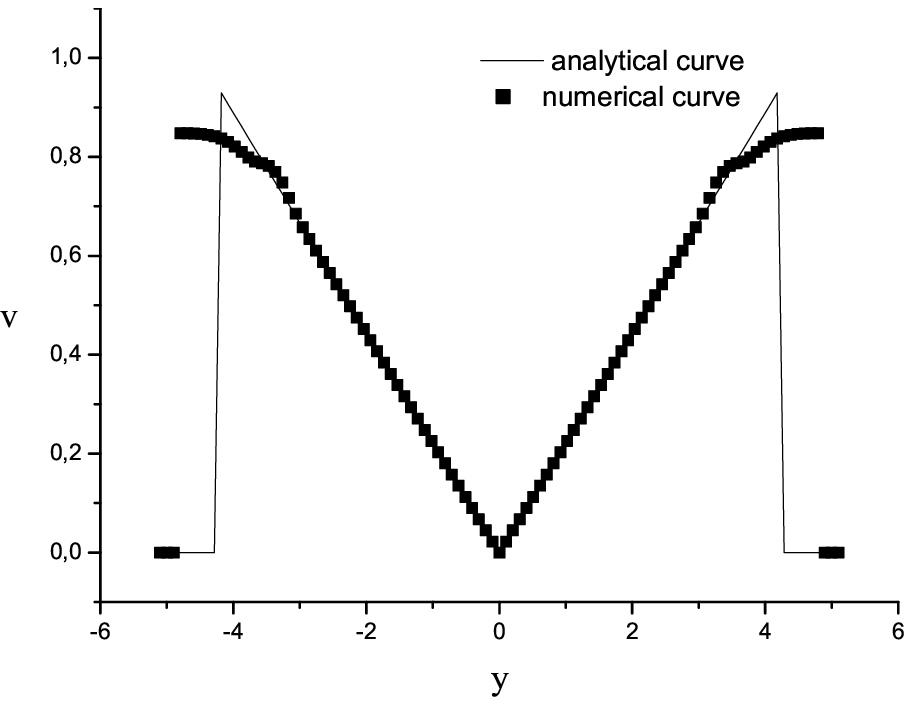} & \epsfysize=4.0cm \epsffile{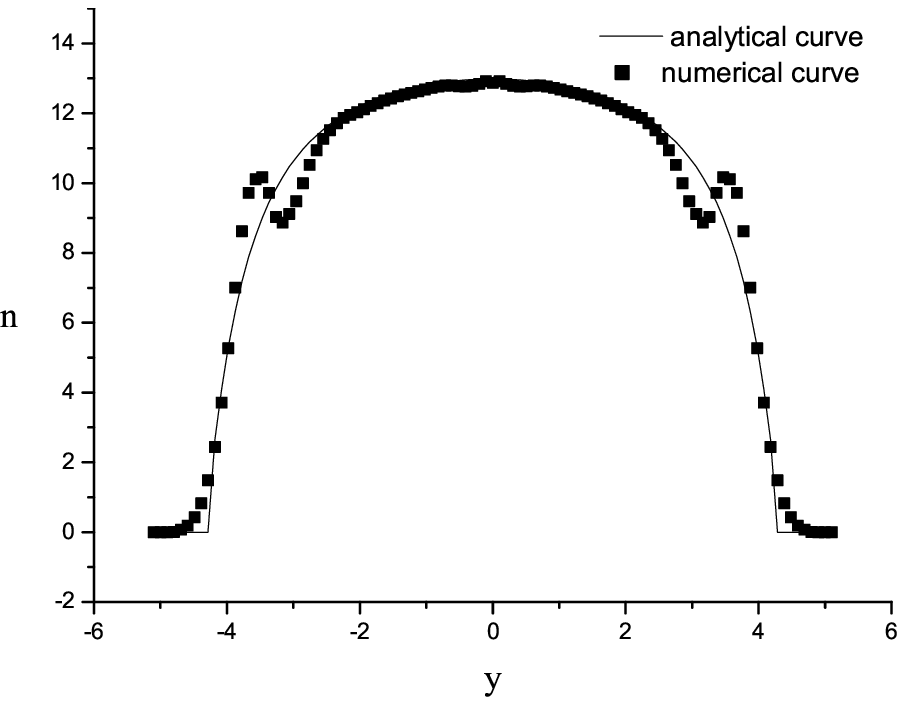} \\
(g) & (h) \\
\epsfysize=4.0cm \epsffile{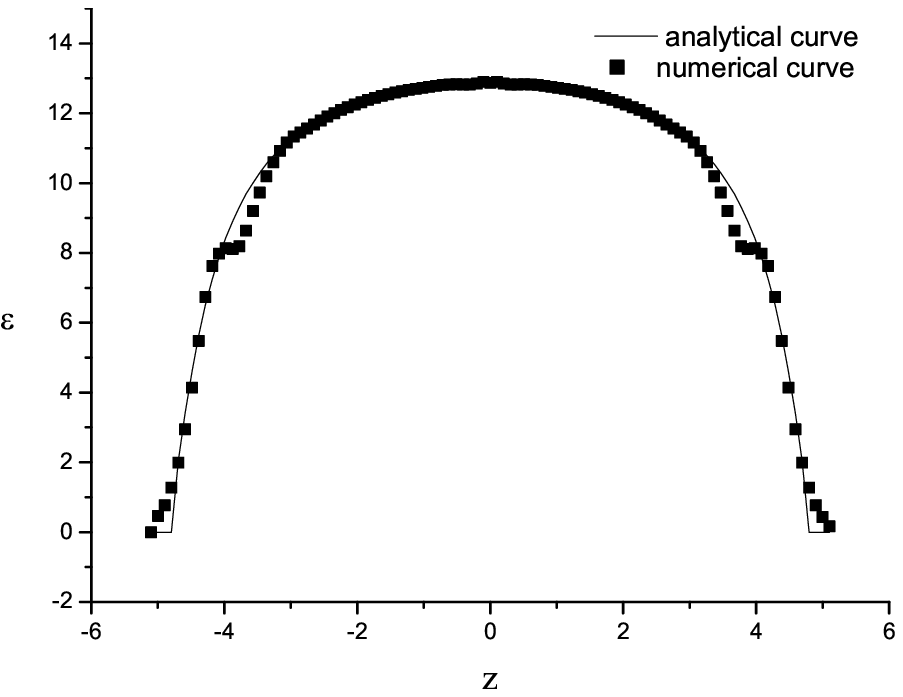} & \epsfysize=4.0cm \epsffile{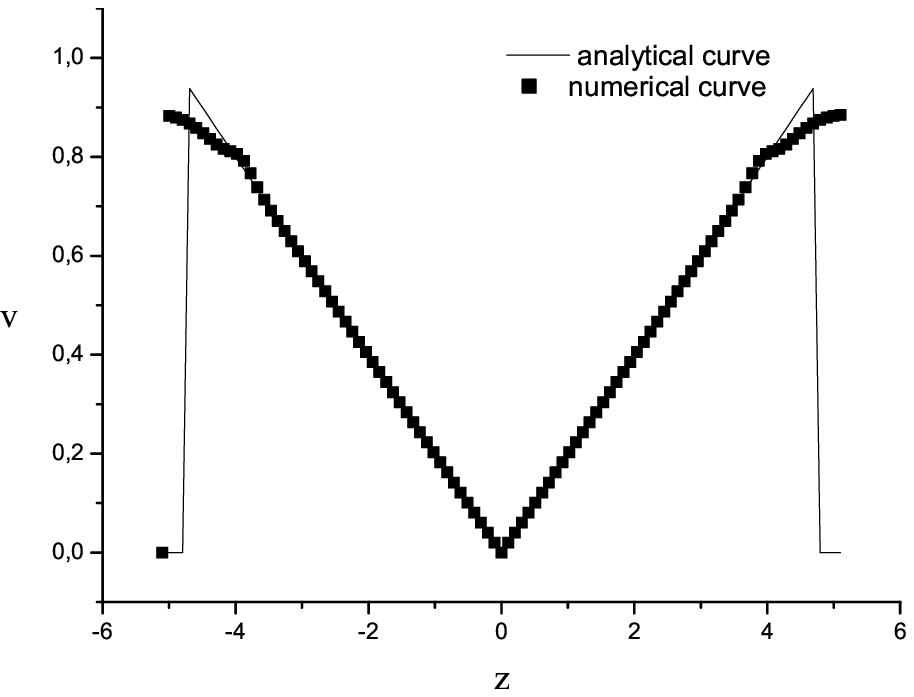} \\
\end{tabular}
\end{center}
\caption{\scriptsize Results of numerical simulations for an elliptical pressureless flow. (a)-(c) are energy density, velocity and baryonic charge density along $\mbox{x}$-axis respectively. (d)-(f) are the same quantities along $\mbox{y}$-direction and (g)-(h) are energy density and velocity along $\mbox{z}$-direction. Baryonic charge is shown on Fig~\ref{f:ell2}. Evolution starts  at time $\mbox{t}_0=0$ and is finished at time $\mbox{t}=3$. All the graphs are shown at time $\mbox{t}$.}
\label{f:ell1}
\end{figure*}

\begin{figure*}
\begin{center}
\begin{tabular}{ll}
(i) & (j) \\ \epsfysize=4.0cm \epsffile{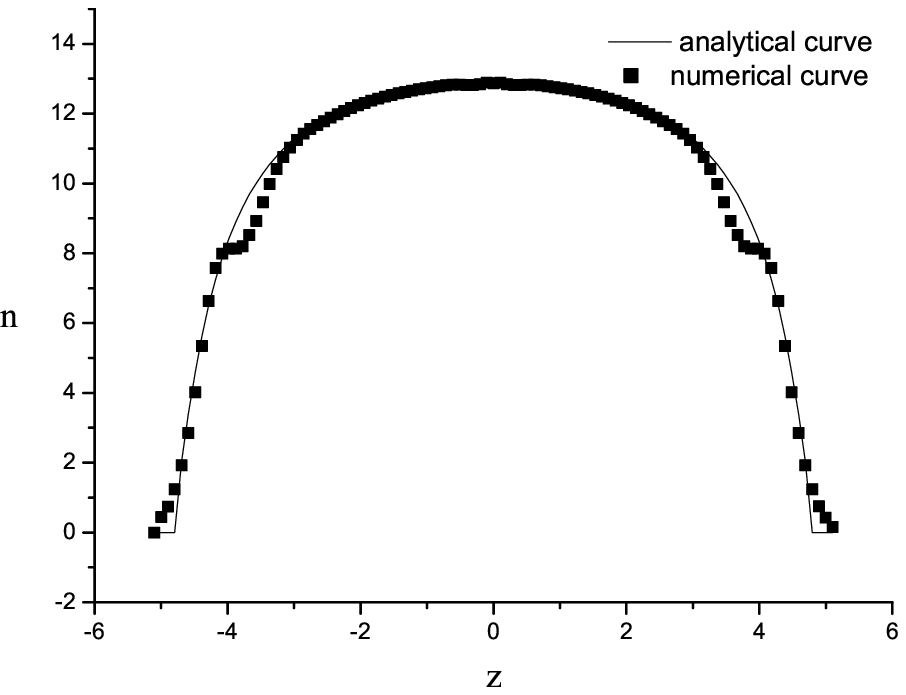} &
\epsfysize=4.0cm \epsffile{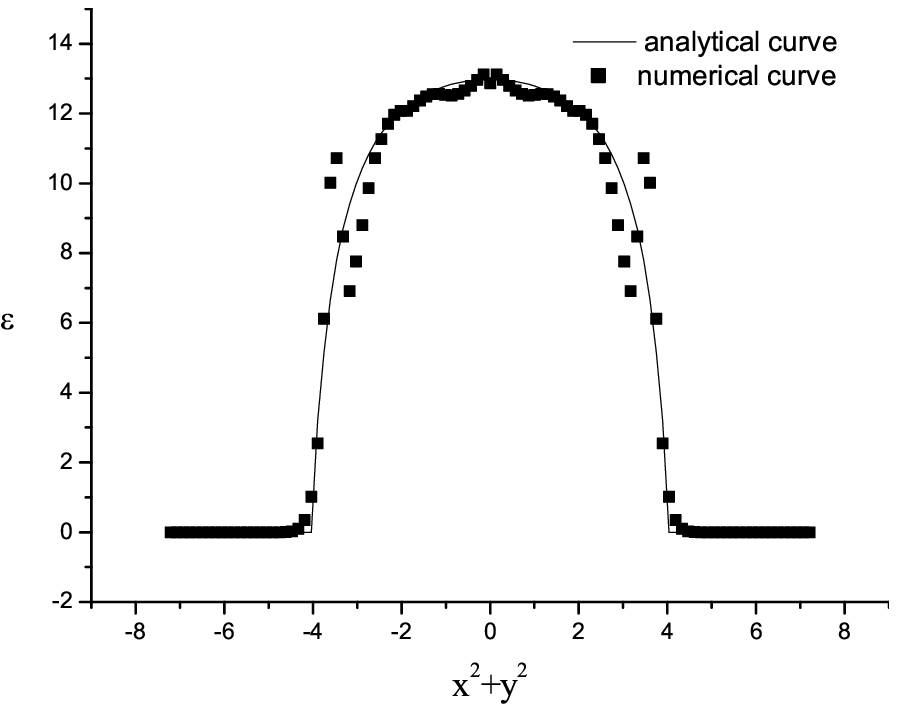} \\
(k) & (l) \\
\epsfysize=4.0cm \epsffile{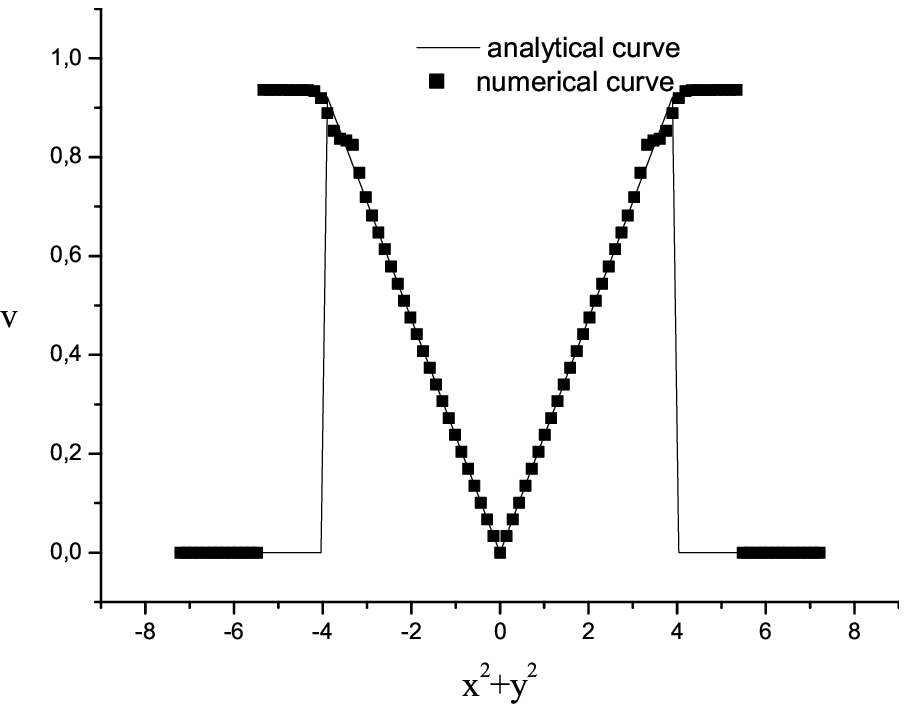} & \epsfysize=4.0cm \epsffile{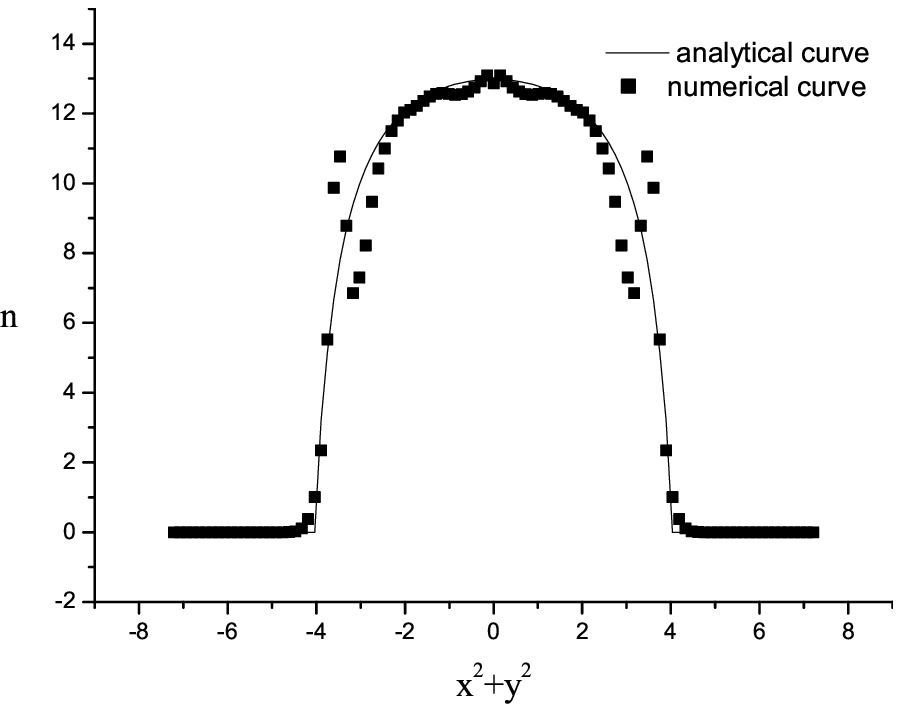} \\
(m) & (n) \\
\epsfysize=4.0cm \epsffile{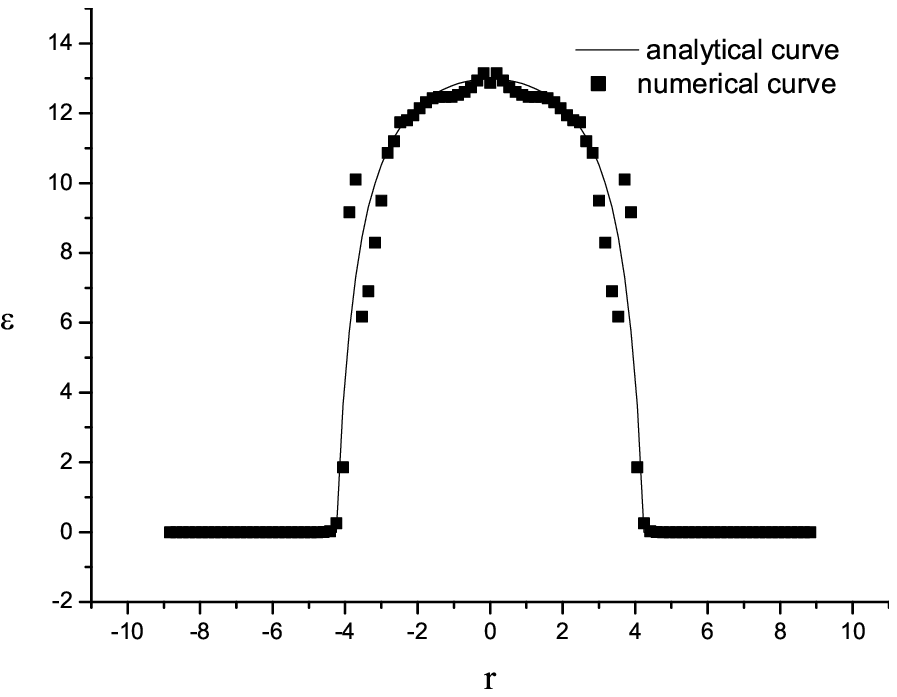} & \epsfysize=4.0cm \epsffile{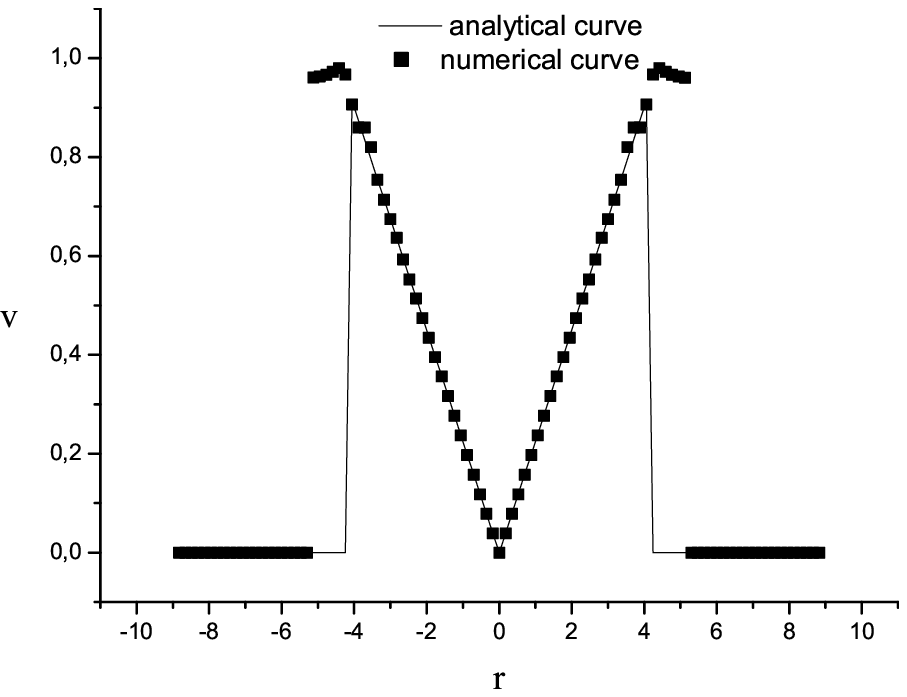} \\
\multicolumn{2}{l}{(o)} \\
\multicolumn{2}{c}{\epsfysize=4.5cm \epsffile{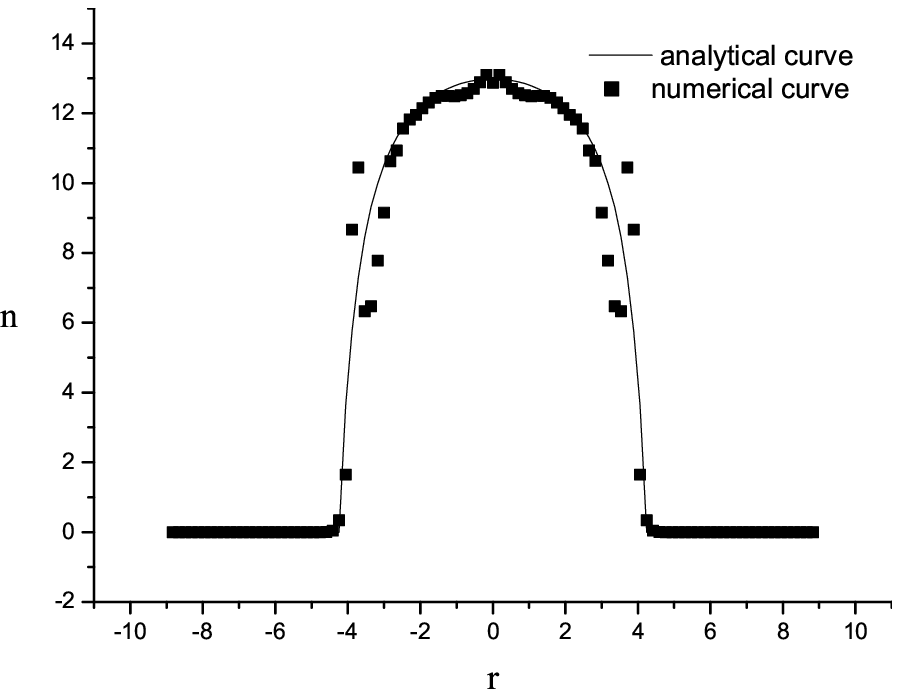}} \\
\end{tabular}
\end{center}
\caption{\scriptsize Results of numerical simulations for an elliptical pressureless flow. (i) is baryonic charge density along $\mbox{z}$-direction. (j)-(l) are energy density, velocity and baryonic charge density in $\mbox{xy}$-plane for the line $\mbox{x}=\mbox{y}$respectively. (m)-(o) are the same quantities for a radial direction for which $\mbox{x}=\mbox{y}=\mbox{z}$. Evolution starts  at time $\mbox{t}_0=0$ and is finished at time $\mbox{t}=3$. All the graphs are shown at time $\mbox{t}$.}
\label{f:ell2}
\end{figure*}

The results of numerical simulations are depicted on Figs~\ref{f:ell1}-\ref{f:ell2}. We take the following parameters: $\mbox{c}_e=1500$, $\mbox{c}_n=1500$, $\mbox{b}_e^2=0.5$, $\mbox{b}_n^2=0.5$, $\mbox{t}_y=1.5$ and $\mbox{t}_z=2$. Initial time is $\mbox{t}_0=0$ and final time is $\mbox{t}=3$. At moment of time $\mbox{t}_0$ we take all the quantities in region $\left| \mbox{x} \right| < \mbox{t}_0+\mbox{t}_z-\alpha$, where $\alpha$ is small enough, being defined by relations (\ref{3se})-(\ref{3sn}). Outside this region all the quantities are put to zero.

\subsection{Hubble-like flow with a source term}

Here we consider solution of the following form
\begin{equation}
\varepsilon=\varepsilon_0 \left( \frac{\tau_0\left( \bv{r}, \mbox{t} \right)}{\tau} \right)^{3\left( 1+\mbox{c}_s^2 \right)}
\label{hse}
\end{equation}
\begin{equation}
\mbox{n}=\mbox{n}_0 \left( \frac{\tau_0\left( \bv{r}, \mbox{t} \right)}{\tau} \right)^3
\label{hsn}
\end{equation}
\begin{equation}
\mbox{v}=\frac{\bv{r}}{\mbox{t}},
\label{hsv}
\end{equation}
where $\displaystyle \tau=\sqrt{\mbox{t}^2-\bv{r}^2}$. Equation of state is $\mbox{p}=\mbox{c}_s^2 \varepsilon$.

\begin{figure*}
\begin{center}
\begin{tabular}{ll}
(a) & (b) \\ \epsfysize=4.0cm \epsffile{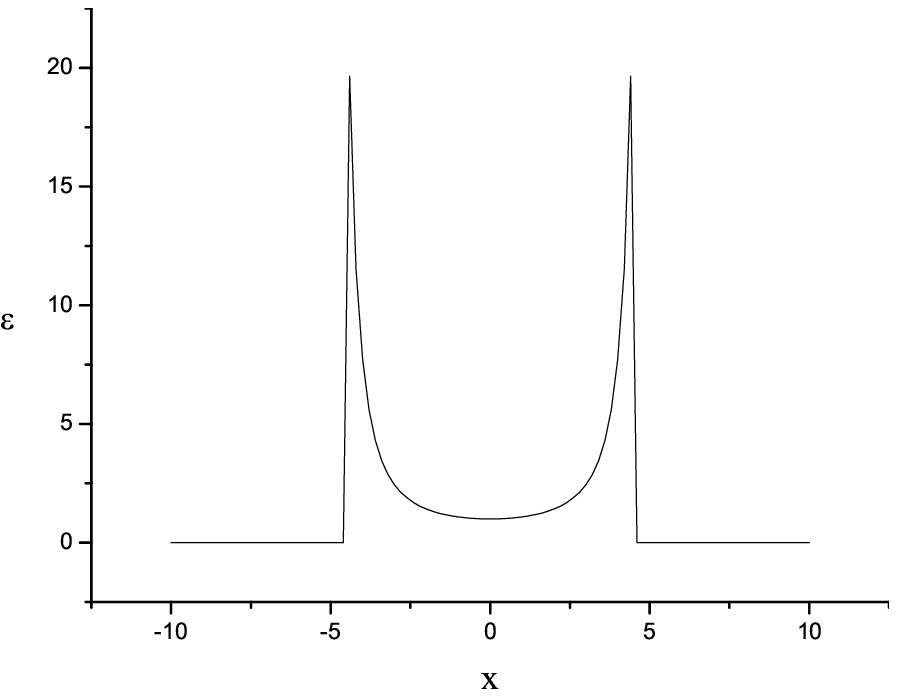} &
\epsfysize=4.0cm \epsffile{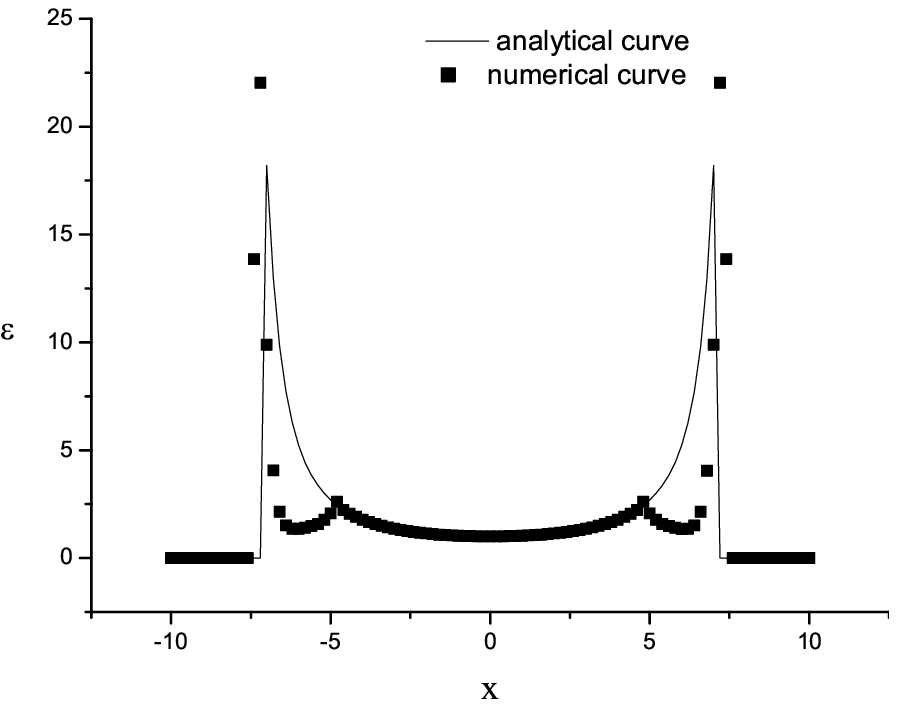} \\
(c) & (d) \\
\epsfysize=4.0cm \epsffile{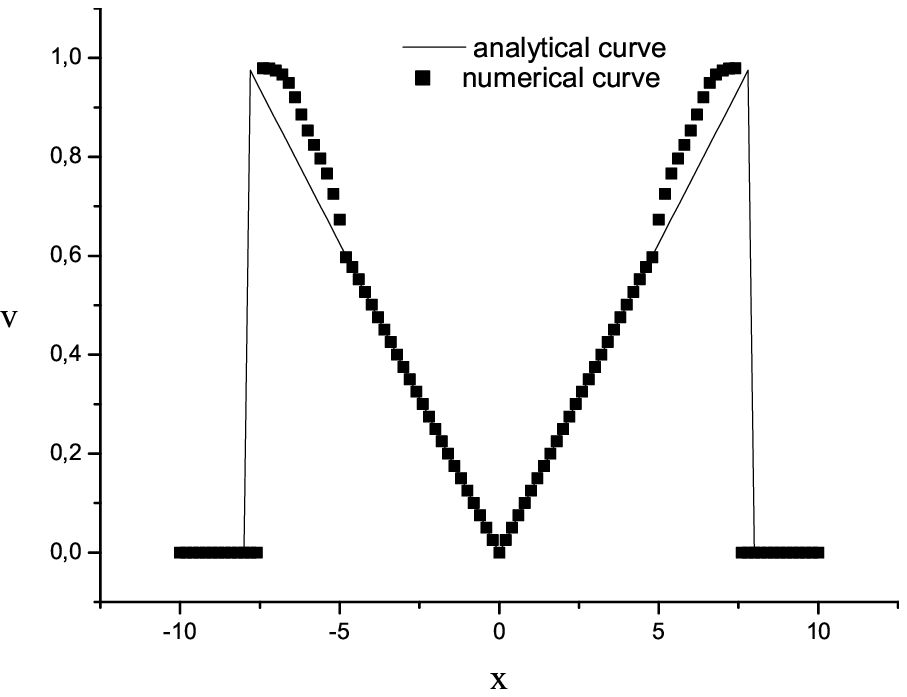} & \epsfysize=4.0cm \epsffile{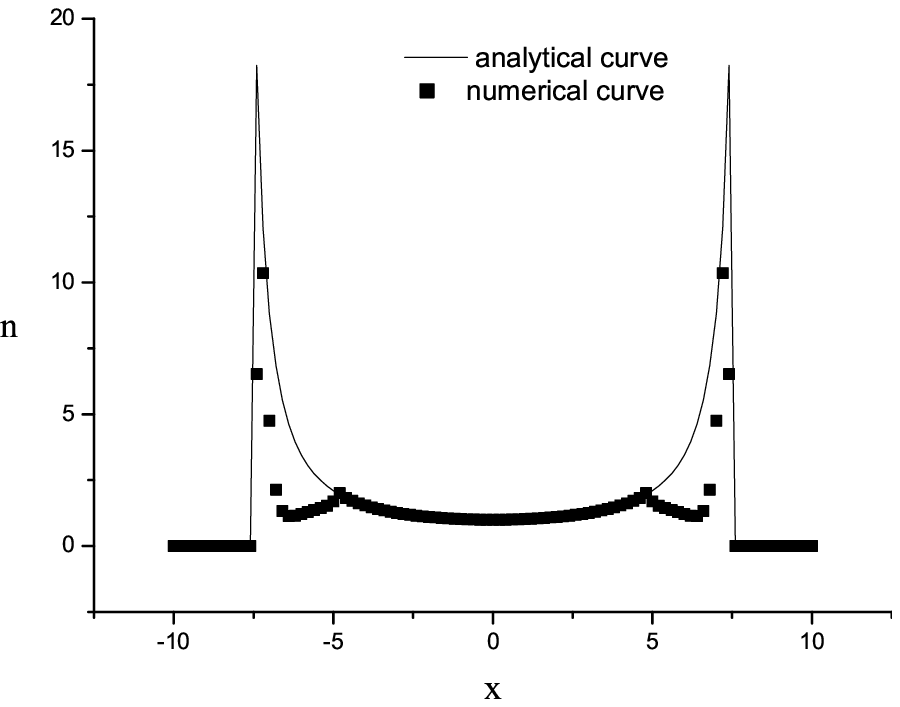} \\
(e) & (f) \\
\epsfysize=4.0cm \epsffile{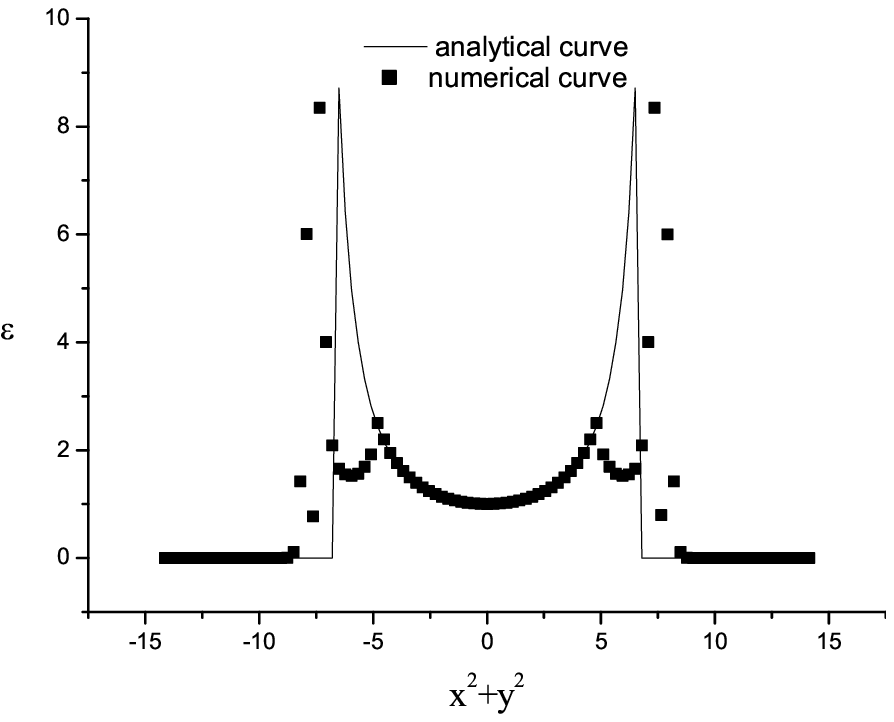} & \epsfysize=4.0cm \epsffile{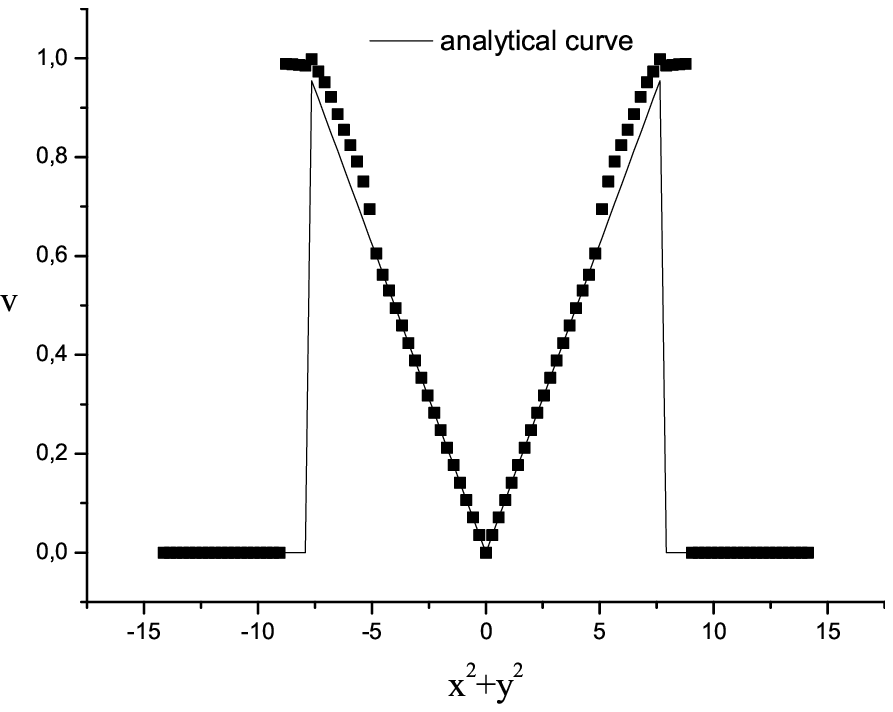} \\
\multicolumn{2}{c}{(g)} \\
\multicolumn{2}{c}{\epsfysize=4.0cm \epsffile{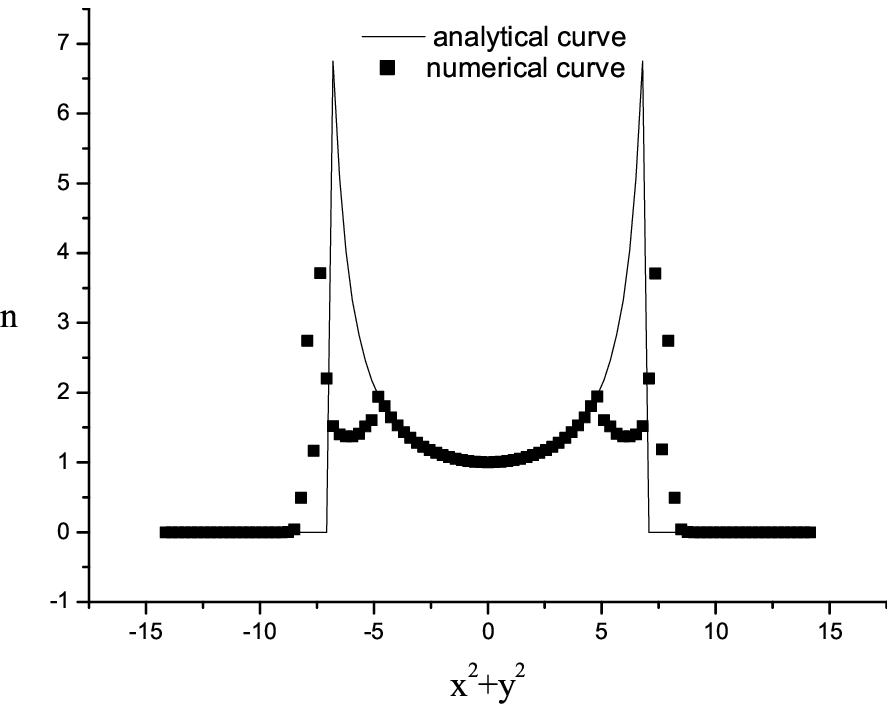}} \\
\end{tabular}
\end{center}
\caption{\scriptsize Results of numerical simulations for a Hubble-like flow with a source term defined by $\tau_0=\mbox{t}$ (see Eq.(\ref{hse})-(\ref{hsv})). (b)-(d) are energy density, velocity and baryonic charge density along $\mbox{x}$-direction. (e)-(g) are the same quantities in $\mbox{xy}$-plane for the line $\mbox{x}=\mbox{y}$. Evolution starts  at time $\mbox{t}_0=5$ and is finished at time $\mbox{t}=8$. All the graphs are taken at time $\mbox{t}$. (a) are initial values of energy density at time $\mbox{t}_0=5$. Outside the region with matter there is no source. It can be seen on (b)-(d) that inside the region with matter numerical and analytical solutions coincide but outside this region solution is distorted. Namely, expansion there is faster that in inner region as choice of $\tau_0=\mbox{t}$ makes expansion in inner regions slower.}
\label{f:3source_t}
\end{figure*}

\begin{figure*}
\begin{center}
\begin{tabular}{ll}
(a) & (b) \\ \epsfysize=4.5cm \epsffile{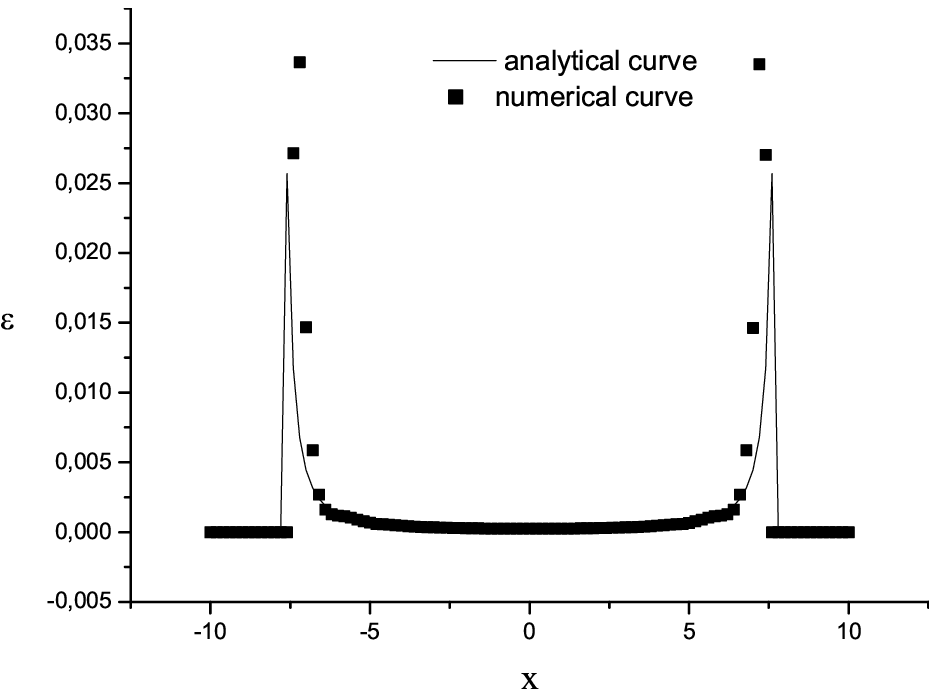} &
\epsfysize=4.5cm \epsffile{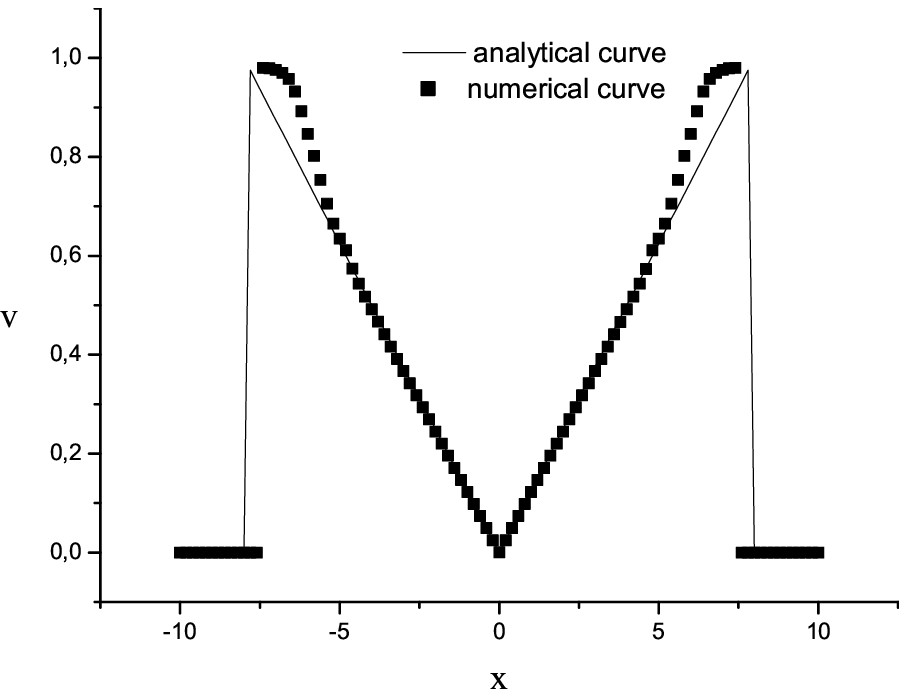} \\
(c) & (d) \\
\epsfysize=4.5cm \epsffile{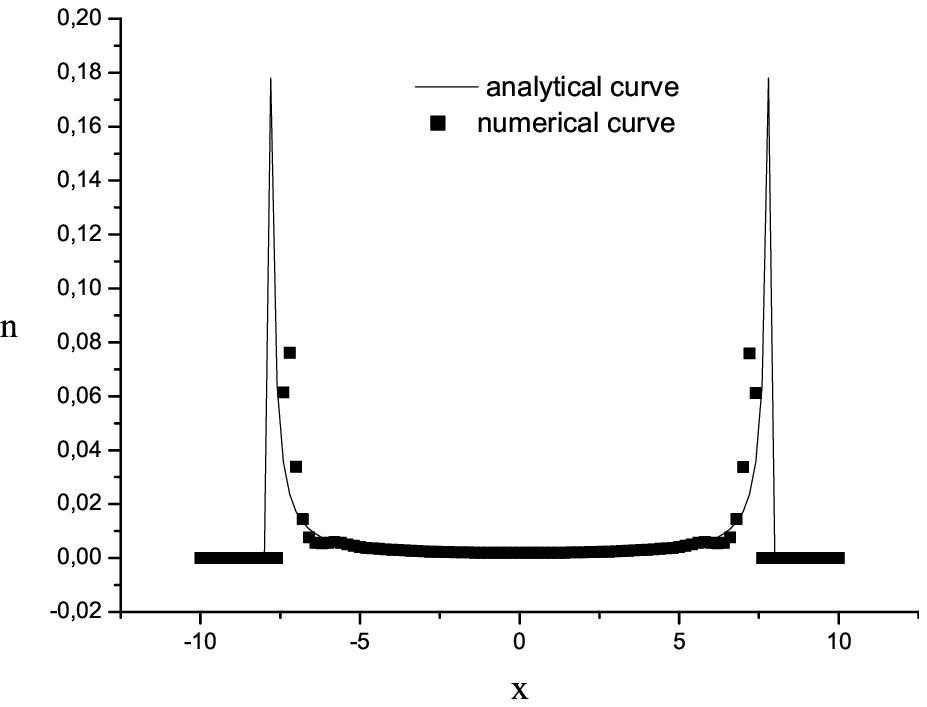} & \epsfysize=4.5cm \epsffile{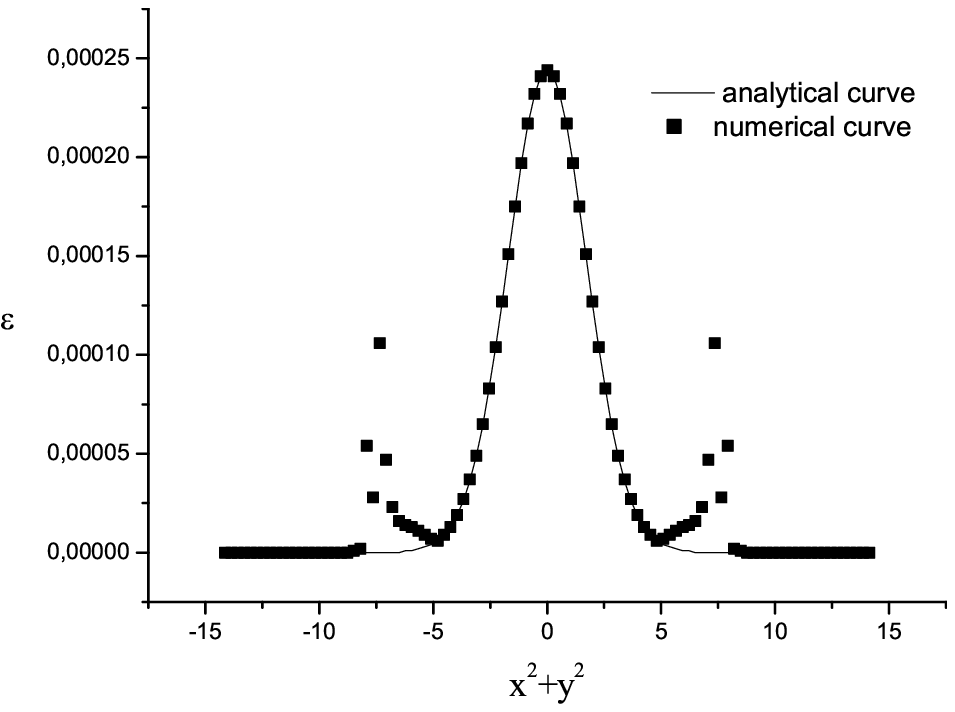} \\
(e) & (f) \\
\epsfysize=4.5cm \epsffile{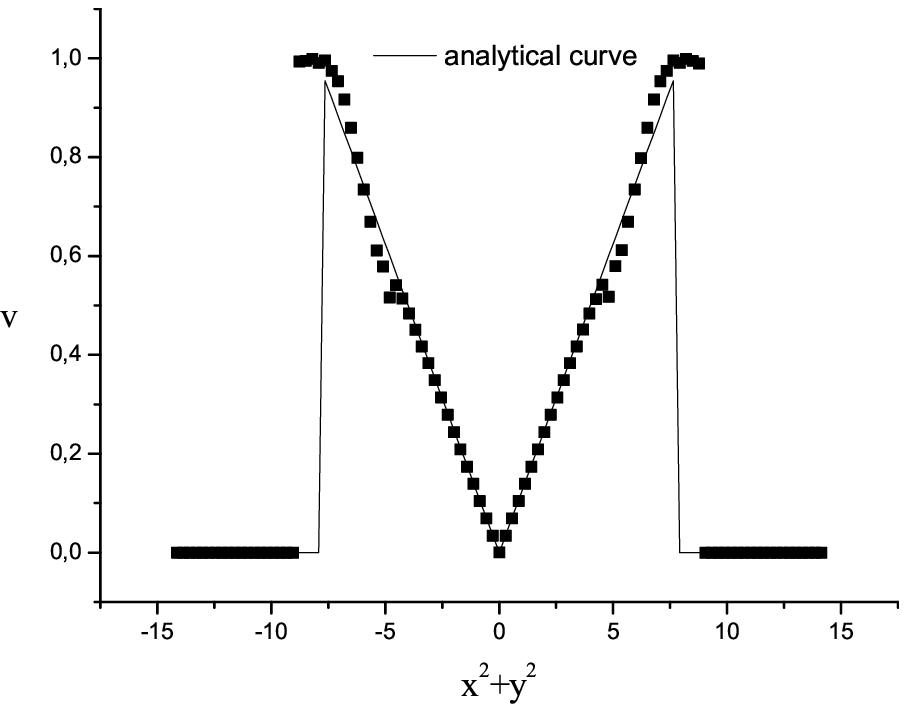} & \epsfysize=4.5cm \epsffile{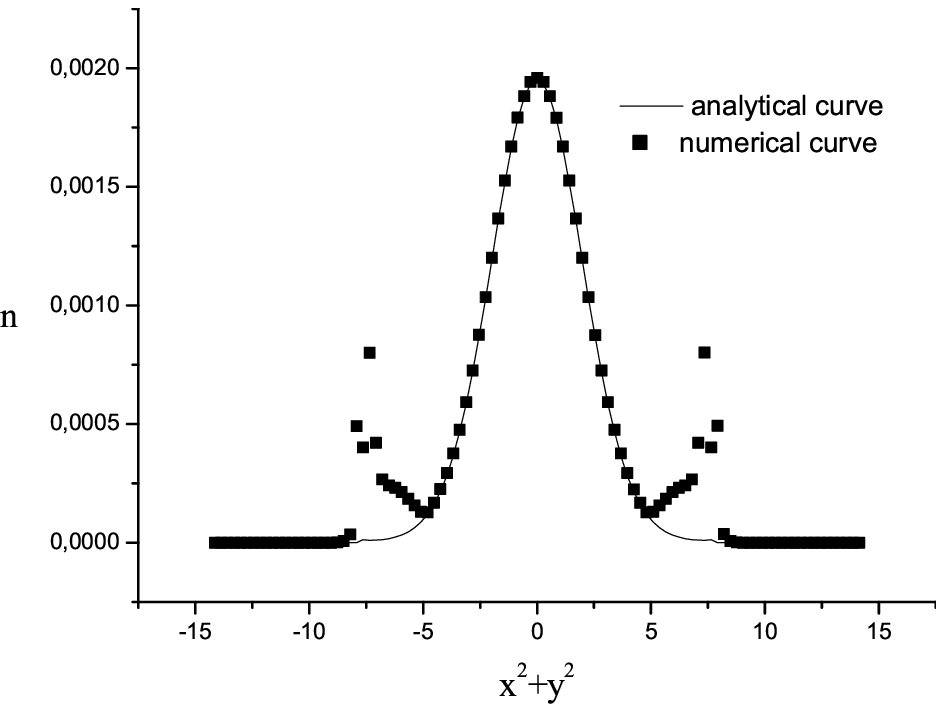} \\
\end{tabular}
\end{center}
\caption{\scriptsize Results of numerical simulations for a Hubble-like flow with a source term defined by $\displaystyle \tau_0=e^{-axy}$ (see Eq.(\ref{hse})-(\ref{hsv})). (a)-(c) are energy density, velocity and baryonic charge density along $\mbox{x}$-direction. (d)-(f) are the same quantities in $\mbox{xy}$-plane for the line $\mbox{x}=\mbox{y}$. Evolution starts at time $\mbox{t}_0=5$ and is finished at time $\mbox{t}=8$. All the graphs are taken at time $\mbox{t}$. Note that solution is extremely asymmetric but numerical solution gives good agreement with analytical one in the region where source term is non-zero (see explanations for Fig.~\ref{f:3source_t}).}
\label{f:3source_exp}
\end{figure*}

The fact that $\tau_0$ is not a constant leads us to an artificial source term $\bv{S}=\left( \mbox{S}_e, \mbox{S}_x, \mbox{S}_y, \mbox{S}_z, \mbox{S}_n \right)$ where
\begin{equation}
\mbox{S}_e=\frac{3\left( 1+\mbox{c}_s^2 \right)\varepsilon \mbox{t}^2}{\tau_0 \tau^2} \left[ \left( 1+\frac{\bv{r}^2}{\mbox{t}^2}\mbox{c}_s^2 \right) \pdif{\tau_0}{\mbox{t}} +\left( 1+\mbox{c}_s^2 \right)\frac{\mbox{x}^j}{\mbox{t}} \pdif{\tau_0}{\mbox{x}^j} \right]
\label{hsse}
\end{equation}
\begin{eqnarray}
\mbox{S}_i=\frac{3\left( 1+\mbox{c}_s^2 \right)\varepsilon \mbox{t}}{\tau_0 \tau} \left[ \frac{\left( 1+\mbox{c}_s^2 \right) \mbox{x}^i}{\tau} \pdif{\tau_0}{\mbox{t}}+\frac{\left( 1+\mbox{c}_s^2 \right) \mbox{x}^i}{\tau_0 \mbox{t}} \mbox{x}^j \pdif{\tau_0}{\mbox{x}^j}+ \frac{\tau}{\mbox{t}} \pdif{\tau_0}{\mbox{x}_i} \right], \nonumber & \\ i=\mbox{x},\mbox{y},\mbox{z}
\label{hssi} &
\end{eqnarray}
\begin{equation}
\mbox{S}_n=\frac{3\mbox{nt}}{\tau_0 \tau} \left[ \pdif{\tau_0}{\mbox{t}}+\frac{\mbox{x}^i}{\mbox{t}} \pdif{\tau_0}{\mbox{x}^i} \right]
\label{hssn}
\end{equation}
We take the following parameters: $\mbox{c}_s^2=1/3$, $\mbox{n}_0=1$ and $\varepsilon_0=1$. Initial time is $\mbox{t}_0=5$ and final time is $\mbox{t}=8$. At moment of time $\mbox{t}_0$ we take all the quantities in region $\left| \bv{r} \right| < \mbox{t}_0-\alpha$, where $\alpha$ is small enough, being defined by relations (\ref{hsse})-(\ref{hssn}). Outside this region all the quantities and source term are put to zero. Function $\tau_0\left( \bv{r}, \mbox{t} \right)$ is taken to be (a) $\tau_0\left( \bv{r}, \mbox{t} \right)=\mbox{t}$ and (b) $\displaystyle \tau_0\left( \bv{r}, \mbox{t} \right)=e^{-axy}$ with $a=0.1$. Results of numerical simulations are depicted for case (a) on Fig.~\ref{f:3source_t} and for case (b) on Fig.~\ref{f:3source_exp}.

\subsection{Source term defined by an arbitrary flow}

Suppose that we have an arbitrary hydrodynamical flow. It is characterized by profiles of energy density $\tilde{\varepsilon}\left( \bv{r}, \mbox{t} \right)$, velocity $\tilde{\bv{v}}\left( \bv{r}, \mbox{t} \right)$, baryonic charge density $\tilde{\mbox{n}}\left( \bv{r}, \mbox{t} \right)$ and by equation of state $\mbox{p} \left( \varepsilon, \mbox{n} \right)$. Then this flow is a solution of hydrodynamics equations with a source term 
\begin{equation}
\pdif{\mbox{T}^{\mu\nu}}{\mbox{x}^{\nu}} = \mbox{S}^\mu,
\label{eq_g}
\end{equation}
\begin{equation}
\pdif{\mbox{nu}^{\mu}}{\mbox{x}^{\mu}}=\mbox{S}_n
\label{eq_n}
\end{equation}
where
\begin{equation}
\mbox{S}^{\mu} =\pdif{\tilde{\mbox{T}}^{\mu\nu}}{\mbox{x}^{\nu}}, \; \mbox{S}_n=\pdif{\tilde{\mbox{n}}\tilde{\mbox{u}}^{\mu}}{\mbox{x}^{\mu}}
\label{g}
\end{equation}
and $\displaystyle \tilde{\mbox{T}}^{\mu\nu}=(\tilde{\varepsilon}+\tilde{\mbox{p}}) \tilde{\mbox{u}}^{\mu} \tilde{\mbox{u}}^{\nu} - \tilde{\mbox{p}} \mbox{g}^{\mu\nu}$

\begin{figure*}
\begin{center}
\begin{tabular}{ll}
(a) & (b) \\ \epsfysize=4.5cm \epsffile{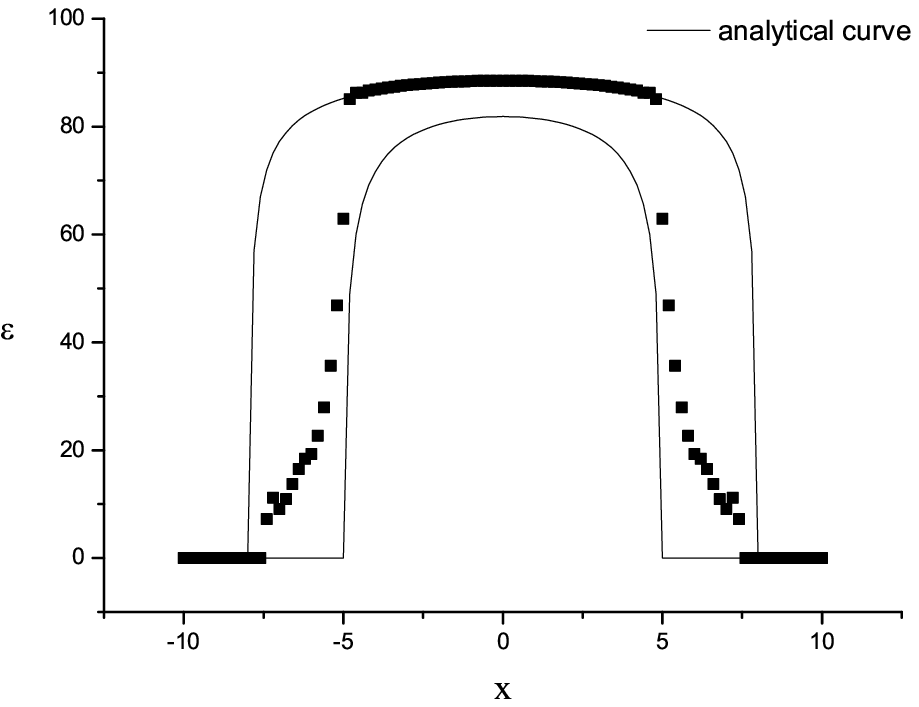} &
\epsfysize=4.5cm \epsffile{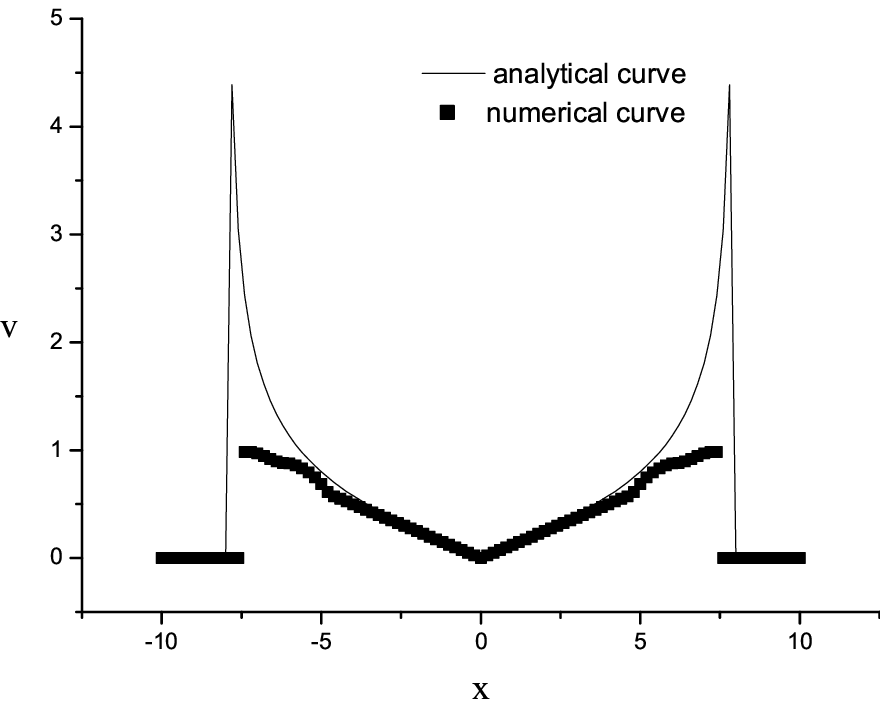} \\
(c) & (d) \\
\epsfysize=4.5cm \epsffile{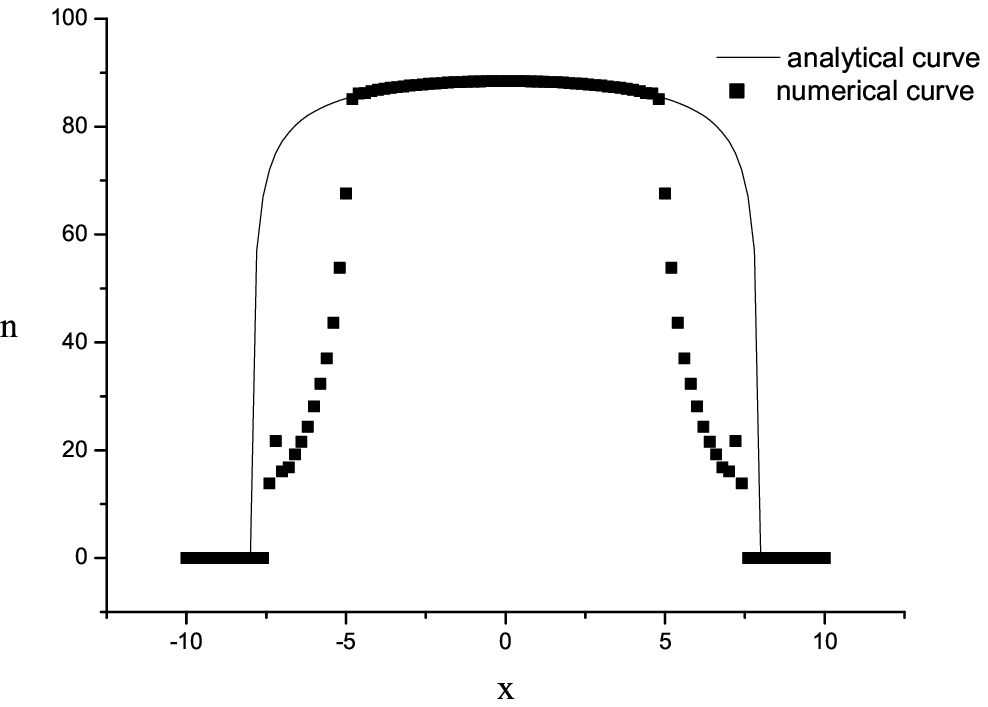} & \epsfysize=4.5cm \epsffile{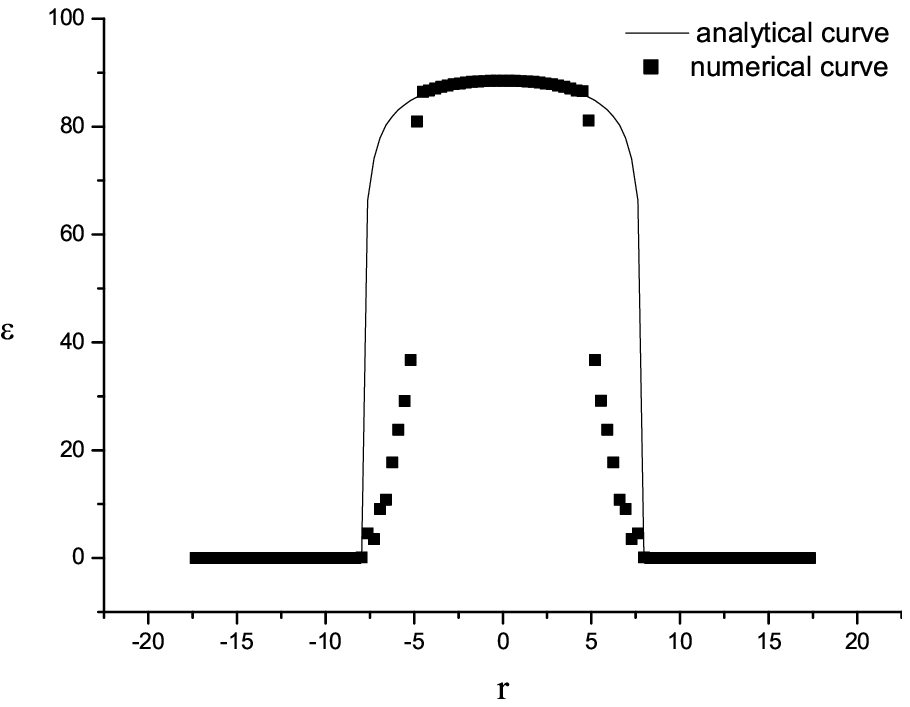} \\
(e) & (f) \\
\epsfysize=4.5cm \epsffile{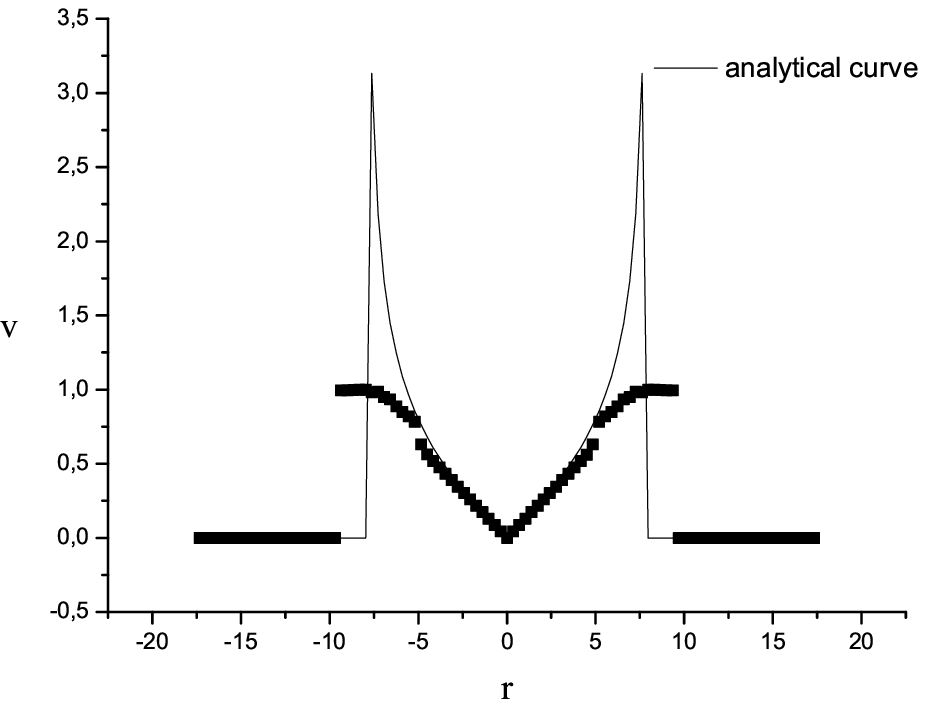} & \epsfysize=4.5cm \epsffile{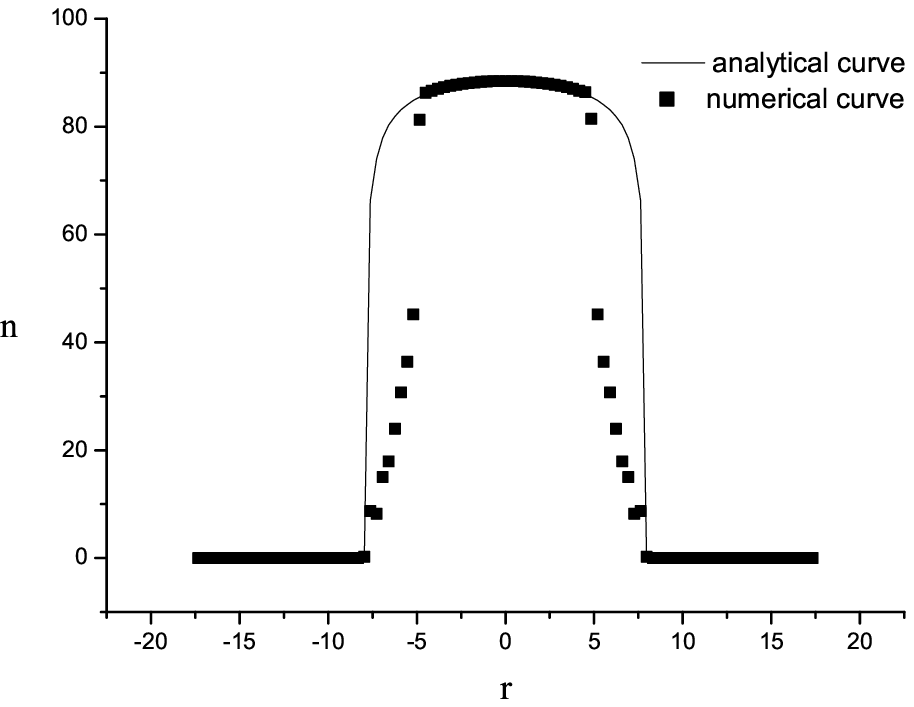} \\
\end{tabular}
\end{center}
\caption{\scriptsize Results of numerical simulations for an arbitrary flow defined source term (see Eq.(\ref{eg})-(\ref{ng})). (a)-(c) are energy density, velocity and baryonic charge density along $\mbox{x}$-direction. (d)-(f) are the same quantities in a radial direction for which $\mbox{x}=\mbox{y}=\mbox{z}$. Evolution starts at time $\mbox{t}_0=5$ and is finished at time $\mbox{t}=8$. All the graphs are shown at time $\mbox{t}$. On (a) we depict also initial values of energy density (lower solid line graph). It is done in order to show the region of non-zero source term (it is non-zero in region where matter is non-zero at initial time). It is seen that in this region analytical and numerical solutions coincide. Outside this region analytical source term solution is distorted as there is no source term.}
\label{f:3source_yura}
\end{figure*}

One of the simplest choices of a flow is
\begin{equation}
\tilde{\varepsilon}=\varepsilon_0 e^{-\frac{\mbox{b}_e}{\displaystyle \tau}}
\label{eg}
\end{equation}
\begin{equation}
\tilde{\bv{v}}=\frac{\bv{r}}{\mbox{t}}
\label{vg}
\end{equation}
\begin{equation}
\tilde{\mbox{n}}=\mbox{n}_0 e^{-\frac{\mbox{b}_n}{\displaystyle \tau}}
\label{ng}
\end{equation}
where $\tau=\sqrt{\mbox{t}^2-\bv{r}^2}$. For such a flow one can obtain
\begin{equation}
\mbox{S}^{\mu} =\frac{\tilde{\mbox{u}}^{\mu}\tilde{\varepsilon}}{\tau} \left( 4+\frac{\mbox{b}_e}{\tau} \right),
\label{smu}
\end{equation}
\begin{equation}
\mbox{S}_n=\frac{\tilde{\mbox{n}}}{\tau} \left( 3+\frac{\mbox{b}_n}{\tau} \right)
\label{sn}
\end{equation}
For numerical simulation we took the following parameters- equation of state $\mbox{p}=\mbox{c}_s^2 \varepsilon$, $\mbox{c}_s^2=1/3$, $\mbox{b}_e=1$, $\varepsilon_0=100$, $\mbox{b}_n=1$, $\mbox{n}_0=100$. Initial time is $\mbox{t}_0=5$ and final time is $\mbox{t}=8$. At moment of time $\mbox{t}_0$ we take all the quantities in region $\left| \bv{r} \right| < \mbox{t}_0-\alpha$, where $\alpha$ is small enough, being defined by relations (\ref{eg})-(\ref{ng}). Outside this region all the quantities and source term are put to zero. Results of numerical simulations are depicted on Fig~\ref{f:3source_yura}.

\section{Conclusions}

We presented a numerical code with a new Riemann solver. This solver is exact for one-dimensional flows. Exactness means that we use real wave patterns in order to obtain fluxes. Such an approach allows us to treat correctly vast range of velocities and configurations. Testing shows that the scheme works correctly in ultrarelativistic regimes and is capable of correct catching strong shock waves as well as strong rarefactions. In this sense our scheme is universal and can be applied to simulations of processes with large gradients of quantities and with large difference in velocities. Also, it account for possible
source terms in right hand side of hydrodynamics equations. And the fact that the method uses physical considerations makes it especially reliable.
On the other hand due to necessity of solving non-linear algebraical equations our code is more numerically expensive than the methods like HLLE or flux splitting even though no matrix decomposition is needed. But drastically increased efficiency of computers makes it possible to apply methods like ours.

\section*{Acknowledgments}

The research has been carried out within the scope of the ERG (GDRE): Heavy ions at ultra-relativistic energies - a European Research Group comprising IN2P3/CNRS, Ecole des Mines de Nantes, Universite de Nantes, Warsaw University of Technology, JINR Dubna, ITEP Moscow and Bogolyubov Institute for Theoretical Physics NAS of Ukraine. This work has been supported, in part, by Award No 017U000396 of Bureau of Physics and Astronomics Division of NASU, Fundamental Research State Fund of Ukraine, Agreement No. F25/718-2007 and Bilateral award DLR(Germany)-MESU (Ukraine) for UKR06/008 Project. PVT thanks Sam Falle and Serguei Komissarov for helpful discussions and for ideas that inspired this work. PVT also thanks Rob Coker and Anatoly Spitkovsky for comments that improved quality of the work. Special thanks are to Valery Zhdanov.

\label{lastpage}

\end{document}